\documentclass[journal]{IEEEtran}
\ifCLASSINFOpdf
\usepackage[dvipsnames]{xcolor}
\usepackage{subcaption}
  \usepackage[pdftex]{graphicx}
  \usepackage{multicol}
  \usepackage{multirow}
	\usepackage{hyperref}
  \graphicspath{{./figs/}, {./jpg/}}
  \DeclareGraphicsExtensions{.pdf,.eps}   
\else
\usepackage[dvips]{graphicx}   
\graphicspath{{../eps/}}    
\fi
\usepackage[cmex10]{amsmath}
\usepackage{mathrsfs}
\usepackage{amsfonts,amssymb,amsthm,calc}
\usepackage{xr}
\usepackage{algorithm}
\usepackage[noend]{algpseudocode}
\usepackage{dsfont}
\usepackage{bbm}
\usepackage{algorithm}
\usepackage[noend]{algpseudocode}
\usepackage{stfloats}
\usepackage{epstopdf}
\usepackage{tabularx}
\usepackage{physics}
\usepackage{mathtools}
\newlength{\mywidth}

\newtheorem{proposition}{Proposition}
\newtheorem{lemma}{Lemma}
\newtheorem{theorem}{Theorem}
\newtheorem{corollary}{Corollary}
\newtheorem{definition}{Definition}
\allowdisplaybreaks
\makeatletter
\newcommand{\multiline}[1]{%
	\begin{tabularx}{\dimexpr\linewidth-\ALG@thistlm}[t]{@{}X@{}}
		#1
	\end{tabularx}
}
\makeatother

\newcommand{\constfkd}{\mathcal{C}_{{F}_{\FDcoef}D}}

\newcommand{\constdd}{\mathcal{C}_{DD}}
\newcommand{\constff}{\mathcal{C}_{FF}}
\newcommand{\constfd}{\mathcal{C}_{FD}}
\newcommand{\constfL}{\mathcal{C}_{F}}
\newcommand{\constdL}{\mathcal{C}_{D}}
\newcommand{\constfLk}{\mathcal{C}_{F,{0_k}}}
\newcommand{\constdLk}{\mathcal{C}_{D,{1_k}}}

\newcommand{\constplusd}{\mathcal{C}_{(1+1)}}
\newcommand{\constonekd}{\mathcal{C}_{1_k}}
\newcommand{\constzerod}{\mathcal{C}_{0}}
\newcommand{\constzeroT}{\mathcal{C}\mathcal{C}_{0}}

\newcommand{\pulseduration}{{\color{black}T_p}}
\newcommand{\chippackt}{{\color{black}\xi}}
\newcommand{\wavepackt}{{\color{black}\xi}}
\newcommand{\fdcoef}{{\color{black}l}}
\newcommand{\FDcoef}{{\color{black}q}}

\begin{document}
\title{Quantum Spread-Spectrum  CDMA \\
Communication Systems: Mathematical Foundations}

\author{Mohammad~Amir~Dastgheib,~\IEEEmembership{Member,~IEEE,}~Jawad~A.~Salehi,~\IEEEmembership{Fellow,~IEEE}
	and~Mohammad Rezai
\thanks{M.~A.~Dastgheib is with the Department of Electrical Engineering, Sharif University of Technology, Tehran, Iran and also with the Sharif Quantum Center (SQC), Sharif University of Technology, Tehran, Iran (e-mail: sma.dastgheib@ee.sharif.edu).}
\thanks{J.~A.~Salehi is with the Department of Electrical Engineering, Sharif University of Technology, Tehran, Iran, Institute for Convergence Science \& Technology (ICST), Sharif University of Technology, Tehran, Iran and also with Sharif Quantum Center (SQC), Sharif University of Technology, Tehran, Iran (e-mail: jasalehi@sharif.edu).}
\thanks{M.~Rezai is with the Institute for Convergence Science \& Technology (ICST), Sharif University of Technology, Tehran, Iran and also with Sharif Quantum Center (SQC), Sharif University of Technology, Tehran, Iran (e-mail: mohammad.rezai@sharif.edu).}
}%



\maketitle
\begin{abstract}
    This paper describes the fundamental principles and mathematical foundations of quantum spread spectrum code division multiple access (QCDMA) communication systems. The evolution of quantum signals through the direct-sequence spread spectrum multiple access communication system is carefully characterized by a novel approach called the decomposition of creation operators. In this methodology, the creation operator of the transmitted quantum signal is decomposed into the chip-time interval creation operators each of which is defined over the duration of a chip. These chip-time interval creation operators are the invariant building blocks of the spread spectrum quantum communication systems. With the aid of the proposed chip-time decomposition approach, we can find closed-form relations for quantum signals at the receiver of such a quantum communication system. Further, the paper details the principles of narrow-band filtering of quantum signals required at the receiver, a crucial step in designing and analyzing quantum communication systems. We show that by employing coherent states as the transmitted quantum signals, the inter-user interference appears as an additive term in the magnitude of the output coherent (Glauber) state, and the output of the quantum communication system is a pure quantum signal. On the other hand, if the transmitters utilize particle-like quantum signals (Fock states) such as single photon states, entanglement and a spread spectrum version of the Hong-Ou-Mandel effect can arise at the receivers. The important techniques developed in this paper are expected to have far-reaching implications for various applications in the exciting field of quantum communication and signal processing.
\end{abstract}


\section{Introduction}
\IEEEPARstart{Q}{uantum} information technologies are emerging due to the unique and fundamental features of the quantum phenomena such as superposition, entanglement, and quantum interference. Similar to the age of conventional information technologies, where classical communications was the key enabler, quantum communications plays a major role in the era of quantum based information technologies \cite{wilde2013quantum,cariolaro2015quantum}. Compelling applications such as the transmission of quantum and classical information within the future quantum communication networks \cite{razavi2018introduction, bathaee2023quantum}, distributed quantum information processing networks \cite{lo2000classical}, quantum signal processing \cite{rezai2022fundamentals} and the ultimate secure communications through quantum key distribution \cite{chen2021integrated, moghaddam2021resource} have made quantum communications a very active field of research. 

Manipulating the shape of the quantum photon wavepackets is the central element of many quantum communication techniques. In this regard, the information can be encoded in quantum states of light, forming a quantum signal that can be transmitted through the quantum communication system \cite{cariolaro2015quantum}. High-speed electro-optical modulators can be utilized to arbitrarily modulate the amplitude and phase of temporal photon wavepackets \cite{keller2004continuous, specht2009phase, kues2019quantum}.

The spread spectrum technology is one of the most adopted techniques in classical communication systems due to multiple access and interference management capability and low probability of intercept \cite{simon2002spread}. The basic concept of spread spectrum communication systems is to broaden the spectrum of the transmitted signals by the processing gain factor utilizing a spreading sequence. This process suppresses the signal and hides it below the noise. Decoding the transmitted signal with the same spreading sequence at the receiver increases the signal-to-noise ratio and reconstructs the signal for the intended user.

Direct-sequence spread spectrum methods have been applied to single photons \cite{belthangady2010hiding}. A quantum spread spectrum multiple access scheme based on optical circulators is described in \cite{garcia2014quantum}. Although the mentioned papers discuss the possibility of applying the spread spectrum techniques to quantum signals, none has rigorously characterized the spread spectrum communication systems in terms of the mathematical language of the quantum mechanics and technology. 
With this background, the essential steps toward describing the quantum mechanical principles of spectral domain quantum CDMA (QCDMA) communication systems are taken in \cite{rezai2021quantum}. 

This paper discusses the mathematical foundations of direct sequence quantum spread spectrum CDMA systems by the temporal domain encoding of the transmitted quantum signals.
The rest of the paper is organized as follows. Section \ref{sec:transmitter} describes the operation of a quantum spread spectrum transmitter, including the description of the quantum signals and the spreading operator. The quantum spread spectrum communication receiver is described in section \ref{sec:receiver}, where the signals are first decoded using the proper signature sequence. After applying the decoding operator, a narrow-band filter (integrator) is used to reject the out-of-band inter-user interference. This section develops a practical description of quantum filters for quantum communication systems from a novel perspective. A photon counting measurement is considered at the final state, giving the photon statistics of the received quantum signal. The overall quantum spread spectrum communication system is described in section \ref{sec:cdma}. This section describes the effect of the quantum broadcasting channel, i.e., a star coupler, and derives the evolution of quantum signals in the considered multiple access quantum communication system. In all of the mentioned sections, the system is carefully analyzed by the proposed chip-time interval decomposition approach that is introduced in subsections \ref{sec::timeintervalcreation} and \ref{sec::timedecomposition}. This approach gives closed-form expressions for the quantum signals that propagate through the system and leads to the chip-time decomposition methodology for the description of the quantum spread spectrum CDMA communication systems in subsection \ref{sec::timedecompositionpicture}. Section \ref{sec:two} applies the proposed methods to a two-user spread spectrum quantum communication system and presents the simulation results. Section \ref{sec:conclusion} concludes the paper. 

\section{Quantum Spread Spectrum Transmitter}\label{sec:transmitter}
\begin{figure}[!tb]
    \centering
    \includegraphics[width=\linewidth]{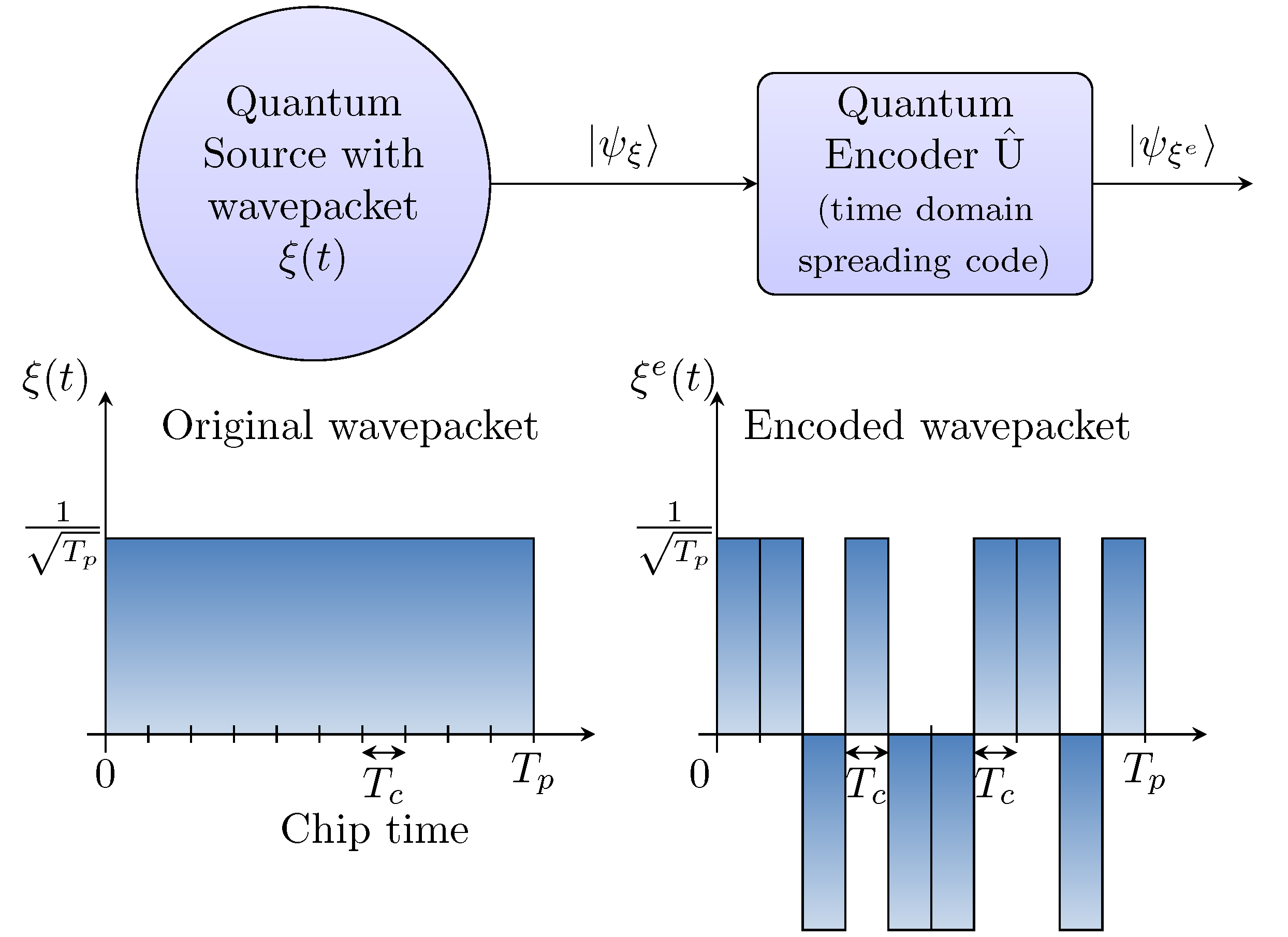}\caption{Effect of quantum encoder on the transmitted quantum communication signal with photon wavepacket $\wavepackt(t)$ with processing gain (code length) of $N_c= \frac{\pulseduration}{T_c}$.}
    \label{fig:encoder}
\end{figure}

In this section, we describe the application of spread spectrum technology to quantum communication signals in terms of Dirac's formalism. A quantum communication signal is mathematically represented as a quantum state of light. At the transmitter, shown in Fig. \ref{fig:encoder}, this quantum light goes through an electro-optical modulator that encodes a pseudo-random sequence into the photon-wavepacket of the transmitted quantum signal. We show that analogous to the classical spread spectrum communication signals, the temporal behavior of this photon-wavepacket can be expressed based on the concept of chip-times. In this regard, we present a temporal decomposition for the creation operators of the transmitted quantum communication signals. 

\subsection{Quantum Signal Description}
Quantum signals generated by a user's transmitter are in the form of quantum light pulses. The temporal shape of these quantum signals can be described by the so-called photon-wavepacket $\wavepackt(t)$.
A pure quantum signal, $\ket{\psi_\wavepackt}$ with photon wavepacket $\wavepackt(t)$ can thus be written as a superposition of number states with photon-wavepacket $\wavepackt(t)$, i.e. $\ket{n_\wavepackt}$ \cite{rezai2021quantum}
\begin{equation}\label{eq:generalState}
    \ket{\psi_\wavepackt} = \sum_n c_n \ket{n_\wavepackt} = f(\hat{a}^\dagger_\wavepackt)\ket{0},
\end{equation} 
with $\ket{0}$ denoting the vacuum state and the creation operator of the quantum light pulse, i.e., $\hat{a}^\dagger_\wavepackt$, is defined based on its corresponding photon-wavepacket
\begin{align}
    \hat{a}^\dagger_\wavepackt&=  \int dt \wavepackt(t) \hat{a}^\dagger(t) = \int d\omega \bar{\wavepackt}(\omega) \hat{a}^\dagger(\omega),
\end{align}  
where  $\hat{a}^\dagger(t)$ and $ \hat{a}^\dagger(\omega)$ are the continuous mode creation operators in time, $t$, and frequency, $\omega$, that are related to each other according to the Fourier transform relationship. It is more appealing to use the engineering convention\footnote{Since, we deal with filters and signal processing throughout this paper, it is more convenient to use the engineering convention for representing the imaginary numbers. The notions are thus consistent with the physics convention using $j=-i$ (see \cite[Page 17]{hagelstein2004introductory}).  } for imaginary numbers throughout this paper. Thus, we may write \cite{loudon2000quantum}
\begin{align}
	\hat{a}(\omega) &= \frac{1}{\sqrt{2\pi}}\int dt \hat{a}(t)e^{-j\omega t}.
\end{align} 
Based on the above definition,  the spectral shape of the quantum signal, i.e. $\bar{\wavepackt}(\omega)$ is the Fourier transform of the temporal wavepacket $\wavepackt(t)$, that is $\bar{\wavepackt}(\omega)=\mathcal{F}\lbrace \wavepackt(t) \rbrace$.

Since $\wavepackt(t)$ and $\bar{\wavepackt}(\omega)$ represent the quantum probability amplitudes, they are normalized
\begin{align}
    \int |\wavepackt(t)|^2 dt = 1, \qquad   \int |\bar{\wavepackt}(\omega)|^2 d\omega = 1.
\end{align} 

Consider a quantum baseband rectangular signal of duration $\pulseduration$. For such a quantum communication signal, we may obtain the corresponding creation operator as 
\begin{equation}
    \hat{a}^\dagger_\wavepackt= \int_{0}^{\pulseduration} dt \frac{1}{\sqrt{\pulseduration}} \hat{a}^\dagger(t).
\end{equation} 

Any pure quantum communication signal with a particular wavepacket such as $\wavepackt(t)$ can be expressed in terms of the general equation in (\ref{eq:generalState}). 

\subsubsection{Example 1} One choice for the transmitted quantum signals is photon number states $\ket{n_\wavepackt}$ with wavepacket $\wavepackt(t)$, and single photon states $\ket{1_\wavepackt}$ as a special case, defined as
\begin{align}
    \ket{n_\wavepackt} \coloneqq  \frac{(\hat{a}_{\wavepackt}^\dagger)^n}{\sqrt{n!}}\ket{0},
\end{align}
with $\ket{1_\wavepackt}=\hat{a}^\dagger_\wavepackt\ket{0}$ as a single photon state.

\subsubsection{Example 2} As another example of quantum light, coherent (Glauber) states of the following form with the same wavepacket, $\wavepackt(t)$, can be used for data transmission
\begin{align}
    \ket{\alpha_\wavepackt}&\coloneqq e^{-\frac{|\alpha|^2}{2}}\sum_{n=0}^{\infty}\frac{\alpha^n}{\sqrt{n!}}\ket{n_\wavepackt}=e^{-\frac{|\alpha|^2}{2}}\sum_{n=0}^{\infty}\frac{\alpha^n(\hat{a}_{\wavepackt}^\dagger)^n}{n!}\ket{0}\nonumber\\
    &= e^{-\frac{|\alpha|^2}{2}}e^{\alpha \hat{a}_{\wavepackt}^\dagger}\ket{0}=e^{-\frac{|\alpha|^2}{2}}e^{\alpha \hat{a}_{\wavepackt}^\dagger}e^{\alpha^* \hat{a}_{\wavepackt}}\ket{0}\nonumber\\
    &= D(\alpha_\wavepackt)\ket{0} ,
\end{align}
where $D(\alpha_\wavepackt)\coloneqq e^{\alpha \hat{a}_{\wavepackt}^\dagger - \alpha^*\hat{a}_{\wavepackt} }$ is the displacement operator and the last equality is obtained using $ \text{exp}(\hat{A}+\hat{B}) = \text{exp} (\hat{A})\text{exp} (\hat{B})\text{exp} (-\frac{1}{2}\comm{\hat{A}}{\hat{B}})$. Also note that $e^{-\alpha^* \hat{a}_{\wavepackt}}\ket{0} = \ket{0}$.

\subsection{Quantum Spreading Operator}
In a quantum communication system, electro-optical modulators can be used to encode user's spreading sequence on the temporal shape of the transmitted quantum signals \cite{hayat2012multidimensional}. The mathematical expression for such a transformation is obtained by applying a unitary operator denoted by $\hat{\mathrm{U}}$ to the original uncoded quantum signal \cite{rezai2021quantum}.
Therefore, we describe the effect of the encoder by the  temporal phase shifting quantum spreading operator
\begin{align}
    \hat{\mathrm{U}} = e^{+{j}{\textstyle\int}_0^\pulseduration dt \theta(t)\hat{a}^\dagger(t)\hat{a}(t)},
\end{align}
where $\theta(t)$ is the encoder's phase shift and the effect of the spreading operator $\hat{\mathrm{U}}$ on the basis states $\ket{n_\wavepackt}$ can be obtained in analogy to \cite{rezai2021quantum} as
\begin{align}
    \hat{\mathrm{U}}\ket{n_\wavepackt} &= \ket{n_{\wavepackt^e}},
\end{align}
where $\wavepackt^e(t)=\wavepackt(t)e^{{j}\theta(t)}$ is the encoded photon-wavepacket. Let us denote the effect of phase shift as $\lambda(t)\coloneqq e^{{j}\theta(t)}$. Hence the quantum phase-shifting operator transforms the temporal shape of the photon wavepacket from $\wavepackt(t)$ to $\wavepackt^e(t) = \lambda(t)\wavepackt(t)$.

Applying the unitary operator to the arbitrary state $\ket{\psi_\wavepackt}$ gives
\begin{align}
    \hat{\mathrm{U}}\ket{\psi_\wavepackt} &=  \sum_n c_n \hat{\mathrm{U}}\ket{n_\wavepackt} = \sum_n c_n \ket{n_{\wavepackt^e}} = \ket{\psi_{\wavepackt^e}}.
\end{align}
In the context of quantum direct sequence spread spectrum systems, we assume that the time-domain phase shifts are designed according to a random binary sequence. We obtain the spread spectrum quantum signals by considering the values of $0$ and $\pi$ for the phase shifts $\theta(t)$.
We have assumed that the random phase shift is constant for the duration of a chip, and the chips are chosen such that the mean absolute square of the time-domain wavepacket is the same for all chips. Hence if the boundaries of the chips are given by $0=t_0, t_1, \cdots, t_{N_c}=\pulseduration$, then we may write:
\begin{equation}\label{eq:normalWavepacket}
    \int_{t_k}^{t_{k+1}}|\wavepackt(t)|^2 dt = \frac{1}{N_c}.
\end{equation}
From a quantum mechanical perspective, the above equation means that the probability that a signal is in interval $[t_k, t_{k+1})$ is $\frac{1}{N_c}$. Assuming that rectangular pulses are used for data transmission, $ t_{k+1}-t_k=T_c$ is called a chip-time. For simplicity, we consider quantum signals with a rectangular pulse shape; therefore, we may write $t_k=kT_c$.

The phase shifts in the spread spectrum encoding are fixed in the chip-time interval of $[t_k, t_{k+1})$. Thus we define $\theta_k(t)$ as
\begin{align}
	\theta_k(t) \coloneqq  \left\lbrace
	\begin{matrix}
		\theta_k & t\in[t_k, t_{k+1})\\
		0 & t\notin[t_k, t_{k+1})
	\end{matrix}\right.,
\end{align}
where $\theta_k\in\lbrace 0, \pi\rbrace$ is the constant phase shift in the chip-time interval of $[t_k, t_{k+1})$.

It is easier to work with amplitude coeffiecients rather than phase shifts. Thus we define
\begin{align}
	\lambda(t) = e^{{j}\theta(t)}.
\end{align}
For $t\in [t_k, t_{k+1})$, $\theta(t)=\theta_k$. Therefore define $\lambda_k \coloneqq \lambda\left(\frac{t_k+t_{k+1}}{2}\right) = e^{{j}\theta_k}$. Hence, for this encoding scheme, we may write:
\begin{equation}
    \lambda(t) = \sum_{k=0}^{N_c-1}\lambda_k \text{rect}\left(\frac{t-t_k}{T_c}\right),
\end{equation}
where $\text{rect}(t)$ is a rectangular pulse of duration one starting at $t=0$ and $\lambda_k\in\lbrace -1,1\rbrace$ for $\theta_k\in\lbrace 0, \pi\rbrace$. Hence, the code sequence can be denoted by $\Lambda=(\lambda_0, \lambda_1,\ldots, \lambda_{N_c-1})$.

\subsection{Chip-time Creation Operators}
\label{sec::timeintervalcreation}
A common technique in classical communications is to describe a spread spectrum signal in terms of the chips that are the system's building blocks. In this subsection, we define the building blocks of a quantum spread spectrum communication system that we call chip-time interval creation operators. 
\begin{figure}[!tb]
    \centering
    \includegraphics[width=\linewidth]{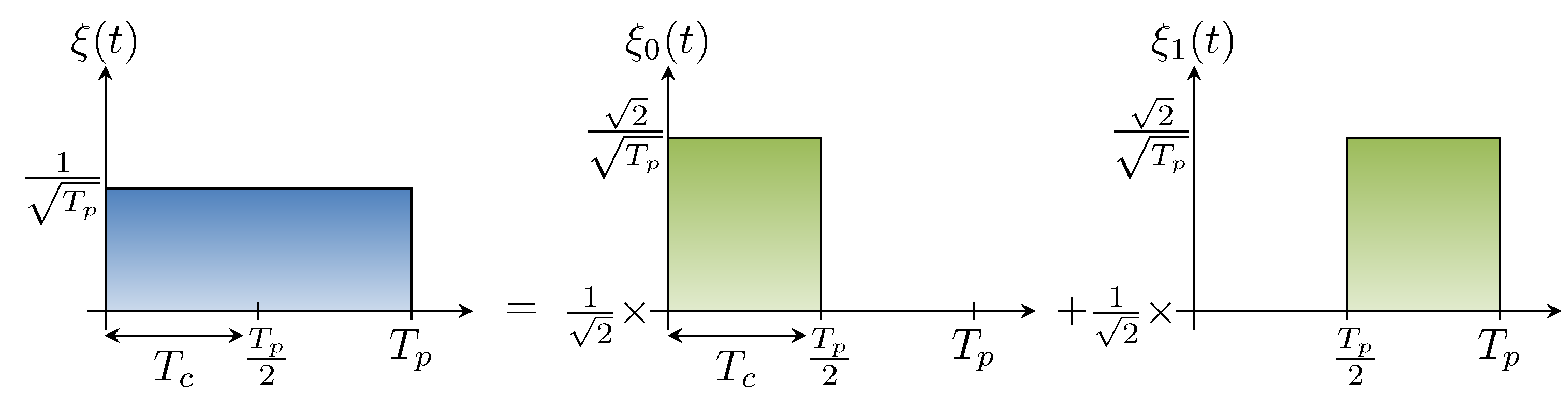}\caption{Temporal decomposition of photon-wavepackets for $N_c=2$. A photon-wavepackets, $\wavepackt(t)$, can be viewed as a combination of chip-time wavepackets, $\chippackt_k(t)$.}
    \label{fig:wavepackets}
\end{figure}

To define the quantum counterpart of the spread spectrum chips, we start by dividing the wavepacket into several intervals of the same probability. Based on (\ref{eq:normalWavepacket}) we have:
\begin{equation}
    \int_{t_k}^{t_{k+1}}N_c|\wavepackt(t)|^2 dt = 1\Rightarrow \int_{t_k}^{t_{k+1}}|\sqrt{N_c}\wavepackt(t)|^2 dt = 1.
\end{equation}
From the above equation, we may define the normalized photon-wavepacket of each chip-time shown in Fig. \ref{fig:wavepackets} and its corresponding chip-time interval creation operators.

\begin{definition}
	The chip-time interval creation operators for a wave-packet, $\wavepackt(t)$ are defined as
	\begin{align}
		\label{eq:timeIntervalDef}
		\hat{a}_{\chippackt_k}^\dagger & \coloneqq  \int_{t_k}^{t_{k+1}} dt \chippackt_k(t) \hat{a}^\dagger(t),
	\end{align}
	where $\chippackt_k(t)$ is a wavepacket constructed from $\wavepackt(t)$ that has value zero for $t\notin [t_k, t_{k+1})$ and is defined as
	\begin{align}
		\chippackt_k(t) = \left\lbrace
		\begin{matrix}
			\sqrt{N_c}\wavepackt(t) & t\in[t_k, t_{k+1})\\
			0 & t\notin[t_k, t_{k+1})
		\end{matrix}\right. .\label{eq:wavepacketTimeIntervalgeneral}
	\end{align}
\end{definition}
As an important example, for the rectangular wavepacket of Fig. \ref{fig:wavepackets}, we have
\begin{align}
	\chippackt_k(t) = \left\lbrace
	\begin{matrix}
		\sqrt{\frac{N_c}{\pulseduration}} & t\in[t_k, t_{k+1})\\
		0 & t\notin[t_k, t_{k+1})
	\end{matrix}\right.\label{eq:wavepacketTimeInterval}
\end{align}

\begin{proposition}
	The chip-time interval creation operators have the following properties
	\begin{enumerate}
		\item The corresponding wavepacket satisfies the required normalization property of a photon-wavepacket
		\begin{align}
			\int_0^\pulseduration |\chippackt_k(t)|^2 dt = 1.\label{eq:normalTimeInterval}
		\end{align}
		\item Nonoverlapping chip-time interval creation operators commute.
		\begin{equation}
			\comm{\hat{a}_{\chippackt_l}}{\hat{a}_{\chippackt_k}^\dagger}=\delta_{lk},
		\end{equation} 
		where $\delta_{lk}$ is the Kronecker delta, where ${\hat{a}_{\chippackt_k}^\dagger = \int_{t_k}^{t_{k+1}} dt \chippackt_k(t) \hat{a}^\dagger(t)}$.
	\end{enumerate}
\end{proposition}
\begin{proof}
	See Appendix \ref{appendix:commutation}.
\end{proof}

Now we can decompose the overall creation operator $\hat{a}_{\wavepackt}^\dagger $ to its temporal building blocks according to the following theorem.

\begin{theorem}[Chip-time interval creation operators]
	A quantum field creation operator, $\hat{a}_{\wavepackt}^\dagger$, with wavepacket $\wavepackt(t)$ can be decomposed into chip-time interval creation operators according to the following relation
	\begin{align}
		\hat{a}_{\wavepackt}^\dagger & = \frac{1}{\sqrt{N_c}}\sum_{k=0}^{N_c-1} \hat{a}_{\chippackt_k}^\dagger.
	\end{align}
\end{theorem}
\begin{proof}
	See Appendix \ref{appendix:commutation}.
\end{proof}

The chip-time interval creation operators give us insight on how the spreading operator acts on different quantum states and how quantum signals evolve through a spread spectrum communication system. This novel interpretation leads to the definition of \textit{chip-time number states}, 
\begin{equation}
    \ket{n_{\chippackt_k}} = \frac{(\hat{a}_{\chippackt_k}^\dagger)^n}{\sqrt{n!}}\ket{0},
\end{equation} 
which indicates a quantum state with exactly $n$ photons in interval $[t_k, t_{k+1})$ and wavepacket $\chippackt_k(t)$.

\begin{figure}[!tb]
    \centering
    \includegraphics[width=0.95\linewidth]{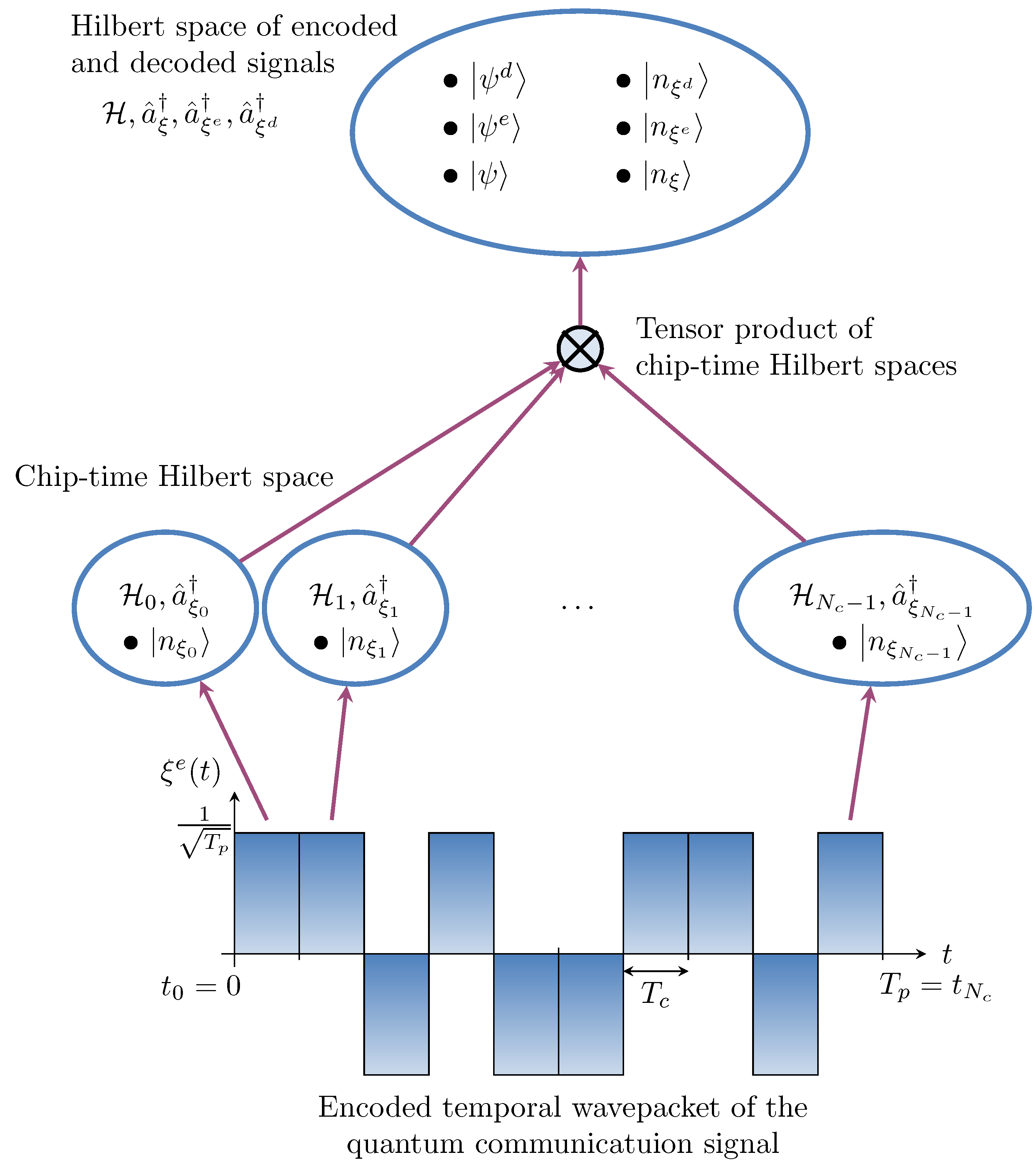}\caption{Hilbert space view of chip-time interval decomposition. A chip-time Hilbert space is associated with each chip-time interval of the quantum communication signal. The encoded and decoded quantum signals can be represented in the tensor product space of these chip-time Hilbert spaces.}
    \label{fig:hilbert}
\end{figure}

\subsection{Chip-Time Hilbert Spaces}
As discussed in the previous subsection, the chip-time interval creation operators are associated with chip-time number states. As we show throughout the paper, these chip-time number states form an invariant basis for the quantum signals' evolution through the spread spectrum communication system. 

The $k$th chip-time number states, $\ket{n_{\chippackt_k}}, \ n=0,1,\cdots$, naturally form a basis for a Hilbert space\footnote{The Hilbert space presented here corresponds to Fock space in the field of quantum many-particle systems \cite{negele1998quantum}.}, $\mathcal{H}_k$. We call this space a chip-time Hilbert space. Number states $\ket{n_{\chippackt_k}}$ associated with the creation  operator $\hat{a}^\dagger_{\chippackt_k}$ form a basis for $\mathcal{H}_k$. Figure \ref{fig:hilbert} shows how the spread spectrum quantum signals can be described in accordance with these chip-time Hilbert spaces. As seen in this figure, a Hilbert space is associated with each chip-time of the quantum spread spectrum signal.  The overall signal's wavepacket can always be decomposed into these chip-time building blocks. Each chip-time Hilbert space describes a single chip of the quantum signal.  Since chip-times are non-overlapping, the overall Hilbert space representing the original, encoded, and decoded  spread spectrum quantum signals is the tensor product of the chip-time Hilbert spaces.

\subsection{Chip-time interval Decomposition of Quantum Signals}
\label{sec::timedecomposition}
With the aid of the chip-time interval creation operators, we can decompose different quantum signals in time. Coherent states and number states are among the most important quantum signals. For coherent states, the decomposition has a tensor product form: 
\begin{proposition}
    The coherent state $\ket{\alpha_\wavepackt}$ can be written as the tensor product of coherent states on different intervals.
    \begin{align}
        D(\alpha_\wavepackt) &= \prod_{k=0}^{N_c-1} D\left(\frac{\alpha_{\chippackt_k}}{\sqrt{N_c}}\right) \\
        \ket{\alpha_\wavepackt} &= \prod_{k=0}^{N_c-1} \ket{\frac{\alpha_{\chippackt_k}}{\sqrt{N_c}}},
    \end{align}
	where $\ket{\frac{\alpha_{\chippackt_k}}{\sqrt{N_c}}}=D\left(\frac{\alpha_{\chippackt_k}}{\sqrt{N_c}}\right)\ket{0}=e^{-\frac{|\alpha|^2}{2N_c}}e^{\frac{\alpha}{\sqrt{N_c}}\hat{a}_{\wavepackt}^\dagger}\ket{0}$.
\end{proposition}
\begin{proof}
    See Appendix \ref{appendix:coherent}.
\end{proof}
For number states, the decomposition results in a superposition.
\begin{proposition}\label{prep:numberState}
    The chip-time interval decomposition of number state $\ket{n_\wavepackt} $ can be expressed in terms of the following superposition of different intervals.
    \begin{align}
        \ket{n_\wavepackt} 
        &= \sum_{n_0+n_1+\cdots+n_{N_c-1}=n} \mathcal{C}_n(n_0,n_1,\ldots,n_{N_c-1})\prod_{k=0}^{N_c-1} \ket{n_{k,\chippackt_k}}\nonumber\\
        &= \sum_{n_0+n_1+\cdots+n_{N_c-1}=n} \mathcal{C}_n(n_0,n_1,\ldots,n_{N_c-1})\\
        &\qquad\qquad\qquad\qquad\times \ket{n_{0,\chippackt_0}, n_{1,\chippackt_1}, \cdots, n_{N_c-1,\chippackt_{N_c-1}}}\nonumber,
    \end{align}
    where $\ket{n_{0,\chippackt_0}, n_{1,\chippackt_1}, \cdots, n_{N_c-1,\chippackt_{N_c-1}}}=\ket{n_{0,\chippackt_0}}\ket{n_{1,\chippackt_1}}\cdots \ket{n_{N_c-1,\chippackt_{N_c-1}}}$ and  $\ket{n_{k,\chippackt_k}}$ means that the state has $n_k$ photons with wavepacket $\chippackt_k(t)$ in interval $[t_k, t_{k+1})$ and its probability in the superposition is $|\mathcal{C}_n(n_0,n_1,\ldots,n_{N_c-1})|^2$, where
    \begin{align}
        &\mathcal{C}_n(n_0,n_1,\ldots,n_{N_c-1}) \coloneqq  \\
        &\qquad\frac{1}{\sqrt{N_c^n}} \sqrt{{n\choose n_0,n_1,\cdots,n_{N_c-1}}},\nonumber
    \end{align}
    and
    \begin{align}
        {n\choose n_0,n_1,\cdots,n_{N_c-1}} = \frac{n!}{n_0!n_1!\cdots n_{N_c-1}!}.
    \end{align}
\end{proposition}
\begin{proof}
    See Appendix \ref{appendix:number}.
\end{proof}

For $n=1$, i.e., the single photon quantum signal, we have:
\begin{align}
    \ket{1_\wavepackt} &= \frac{1}{\sqrt{N_c}} \sum_{k=0}^{N_c-1} \ket{1_{\chippackt_k}},
\end{align}
which means that a single photon with wavepacket $\wavepackt(t)$ can be viewed as a superposition of single photons at different chips with probability $\frac{1}{N_c}$.

Since number states $\ket{n_\wavepackt}$ form a basis for the Hilbert space of quantum signals with photon-wavepacket $\wavepackt(t)$, one can obtain the temporal decomposition of any arbitrary quantum signal by substituting Proposition \ref{prep:numberState} in $\ket{\psi_\wavepackt} = \sum_n c_n \ket{n_\wavepackt}$.
\subsection{Chip-time interval Decomposition of the Spreading Operator}
In order to further investigate the effect of the spreading operator on the quantum signals, we decompose $\hat{\mathrm{U}}$ into its constructing chip-time interval operators. The following theorem shows that the spreading operator can be decomposed into operators that each act on an individual chip-time.

\begin{theorem}
	The chip-time interval decomposition of quantum spreading operator, $\hat{\mathrm{U}}$, results in the following expression\footnote{Some authors use $\otimes$ instead of $\prod$ to represent the tensor-product. See Appendix \ref{appendix:tensor} for related discussions.}
	\begin{align}
		\hat{\mathrm{U}} &= \prod_{k=0}^{N_c-1}\hat{\mathrm{U}}_k,
	\end{align}
	where
	\begin{equation}
		\hat{\mathrm{U}}_k \coloneqq  \exp({j}\int_{t_k}^{t_{k+1}} dt \theta_k\hat{a}^\dagger(t)\hat{a}(t))
	\end{equation}
	has the following property
	\begin{align}
		\comm{\hat{\mathrm{U}}_k}{f(\hat{a}_{\chippackt_l}^\dagger)} &= 0,\qquad l\neq k.
	\end{align}
\end{theorem}
\begin{proof}
	See Appendix \ref{appendix:unitary:commute}.
\end{proof}

\subsection{Chip-time interval Decomposition of Spread Spectrum Quantum Signals}
In this subsection, we obtain the chip-time interval decomposition of the spread spectrum quantum signals by applying the spreading operator to the transmitted quantum signals. As a result, we demonstrate that a spread spectrum quantum signal can be characterized by chip-time interval creation operators.

\begin{figure}[!tb]
	\centering
	\includegraphics[width=0.7\linewidth]{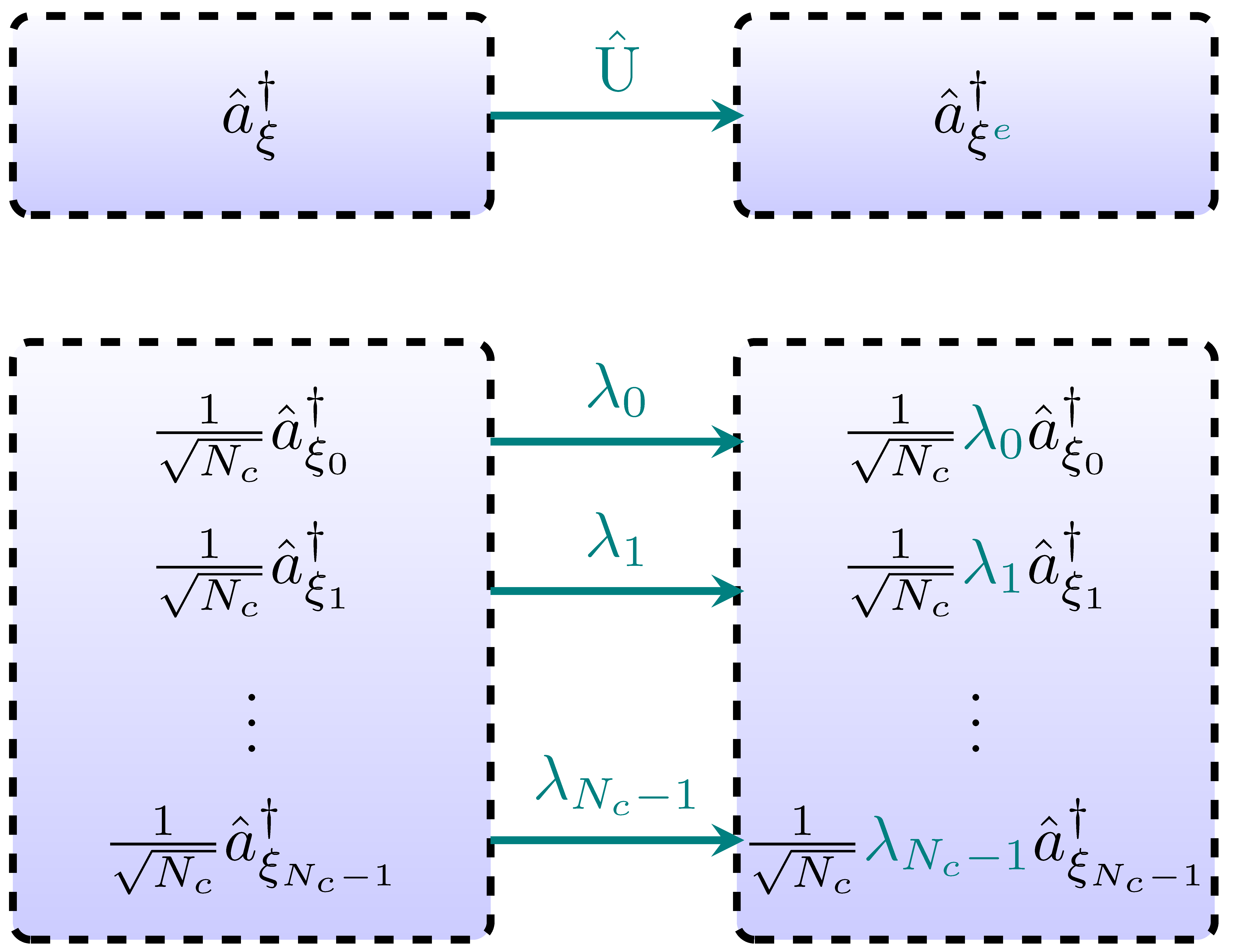}\caption{Effect of spreading operator on the creation operators. The spreading operator converts $\wavepackt(t)$ to $\wavepackt^e(t)$. According to Theorem \ref{thm:spreading}, this evolution is equivalent to applying the code sequence on the individual chip time interval creation operators.}
	\label{fig:spreading}
\end{figure}
\begin{theorem}
	\label{thm:spreading}
	The effect of spreading operator on the creation operator of a quantum signal with wavepacket $\wavepackt(t)$ is given by the following chip-time interval decomposition
	\begin{align}
		\hat{a}_{\wavepackt^e}\coloneqq\hat{\mathrm{U}}\hat{a}^\dagger_\wavepackt \hat{\mathrm{U}}^\dagger= \frac{1}{\sqrt{N_c}}\sum_{k=0}^{N_c-1} \lambda_k \hat{a}_{\chippackt_k}^\dagger.
	\end{align}
\end{theorem}
\begin{proof}
	See Appendix \ref{appendix:unitary:creation}.
\end{proof}

From Theorem \ref{thm:spreading}, illustrated in Fig. \ref{fig:spreading}, we obtain the chip-time interval decomposition of quantum signals in the following propositions.
\begin{proposition}
    The spread spectrum coherent state obtained by applying the spreading unitary transformation $\hat{\mathrm{U}}$ on $\ket{\alpha_\wavepackt}$, i.e., $\ket{\alpha_{\wavepackt^e}}$, can be written in the following tensor product form
    \begin{align}
        \hat{\mathrm{U}}\ket{\alpha_{\wavepackt}} = \ket{\alpha_{\wavepackt^e}} &= \prod_{k=0}^{N_c-1} \ket{\frac{\lambda_k\alpha_{\chippackt_k}}{\sqrt{N_c}}}.
    \end{align}
\end{proposition}
\begin{proof}
    See Appendix \ref{appendix:sscoherentEn}.
\end{proof}
\begin{proposition}
	\label{prop:spreadnumber}
    The spread spectrum number state obtained by applying the spreading unitary transformation $\hat{\mathrm{U}}$ on $\ket{n_\wavepackt}$, i.e. $\ket{n_{\wavepackt^e}}$, can be written in the following superposition form
    \begin{align}
        \hat{\mathrm{U}}\ket{n_\wavepackt} &=  \ket{n_{\wavepackt^e}}= \sum_{n_0+n_1+\cdots+n_{N_c-1}=n} \mathcal{C}_n( n_0,n_1,\ldots,n_{N_c-1})\nonumber\\
        &\qquad\qquad\qquad\qquad\times\prod_{k=0}^{N_c-1}(\lambda_k)^{n_k}\ket{n_{\chippackt_k}}.
    \end{align}
\end{proposition}
\begin{proof}
    See Appendix \ref{appendix:ssnumber:encoding}.
\end{proof}
As an example, we have simplified the results of Proposition \ref{prop:spreadnumber} for $n=1,2,3$ in Appendix \ref{appendix:ssnumber:encoding}.

For $n=1$ we have:
\begin{align}
    \hat{\mathrm{U}}\ket{1_\wavepackt} &= \ket{1_{\wavepackt^e}}= \frac{1}{\sqrt{N_c}} \sum_{k=0}^{N_c-1} \lambda_k\ket{1_{\chippackt_k}}.
\end{align}

For $n=2$ we obtain:
\begin{align}
    \hat{\mathrm{U}}\ket{2_\wavepackt} = \ket{2_{\wavepackt^e}}= \frac{1}{N_c}\sum_{k_0=0}^{N_c-1}&\ket{2_{\chippackt_{k_0}}}\\
	+\frac{1}{N_c} \underset{\text{s.t. }k_1>k_0}{\sum_{k_1=0}^{N_c-1}\sum_{k_0=0}^{N_c-1}} \sqrt{2}\lambda_{k_0}\lambda_{k_1}&\ket{1_{\chippackt_{k_0}}}\ket{1_{\chippackt_{k_1}}}.\nonumber
\end{align}

For $n=3$:
\begin{align}
    \hat{\mathrm{U}}\ket{3_\wavepackt} = \ket{3_{\wavepackt^e}}= \frac{1}{\sqrt{N_c^3}}\sum_{k_0=0}^{N_c-1}\lambda_{k_0} &\ket{3_{\chippackt_{k_0}}}\\
	+\frac{1}{\sqrt{N_c^3}} \sum_{k_1=0}^{N_c-1}\sum_{k_0=0}^{N_c-1} \sqrt{3} \lambda_{k_0}&\ket{1_{\chippackt_{k_0}}}\ket{2_{\chippackt_{k_1}}}\nonumber\\
	+\frac{1}{\sqrt{N_c^3}} \underset{\text{s.t. }k_2>k_1>k_0}{\sum_{k_2=0}^{N_c-1}\sum_{k_1=0}^{N_c-1}\sum_{k_0=0}^{N_c-1}} \sqrt{3!} \lambda_{k_0}\lambda_{k_1}\lambda_{k_2}&\ket{1_{\chippackt_{k_0}}}\ket{1_{\chippackt_{k_1}}}\ket{1_{\chippackt_{k_2}}}.\nonumber
\end{align}

We can observe that for number states, the spreading operator encodes the random sequence into the superposition coefficients of the chip-time interval number states. On the other hand, the same spreading operator results in separable chip-time encoded states for the coherent transmitted quantum signal.

\begin{figure*}[!tb]
    \centering
    \includegraphics[width=\linewidth]{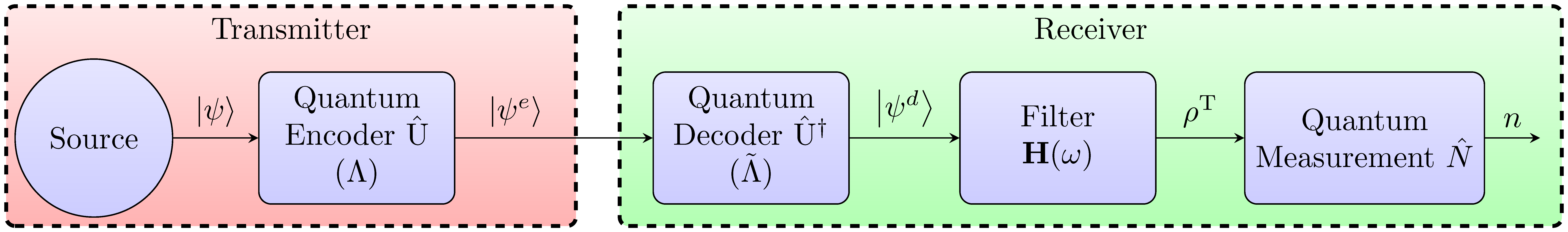}\caption{A point-to-point quantum spread spectrum communication system.}
    \label{fig:system1}
\end{figure*}

\section{Quantum Spread Spectrum Receiver}\label{sec:receiver}
In order to investigate the basic operation of the quantum spread spectrum receiver, consider a point-to-point communication scenario depicted in Fig. \ref{fig:system1}. The sequence $\Lambda$ encodes the transmitted quantum signals in this scenario. Post encoding, the signals are decoded by another sequence $\tilde{\Lambda}$. Then a narrow-band filter is applied so that the correctly decoded signal arrives at the receiver end.  The narrow-band filter rejects the out of band incorrectly decoded signals, as they have higher bandwidth due to the use of spread spectrum technology.

The typical point-to-point quantum spread spectrum communication system of Fig. \ref{fig:system1} comprises a quantum transmitter, quantum spreading operator (encoder)  with sequence $\Lambda$, quantum despreading operator (decoder)  with sequence $\tilde{\Lambda}$, quantum filter, and a photo-detector (quantum measurement). The quantum transmitter associates a pure quantum state to each transmitted data symbol. After that, the spreading operator $\hat{\mathrm{U}}$ applies a unitary transformation to the transmitted quantum signal. This direct sequence code multiplier may be realized using electro-optical modulators to apply the required random temporal phase shifts. The signal is then despread (decoded) at the receiver by applying the inverse of the quantum spreading operator $\hat{\mathrm{U}}^\dagger$. The signal then passes through an optical narrow-band filter. The desired output of the filter is possibly a mixed quantum signal and is denoted by $\rho^{\text{T}}$. In order to obtain the photon statistics of the received signal, we consider an ideal photo-detector after the filter. The photon counting detector is described by the number operator $\hat{N}$. We show that decoding the quantum signal with a wrong code sequence results in noise due to unmatched spreading sequences at the receiver, while the original quantum signal can be recovered by decoding with the correct code. 

\subsection{Decoding the Spread Spectrum Quantum Signals}
The description of decoded sequence can be obtained according to the proposed temporal decomposition framework.
\begin{proposition}
    Decoding a spread spectrum coherent state $\ket{\alpha_{\wavepackt^e}} = \prod_{k=0}^{N_c-1} \ket{\frac{\lambda_k\alpha_{\chippackt_k}}{\sqrt{N_c}}}$ with spreading sequence $\Lambda $ and despreading sequence $\tilde{\Lambda}$ gives the following state
    \begin{align}
        \ket{\alpha_{\wavepackt^d}} &= \prod_{k=0}^{N_c-1} \ket{\frac{\tilde{\lambda}_k\lambda_k\alpha_{\chippackt_k}}{\sqrt{N_c}}}.
    \end{align}
\end{proposition}
\begin{proof}
    See Appendix \ref{appendix:sscoherentDe}.
\end{proof}

\begin{proposition}
    Decoding a spread spectrum number state $\ket{n_{\wavepackt^e}}$ with spreading sequence $\Lambda $ and decoding sequence $\tilde{\Lambda}$ gives the following state
    \begin{align}
        \ket{n_{\wavepackt^d}} &= \sum_{n_0+n_1+\cdots+n_{N_c-1}=n} \mathcal{C}_n( n_0,n_1,\ldots,n_{N_c-1})\\ &\qquad\qquad\qquad\qquad\times\prod_{k=0}^{N_c-1}(\tilde{\lambda}_k\lambda_k)^{n_k}\ket{n_{\chippackt_k}}.\nonumber
    \end{align}
\end{proposition}
\begin{proof}
    See Appendix \ref{appendix:ssnumber:decoding}.
\end{proof}
If the encoded quantum signal is decoded with the correct sequence, i.e., $\tilde{\Lambda}= \Lambda$, then the receiver will obtain the same signal as the transmitted one.

\subsection{Filtering Quantum Signals at the Receiver}
\begin{figure}[!tb]
    \centering
    \includegraphics[width=0.7\linewidth]{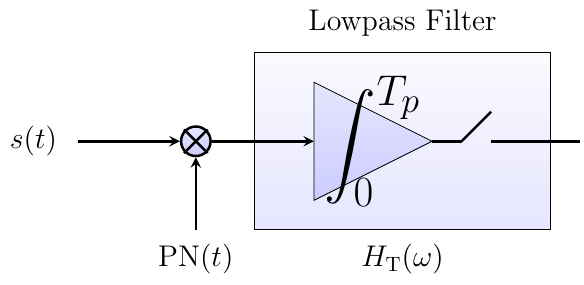}\caption{A typial receiver structure for classical baseband direct-sequence spread spectrum communication systems. The received classical communication signal, $s(t)$ is a pulse encoded with pseudo random sequence PN$(t)$. This signal is multplied by the despreading code PN$(t)$ at the receiver. The output passes through a lowpass filter (an integrator). }
    \label{fig:classicalReceiver}
\end{figure}
\begin{figure}[!tb]
    \centering
    \includegraphics[width=0.7\linewidth]{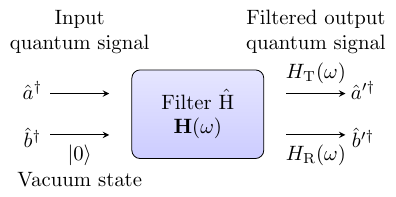}\caption{Filtering a quantum signal using a general structure with input two ports corresponding to $\hat{a}^\dagger$ and $\hat{b}^\dagger$ and two output ports corresponding to $\hat{a}'^\dagger$ and $\hat{b}'^\dagger$. At the receiver the quantum signals is considered in the port corresponding to $\hat{a}^\dagger$, while the auxiliary mode $\hat{b}^\dagger$ is in vacuum state. The output of the filter corresponds to mode $\hat{a}'^\dagger$ with a narrow-band response for $H_{\text{T}}(\omega)$.}
    \label{fig:filter}
\end{figure}
Figure \ref{fig:classicalReceiver} shows a typial receiver structure for classical baseband direct-sequence spread spectrum communication systems. In QCDMA the decoding (despreading) operator plays the role of the classical multiplier in the quantum domain. After decoding the quantum signal, similar to classical spread spectrum CDMA communication systems, a narrow-band filter (lowpass filter for equivalent baseband signals) is required to reject the out-of-band signal. In this paper, we consider a general two-port passive time-invariant structure for the filter \cite{moslehi1984fiber,salehi1988low, razavi2002performance} as is shown in Fig. \ref{fig:filter}. A time-independent quantum filter is a passive (energy-preserving)  device that can be described by a unitary transformation on two sets of input modes described by creation operators $\hat{a}^\dagger(\omega)$ and $\hat{b}^\dagger(\omega)$. The input modes of interest  $\hat{a}^\dagger(\omega)$ correspond to the quantum signal that is being filtered. In contrast, $\hat{b}^\dagger(\omega)$ belongs to auxiliary modes that are assumed initially to be in the vacuum state. Hence a two-port quantum filter is a unitary operator  $\mathbf{H}(\omega)$ that acts on two modes and generate new transmitted and reflected modes $\hat{a}'^\dagger(\omega)$ and $\hat{b}'^\dagger(\omega)$, respectively.
\begin{figure*}[!htb]
	\centering
	\includegraphics[width=\linewidth]{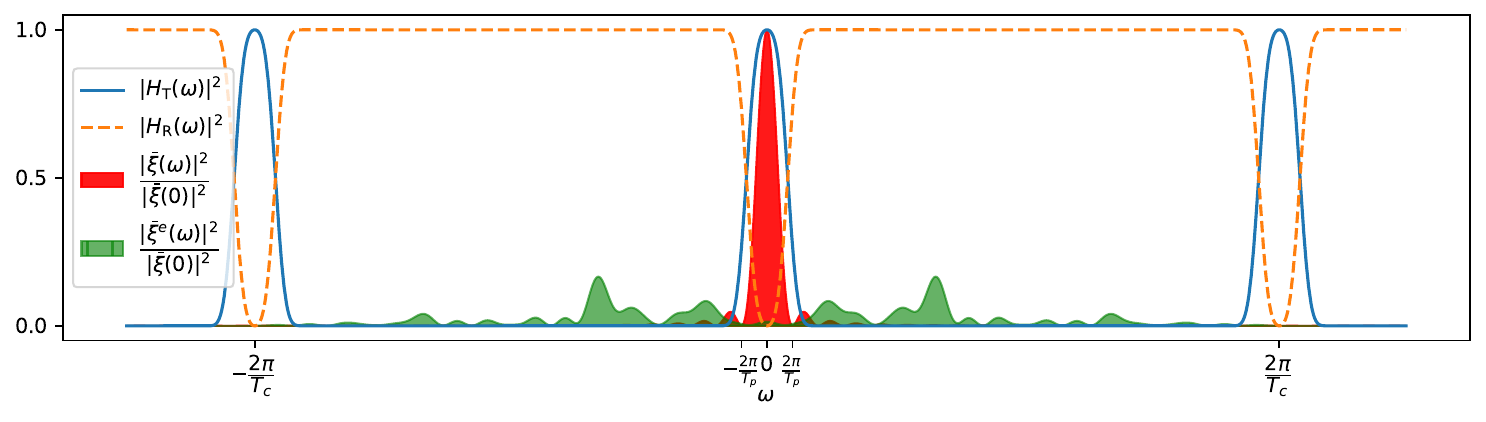}\caption{The frequency response of a typical lowpass baseband filter. The 3-db bandwidth of the transmission filter $H_{\text{T}}(\omega)$ is equal to the bandwidth of the original quantum signal and $N_c=20$. The quantum spread spectrum signal has a higher bandwidth and does not significantly pass through $H_{\text{T}}(\omega)$ but significantly passes through $H_{\text{R}}(\omega)$. Also note that $|H_{\text{T}}(\omega)|^2+|H_{\text{R}}(\omega)|^2 = 1$ (all-pass filter).}
	\label{fig:filterlarge}
\end{figure*}
Assuming that the evolution operator of the filter is denoted by $\hat{\mathrm{H}}$, we may write \cite{van2017time}
\begin{align}
    \hat{\mathrm{H}}\begin{pmatrix}
        \hat{a}^\dagger(\omega)\\
        \hat{b}^\dagger(\omega)
    \end{pmatrix}\hat{\mathrm{H}}^\dagger &= \mathbf{H}(\omega)\begin{pmatrix}
        \hat{a}'^\dagger(\omega)\\
        \hat{b}'^\dagger(\omega)
    \end{pmatrix}\nonumber\\
&=\begin{pmatrix}
        H_{\text{T}}(\omega) & H_{\text{R}}(\omega)\\
        H_{\text{R}}(\omega) & H_{\text{T}}(\omega)
    \end{pmatrix}\begin{pmatrix}
        \hat{a}'^\dagger(\omega)\\
        \hat{b}'^\dagger(\omega)
    \end{pmatrix}, \label{eq:filter:omega}
\end{align}
where $H_{\text{T}}(\omega)$ and $H_{\text{R}}(\omega)$ be the complex frequency response of transmission (through
the desired port of the filter) and reflection (of the other filter port) components. The unitary property of the transformation means that the filter coefficients must satisfy the following relationships \cite{barnett1998quantum}
\begin{align}
    |H_{\text{T}}(\omega)|^2&+|H_{\text{R}}(\omega)|^2 = 1,\\
    H_{\text{T}}(\omega)H_{\text{R}}^*(\omega)&+H_{\text{R}}(\omega)H_{\text{T}}^*(\omega)=0.
\end{align}

To obtain the filter's effect on the quantum signal, we write the evolution of the creation operator of the transmitted mode, which describes the received quantum signal.
\begin{align}
    \hat{\mathrm{H}}\hat{a}^\dagger_\wavepackt\hat{\mathrm{H}}^\dagger&=  \int d\omega \bar{\wavepackt}(\omega) \hat{\mathrm{H}}\hat{a}^\dagger(\omega)\hat{\mathrm{H}}^\dagger \\
    &= \int d\omega H_{\text{T}}(\omega)\bar{\wavepackt}(\omega)  \hat{a}'^\dagger(\omega) +  \int d\omega H_{\text{R}}(\omega)\bar{\wavepackt}(\omega) \hat{b}'^\dagger(\omega)\\
    &= \int dt (h_{\text{T}}(t)*\wavepackt(t))  \hat{a}'^\dagger(t) +  \int dt (h_{\text{R}}(t)*\wavepackt(t)) b'^\dagger(t)\\
    &\coloneqq w_{\text{T}}\hat{a}'^\dagger_{\wavepackt_{\text{T}}}+{w_{\text{R}}}\hat{b}'^\dagger_{\wavepackt_{\text{R}}},
\end{align} 
where $*$ stands for convolution, $h_{\text{T}}(t) = \mathcal{F}^{-1}\lbrace H_{\text{T}}(\omega) \rbrace$, $h_{\text{R}}(t) = \mathcal{F}^{-1}\lbrace H_{\text{R}}(\omega) \rbrace$,  $H_{\text{T}}(\omega)\bar{\wavepackt}(\omega)= \mathcal{F}\lbrace h_{\text{T}}(t)*\wavepackt(t)\rbrace$ and $H_{\text{R}}(\omega)\bar{\wavepackt}(\omega)= \mathcal{F}\lbrace h_{\text{R}}(t)*\wavepackt(t)\rbrace$. Also,
\begin{align}
    \hat{a}'^\dagger_{\wavepackt_{\text{T}}} &\coloneqq  \frac{1}{w_{\text{T}}}\int dt (h_{\text{T}}(t)*\wavepackt(t))  \hat{a}'^\dagger(t),\\
    \hat{b}'^\dagger_{\wavepackt_{\text{R}}} &\coloneqq \frac{1}{{w_{\text{R}}}}\int dt (h_{\text{R}}(t)*\wavepackt(t))  \hat{b}'^\dagger(t).
\end{align} 
Note that for the quantum filter, transmission probability is 
\begin{align}
    w_{\text{T}}^2 \coloneqq  \int dt |h_{\text{T}}(t)*\wavepackt(t)|^2,
\end{align}
and the reflection probability
\begin{align}
    w_{\text{R}}^2 \coloneqq  \int dt |h_{\text{R}}(t)*\wavepackt(t)|^2.
\end{align}
Indeed (using Parseval's theorem) we have $w_{\text{T}}^2 +w_{\text{R}}^2 =1$, that is the probabilities of reflection and transmission should sum to one.

\subsection{Chip-time interval Decomposition of the Quantum Filter}
\subsubsection{A General Two Port Filter}
We can represent a general optical filter  \cite{madsen1999optical} by the Fourier series of its transfer function. Thus for a causal optical filter \cite[Section~10.7.1]{oppenheim1997signal} we may write
\begin{align}
    H_{\text{T}}(\omega) &= \sum_{\fdcoef=0}^{\infty} d_\fdcoef e^{-{j}\fdcoef\omega \tau},\\
    H_{\text{R}}(\omega) &= \sum_{\fdcoef=0}^{\infty} f_\fdcoef e^{-{j}\fdcoef\omega \tau}.
\end{align}

The case of $\tau=T_c$ is of special interest for the quantum spread spectrum communication system, where $T_c$ is the chip-time. For this case, we obtain the temporal decomposition of the creation operators for rectangular wavepackets.

As shown in Fig. \ref{fig:filterlarge}, in a spread spectrum communication system, $H_{\text{T}}(\omega)$ has effectively a lowpass baseband frequency response. The transmission response of the filter should have a bandwidth equal to the signal bandwidth, i.e. $\frac{2\pi}{\pulseduration}$. The response of the filter is periodic with period $\frac{2\pi}{T_c}$, which is larger than the bandwidth of the spread spectrum quantum signal.
\begin{proposition}
    The chip-time interval creation operators of the input mode to a quantum filter with a rectangular wavepacket  can be expressed in terms of the chip-time interval creation operators as follows
        \begin{align}
            \hat{\mathrm{H}}\hat{a}^\dagger_{\chippackt_k} \hat{\mathrm{H}}^\dagger= 
            \sum_{\fdcoef=0}^{\infty} (d_{\fdcoef} \hat{a}'^\dagger_{\chippackt_{k+\fdcoef}}+  f_{\fdcoef}  \hat{b}'^\dagger_{\chippackt_{k+\fdcoef}}).
        \end{align}
\end{proposition}
\begin{proof}
    See Appendix \ref{appendix:filter}.
\end{proof}

We further track the filter's effect on a creation operator with wavepacket $\wavepackt(t)$ utilizing the temporal decomposition of creation operators.

\begin{theorem}
	\label{thm:filter}
    The creation operator of the input mode to a quantum filter with a rectangular wavepacket can be expressed in terms of the chip-time interval creation operators as follows
    \begin{align}
        \hat{\mathrm{H}}\hat{a}^\dagger_\wavepackt \hat{\mathrm{H}}^\dagger=\sum_{\FDcoef=0}^{\infty}&(D_{\FDcoef}\hat{a}^{'\dagger}_{\chippackt_\FDcoef}+ {F}_{\FDcoef}\hat{b}^{'\dagger}_{\chippackt_\FDcoef}),
    \end{align}
	where the transmitted and reflected probability amplitude coefficients at time interval $\FDcoef$ are respectively  
	\begin{align}
		D_{\FDcoef} &\coloneqq  \sum_{\fdcoef=\max(0,\FDcoef-N_c+1)}^{\FDcoef} \frac{1}{\sqrt{N_c}}d_{\fdcoef},\\
		{F}_{\FDcoef}  &\coloneqq  \sum_{\fdcoef=\max(0,\FDcoef-N_c+1)}^{\FDcoef} \frac{1}{\sqrt{N_c}}f_{\fdcoef}.
	\end{align}
\end{theorem}
\begin{proof}
    See Appendix \ref{appendix:filter}.
\end{proof}
Note that the unitary condition implies that
\begin{align}
	\sum_{\FDcoef=0}^{\infty} \left(\left|{D}_{\FDcoef} \right|^2 + \left|{F}_{\FDcoef} \right|^2\right) = 1.
\end{align}

\begin{figure}[!tb]
	\centering
	\includegraphics[width=\linewidth]{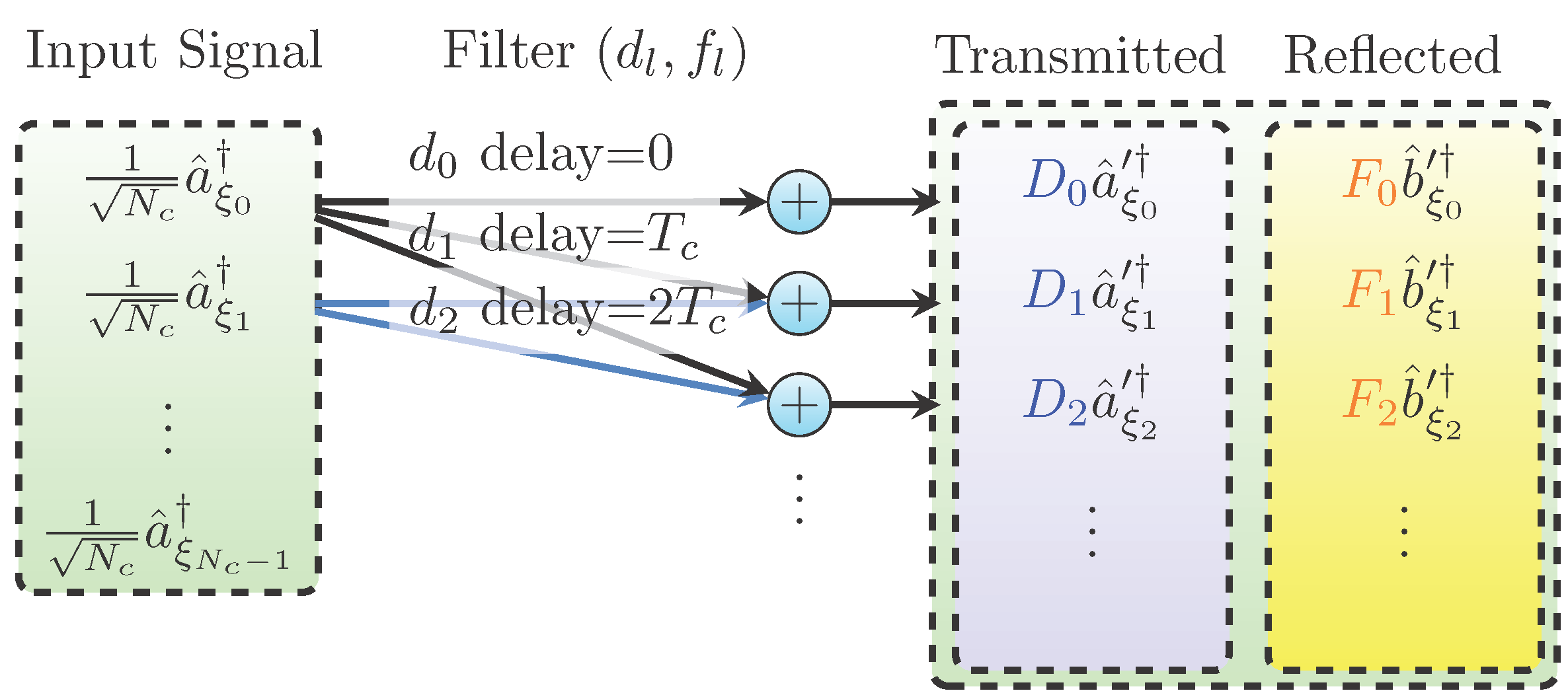}\caption{Filter interleaves the chip-time interval creation operators. Only transmission coefficients $d_\fdcoef $ are illustrated. Note that the time-interval creation operators do not backtrack due to the causality of the filter.}
	\label{fig:filtercoef}
\end{figure}

Figure \ref{fig:filtercoef} illustrates the results of Theorem \ref{thm:filter}.
\subsubsection{Finite Iimpulse Response Filter}
The filtering coefficients reduce to the following expression for finite impulse response (FIR) filters with a maximum delay of $L T_c$, where $L$ is the number of FIR filter coefficients.
\begin{align}
    D_{\FDcoef} &= \sum_{\fdcoef=\max(0,\FDcoef-N_c+1)}^{\min(\FDcoef, L)} \frac{1}{\sqrt{N_c}}d_{\fdcoef}, \\
    {F}_{\FDcoef}  &= \sum_{\fdcoef=\max(0,\FDcoef-N_c+1)}^{\min(\FDcoef, L)} \frac{1}{\sqrt{N_c}}f_{\fdcoef}.
\end{align} 
Figure \ref{fig:asymptotic}a depicts typical asymptotic behavior of values $D_{\FDcoef}$ and $F_{\FDcoef}$ for near ideal filters.

\subsection{Spread Spectrum Signal After Filter}
To further investigate the effect of filter on the encoded (spreaded) and decoded (despreaded) spread spectrum quantum signals, we first define the corresponding filter coefficients as follows
\begin{align}
    \tilde{D}_{\FDcoef} &\coloneqq  \sum_{\fdcoef=\max(0,\FDcoef-N_c+1)}^{\FDcoef} \frac{1}{\sqrt{N_c}}d_{\fdcoef} \tilde{\lambda}_{\FDcoef-\fdcoef}\lambda_{\FDcoef-\fdcoef},\label{eq:filterDtilde}\\
    \tilde{F}_{\FDcoef}  &\coloneqq  \sum_{\fdcoef=\max(0,\FDcoef-N_c+1)}^{\FDcoef} \frac{1}{\sqrt{N_c}}f_{\fdcoef} \tilde{\lambda}_{\FDcoef-\fdcoef}\lambda_{\FDcoef-\fdcoef}.\label{eq:filterFtilde}
\end{align}
From a signal processing point of view, we can also describe the above coefficients as the result of a discrete-time convolution \cite[Section~2.9.6]{oppenheim2009discrete}, i.e.,
\begin{align}
    \tilde{D}_{\FDcoef} &\coloneqq  d_\FDcoef*\Pi^{\tilde{\Lambda}\Lambda}_\FDcoef,\\
    \tilde{F}_{\FDcoef}  &\coloneqq  f_\FDcoef*\Pi^{\tilde{\Lambda}\Lambda}_\FDcoef,
\end{align}
where 
\begin{align}
    \Pi^{\tilde{\Lambda}\Lambda}_\FDcoef &\coloneqq \left\lbrace
	\begin{matrix}
		\frac{1}{\sqrt{N_c}}\tilde{\lambda}_{\FDcoef}\lambda_{\FDcoef}, && 0\le \FDcoef\le N_c-1\\
		0, && \text{o.w.}
	\end{matrix}\right. .
\end{align}

The following proposition describes the evolution of creation operators through the filter using the proposed chip-time interval decomposition methodology.
\begin{proposition}
	\label{prop:filter}
    The creation operator of a quantum signal encoded with $\Lambda$ and decoded with $\tilde{\Lambda}$ can be expressed in terms of the chip-time interval creation operators as follows
    \begin{align}
        \hat{\mathrm{H}}\hat{a}^\dagger_{\wavepackt^d}\hat{\mathrm{H}}^\dagger=\sum_{\FDcoef=0}^{\infty}&(\tilde{D}_{\FDcoef}\hat{a}^{'\dagger}_{\chippackt_{\FDcoef}}+ \tilde{F}_{\FDcoef}\hat{b}^{'\dagger}_{\chippackt_{\FDcoef}}).
    \end{align}
\end{proposition}
\begin{proof}
    See Appendix \ref{appendix:filter}.
\end{proof}
Figure \ref{fig:asymptotic}b depicts typical asymptotic behavior of values $\tilde{D}_{\FDcoef}$ and $\tilde{F}_{\FDcoef}$ when the random codes $\Lambda$ and $\tilde{\Lambda}$ does not match, i.e., $\Lambda\ne\tilde{\Lambda}$.

If the decoding sequence is different from the encoding sequence, the output of the decoder is wideband and thus is rejected by the filter. In other words, the values of $\tilde{F}_{\FDcoef}$ and $D_{\FDcoef}$ are large over the duration of the transmitted quantum signal, while the values of ${F}_{\FDcoef}$ and $\tilde{D}_{\FDcoef}$ are small.

As discussed in Appendix \ref{appendix:asymptotic} the shape of the transmitted wavepacket $\wavepackt(t)$ is asymptotically preserved after correct decoding of the quantum signals. Therefore, the transmitted wavepacket is almost equal to the signal wavepacket, i.e., $\braket{\wavepackt(t)}{\wavepackt_{\text{T}}(t-T_{\text{delay}})}\approx 1$, where $T_{\text{delay}} = \FDcoef_\text{delay}\times T_c$ is the delay imposed by the causal filter and $\FDcoef_\text{delay}$ is the number of chip-times corresponding to $T_{\text{delay}}$.

\begin{figure}[!tb]
	\centering
	\begin{picture}(260,165)
		\put(0,0){\includegraphics[width=0.99\linewidth]{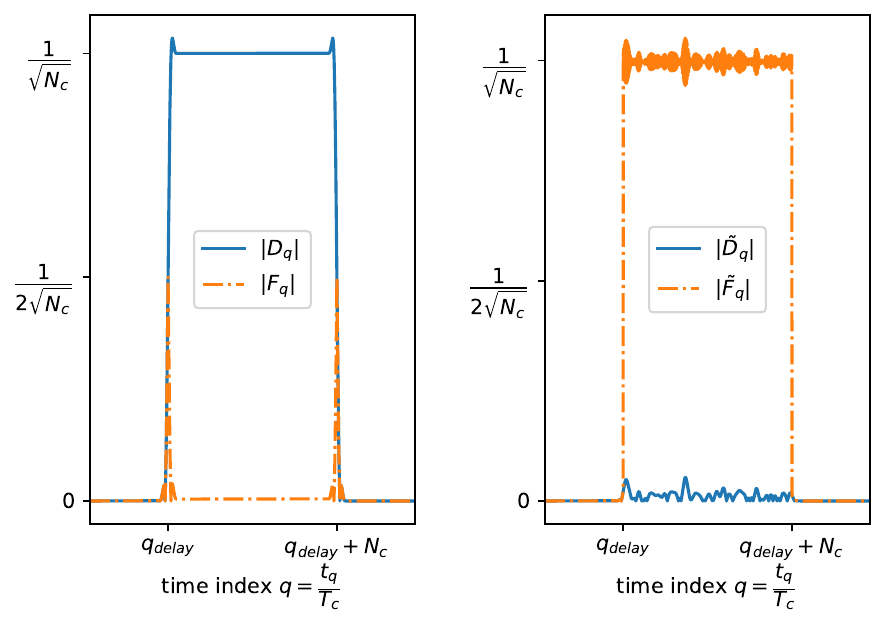}}
		\put(30,155){(a)}
		\put(160,155){(b)}
	\end{picture}
	\caption{Asymptotic behavior of the discrete coefficients of the filter's output, where the filter's bandwidth is sufficiently greater than the quantum signal's bandwidth and a high processing gain is used, i.e., $T_c\ll T$. To investigate the asymptotic behavior, we choose $N_c=100000$ and the bandwidth of filter is $20\times \frac{2\pi}{\pulseduration}$. Both filters, $H_\text{T}(\omega)$ and $H_\text{R}(\omega)$, are approximated with type I FIR filters with odd number of taps $L=10001$ and designed using the windowing method \cite{oppenheim2009discrete}.}
	\label{fig:asymptotic}
\end{figure}
The evolution of specific quantum signals can also be expressed in terms of chip-time intervals as follows.
\begin{proposition}
    The coherent quantum signal encoded with sequence $\Lambda$ and decoded with sequence $\tilde{\Lambda}$ results in the following pure quantum signal after the filter
    \begin{align}
        \ket{\alpha_{\wavepackt_{\emph{\text{T}}}}} = \prod_{\FDcoef=0}^{\infty} \ket{\tilde{D}_{\FDcoef}\alpha_{\chippackt_\FDcoef}}  ,
    \end{align}
	where $\emph{\text{T}}$ in $\ket{\alpha_{\wavepackt_{\emph{\text{T}}}}}$ means that the output state is in the transmitted mode of the filter, and $\ket{\alpha_{\chippackt_\FDcoef}}$ corresponds to the related chip-time interval coherent state.
\end{proposition}
\begin{proof}
    See Appendix \ref{appendix:filterEffect}.
\end{proof}

At the receiver, a photo-detector can be used to perform a measurement on the signal after filtering. Each term in the tensor product results in a Poisson distribution for the number of photons in that specific chip-time, i.e., $n_{\FDcoef,\chippackt_\FDcoef}$. Therefore, the output statistics can be obtained by noticing that sum of independent Poisson random variables is also a Poisson random variable
\begin{align}
    \mathbb{P}(n)
    &=\sum_{n_0+n_1+\cdots=n} \left|\braket{n_{0,\chippackt_0}, n_{1,\chippackt_1}, \cdots}{\alpha_{\wavepackt_{\text{T}}}}\right|^2\\
    &=\text{Poisson}\left(|\alpha|^2 \sum_{\FDcoef=0}^{\infty}\left|\tilde{D}_{\FDcoef}\right|^2 \right).
\end{align}

\begin{proposition}
    A single photon quantum signal encoded with sequence $\Lambda$ and decoded with sequence $\tilde{\Lambda}$ results in a mixed quantum signal after filter with a density operator as follows
    \begin{align}
        \rho^{\emph{\text{T}}} &= \left(\sum_{\FDcoef=0}^{\infty}\left|\tilde{F}_{\FDcoef}\right|^2\right)\ketbra{0}+\left(\sum_{\FDcoef=0}^{\infty}\left|\tilde{D}_{\FDcoef}\right|^2 \right)\ketbra{1_{\wavepackt_{\emph{\text{T}}}}},
    \end{align}  
	where 
	\begin{align}
		\ket{1_{\wavepackt_{\emph{\text{T}}}}}\coloneqq \frac{1}{\sqrt{\sum_{\FDcoef=0}^{\infty}\left|\tilde{D}_{\FDcoef}\right|^2 }}\sum_{\FDcoef=0}^{\infty}\tilde{D}_{\FDcoef}\ket{1_{\emph{\text{T}},\chippackt_\FDcoef}}.
	\end{align}
\end{proposition}
\begin{proof}
    See Appendix \ref{appendix:filterEffect}.
\end{proof}

The term $\sum_{\FDcoef=0}^{\infty}\left|\tilde{F}_{\FDcoef}\right|^2$ shows the probability of reflection from the filter.  On the other hand, the term $\sum_{\FDcoef=0}^{\infty}\left|\tilde{D}_{\FDcoef}\right|^2$ implies that the properly decoded quantum signal is recovered at the receiver. From the discussion of Appendix \ref{appendix:asymptotic}, we observe that when the codes match, i.e. $\Lambda=\tilde{\Lambda}$, we have $\rho^{\text{T}}\approx \ketbra{1_{\wavepackt_{\text{T}}}}$ asymptotically. Also for random codes  $\Lambda\ne\tilde{\Lambda}$, the asymptotic behavior of $\rho^{\text{T}}$ is $\rho^{\text{T}}\approx \ketbra{0}$.

\section{Quantum Spread Spectrum CDMA}
\label{sec:cdma}
\begin{figure*}[!tb]
    \centering
    \includegraphics[width=0.9\linewidth]{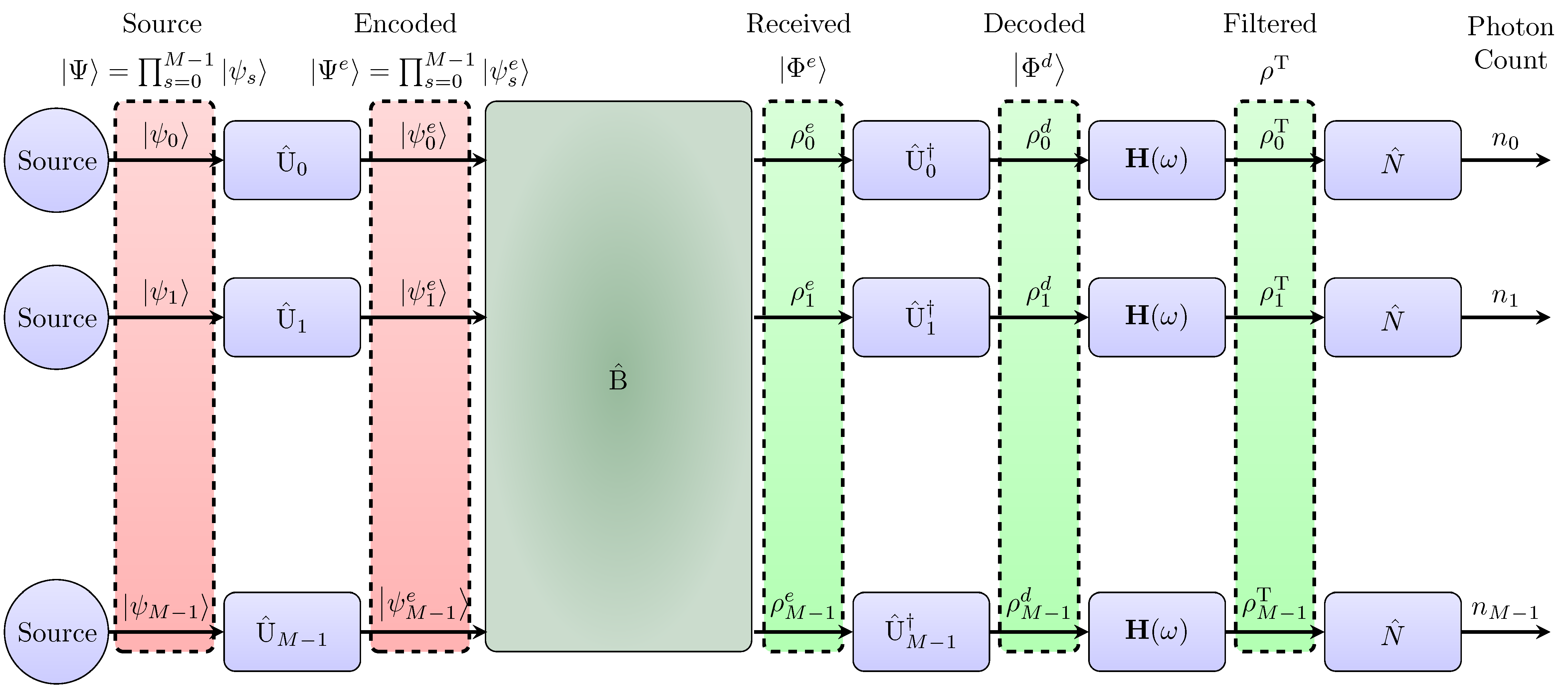}\caption{A multiple access quantum spread spectrum CDMA communication system in Schrodinger's picture. In this picture, the quantum communication signals evolve through the quantum communication system. A transmitted pure quantum signal may mix with other transmitters and is expressed generally by a density matrix on the receivers’ end. In this figure, we assume that receiver $r$ is using the decoding (despreading) sequence corresponding to user $s=r$.}
    \label{fig:system2}
\end{figure*}
The quantum spread spectrum CDMA communication system utilizes the quantum spread spectrum transmitter and receiver described in the previous sections to transmit the classical or quantum information over a quantum broadcasting interference channel. Similar to quantum CDMA systems with spectral encoding \cite{rezai2021quantum}, the quantum broadcasting channel can be modeled as a passive star-coupler. In this model, the encoded spread spectrum quantum signals are broadcasted by an $M\times M $ star-coupler. The broadcasting channel performs a unitary transformation on the transmitted signals according to a unitary matrix $\mathrm{B}$.  

\subsection{Schrodinger's picture of Quantum Spread Spectrum CDMA}
Figure \ref{fig:system2} depicts the Schrodinger's picture of the quantum spread spectrum code division multiple access communication system. From this viewpoint, quantum signals evolve as they propagate through the quantum communication system. The transmitted pure quantum signals from $M$ different users, i.e. $\ket{\psi_0}, \ket{\psi_1}, \cdots ,\ket{\psi_{M-1}}$ form a general tensor product state 
\begin{align}
    \ket{\Psi} = \prod_{s=0}^{M-1}\ket{\psi_s}.
\end{align}

Each transmitter $s$ utilizes an electro-optical modulator to construct the spread spectrum signal, $\ket{\psi^e_s}$. The encoder acts as a unitary operator $\hat{\mathrm{U}}_s$ to modulate the wavepacket of the transmitted quantum signal with pseudo-random code sequence $\Lambda_s=(\lambda_{s,0}, \lambda_{s,1}, \cdots, \lambda_{s,N_c-1})$. The resulting quantum state of the system before the broadcasting channel is thus
\begin{align}
    \ket{\Psi^e} = \hat{\mathrm{U}}\ket{\Psi}= \prod_{s=0}^{M-1}\ket{\psi^e_s},
\end{align}
where $\hat{\mathrm{U}}= \prod_{s=0}^{M-1}\hat{\mathrm{U}}_s$ considers the effect of all encoders of all users.

The tensor product state $\ket{\Psi^e}$ passes through the broadcasting channel evolving into pure state $\ket{\Phi^e}$. The received signal at each receiver can be obtained by taking the partial trace of $\ket{\Phi^e}$ with respect to all other receivers. As we show, the received quantum signal includes interference from other users and can potentially be an entangled quantum state. The entanglement results in a mixed quantum signal after taking the partial trace with respect to other users. The receivers then apply the decoding operation utilizing an electro-optical modulator to recover their intended signals. An optical narrow-band filter is then used to discard the interference. At this point, we show that for a typical optical filter, the system's overall state can be a mixed state. The quantum signal representing the output of all receivers is denoted by the density $\rho^{\text{T}}$ (T~stands for transmission of the filter) and is obtained by taking the partial trace of the filter expression with respect to the reflected portion of the incoming quantum signal.

In the following subsection, we describe the evolution of quantum signals through the quantum spread spectrum multiple access communication system utilizing our proposed chip-time interval decomposition approach.

\subsection{Quantum Broadcasting Channel}
\begin{figure}[!tb]
	\centering
	\includegraphics[width=0.9\linewidth]{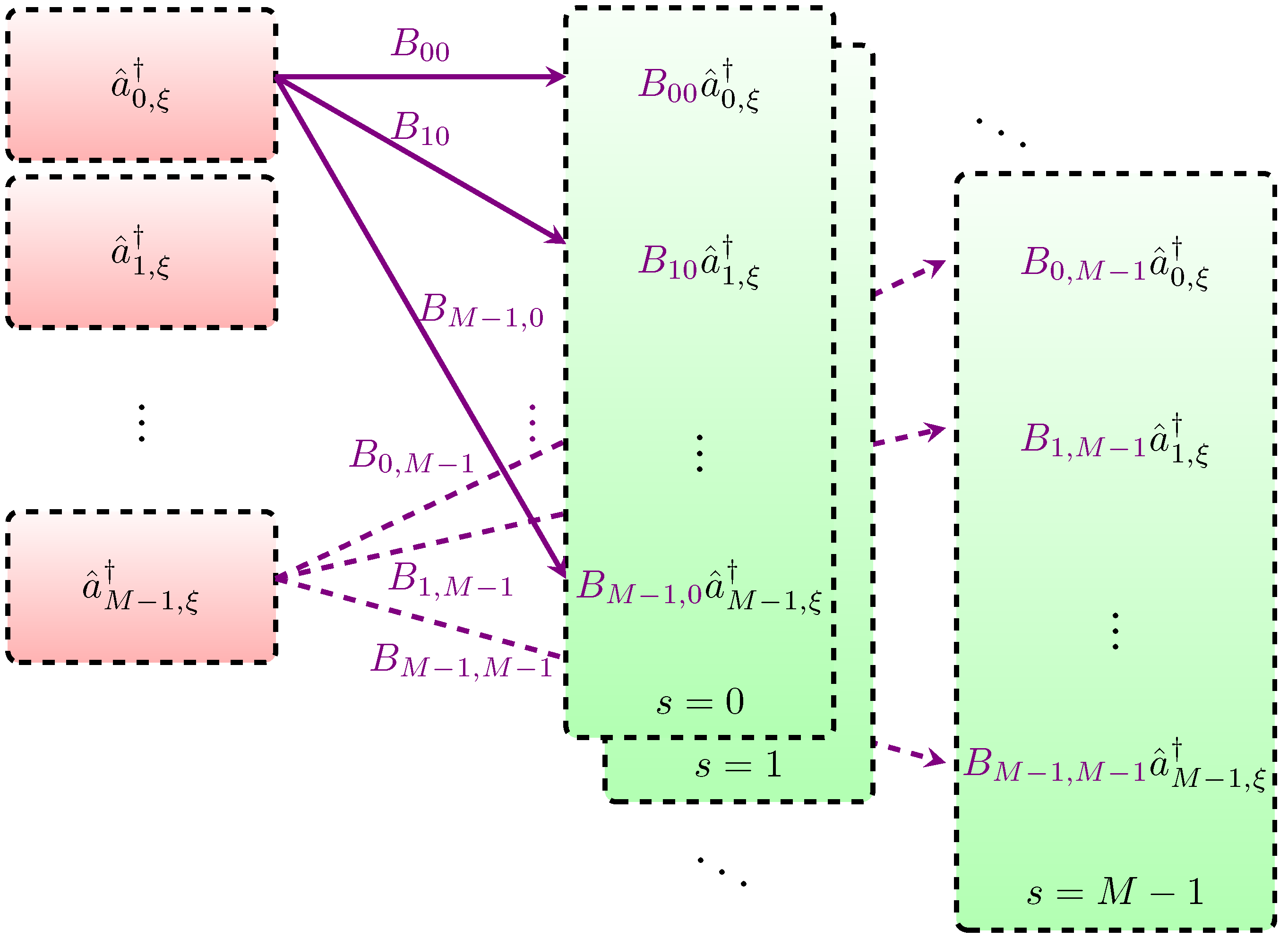}\caption{The effect of quantum broadcasting channel on the creation operators. The creation operator corresponding to each transmitter $s$ is broadcasted toward all receivers resulting in possible entanglement between them.  For simplicity, only the connections regarding transmitter $s=0$  and  $s=M-1$ are illustrated. The effect of other transmitters given by (\ref{eq:broadcast:tensor}) is in terms of a tensor product, which is demonstrated using indices $s=0$, $s=1$ , $\cdots$, $s=M-1$. Inspired by the representation of tensor networks from classical and quantum machine learning \cite{huggins2019towards}, the tensor product is shown as a stack of blocks.}
	\label{fig:broadcast}
\end{figure}
The broadcasting star-coupler channel performs the following transformation from the input signal space with subscript $s$ to the received signal space with subscript $r$ \cite{rezai2021quantum}:
\begin{align}
     \hat{\mathrm{B}}\hat{a}^\dagger_{s,\wavepackt} \hat{\mathrm{B}}^\dagger&= \sum_{r=0}^{M-1} B_{rs}\hat{a}_{r,\wavepackt}^\dagger\label{eq:broadcast}.
\end{align}
This article assumes that all of the transmitters are synchronized and utilize the same wavepacket $\wavepackt(t)$. Figure \ref{fig:broadcast} depicts the broadcasting channel of (\ref{eq:broadcast}). 

In the sense of the proposed temporal decomposition approach, we may write:
\begin{align}
    \hat{\mathrm{B}}\hat{a}^\dagger_{s,\wavepackt} \hat{\mathrm{B}}^\dagger &= \frac{1}{\sqrt{N_c}}\sum_{k=0}^{N_c-1} \sum_{r=0}^{M-1} B_{rs}\hat{a}_{r,\chippackt_k}^\dagger .
\end{align}

Hence, 
\begin{align}
    \ket{\Phi}  &= \hat{\mathrm{B}}\ket{\Psi} = \prod_{s=0}^{M-1} f_s(\hat{\mathrm{B}}\hat{a}^\dagger_{s,\wavepackt} \hat{\mathrm{B}}^\dagger)\ket{0}\label{eq:broadcast:tensor}\\ 
    &= \prod_{s=0}^{M-1} f_s(\frac{1}{\sqrt{N_c}}\sum_{k=0}^{N_c-1} \hat{\mathrm{B}}\hat{a}_{s,\chippackt_k}^\dagger\hat{\mathrm{B}}^\dagger)\ket{0}\\
    &= \prod_{s=0}^{M-1} f_s(\frac{1}{\sqrt{N_c}}\sum_{k=0}^{N_c-1} \sum_{s=0}^{M-1} B_{rs}\hat{a}_{r,\chippackt_k}^\dagger)\ket{0}.
\end{align}
The above relation is the general form of the decomposed quantum signals after the broadcasting channel. In the rest of this subsection, we derive the exact relations for coherent and number states.

\subsubsection{Coherent State}
\begin{proposition}
    The effect of the quantum broadcasting channel (star-coupler) on the coherent state inputs turns to the tensor product of each chip of  receivers' decomposed states.
    \begin{align}
        \ket{\Phi}  &= \hat{\mathrm{B}}\ket{\Psi} = \hat{\mathrm{B}} \prod_{s=0}^{M-1} \ket{\alpha_{s,\wavepackt}}= \prod_{r=0}^{M-1}\prod_{k=0}^{N_c-1}\ket{\sum_{s=0}^{M-1}\frac{B_{rs}}{\sqrt{N_c}}\alpha_{s,\chippackt_k}}.
    \end{align}
\end{proposition}
\begin{proof}
    See Appendix \ref{appendix:coherentCoupler}.
\end{proof}

From the above proposition, we obtain the received quantum state at a particular receiver, $r$, by taking the partial trace of the density operator $\ketbra{\Phi}$ with respect to the other receivers, which gives
\begin{align}
    \ket{\phi_{r}}    &= \prod_{k=0}^{N_c-1}\ket{\sum_{s=0}^{M-1}\frac{B_{rs}}{\sqrt{N_c}}\alpha_{s,\chippackt_k}}.
\end{align}

Let us consider the effect of encoding and decoding.  The encoding operator at transmitter $s$, $\hat{\mathrm{U}}_s$, converts the original wavepacket, $\wavepackt(t)$, into the encoded (spreaded) wavepacket, $\wavepackt^{e_s}(t)$. Receiver $r$ applies its decoding operator $\hat{\mathrm{U}}^\dagger_r$, which turns the wavepacket $\wavepackt^{e_s}(t)$ to the decoded (despreaded) wavepacket $\wavepackt^{e_sd_r}(t)$. Since we assume that receiver $r=s$ is decoding the data from transmitter $s$, we have $\wavepackt^{e_sd_{r=s}}(t)= \wavepackt(t)$.

\begin{proposition}
    The output of the spread spectrum quantum CDMA communication system utilizing coherent states at receiver $r$ after decoding the data of transmitter $s=r$ is 
    \begin{align}
        \ket{\phi^d_{r }}  
        &= \prod_{k=0}^{N_c-1}\ket{\frac{B_{r r}}{\sqrt{N_c}}\alpha_{s=r,\chippackt_k}+\frac{1}{\sqrt{N_c}} \sum_{s\neq r}\lambda_{r ,k}\lambda_{s,k}B_{r s}\alpha_{s,\chippackt_k}}\\
        &=\ket{B_{r r}\alpha_{s=r,\wavepackt}+\sum_{s\neq r}B_{r s}\alpha_{s,\wavepackt^{e_sd_r}}}.
    \end{align}
\end{proposition}
\begin{proof}
    See Appendix \ref{appendix:coherentCoupler}.
\end{proof}

Similar to point-to-point communication, (\ref{eq:filterDtilde}) and (\ref{eq:filterFtilde}), we define the following coefficients that track the effect of encoding and decoding operators and quantum filter
\begin{align}
    D_{\FDcoef}^{s, r} &\coloneqq  \sum_{\fdcoef=\max(0,\FDcoef-N_c+1)}^{\FDcoef} \frac{1}{\sqrt{N_c}}d_{\fdcoef}\lambda_{{s},\FDcoef-\fdcoef}\lambda_{{r},\FDcoef-\fdcoef },\label{eq:Dksr}\\
    {F}_{\FDcoef}^{s, r}   &\coloneqq  \sum_{\fdcoef=\max(0,\FDcoef-N_c+1)}^{\FDcoef} \frac{1}{\sqrt{N_c}}f_{\fdcoef}\lambda_{{s},\FDcoef-\fdcoef}\lambda_{{r},\FDcoef-\fdcoef }.\label{eq:Fksr}
\end{align}
For multiple access communication systems, this article uses superscript $\lbrace s,r\rbrace$ for $D$ and $F$ coefficients to specify the corresponding sender and receiver and their encoding and decoding code sequences.
\begin{proposition}
	\label{prop:coherentFiltered}
    The coherent quantum signal decoded at the receiver $r$ from transmitter $s=r$  results in the following pure quantum signal after the filter
    \begin{align}
        \ket{\alpha_{r, \wavepackt_{\text{T}}}} & = \prod_{\FDcoef=0}^{\infty}\Bigg|B_{rr}D_{\FDcoef} \alpha_{s=r,\chippackt_\FDcoef}+ \sum_{s\neq r}B_{rs}D_{\FDcoef}^{s, r}\alpha_{s,\chippackt_\FDcoef}\Bigg\rangle.
    \end{align}
\end{proposition}

\subsubsection{Number State}
Unlike the spread spectrum system with coherent quantum signals, the system with continuous mode Fock states does not result in a separable state at each receiver. For this case, we write the overall state of encoded transmitted signals as
\begin{align}
    \ket{\Psi^e} &= \prod_{s=0}^{M-1} \ket{\psi^e} = \prod_{s=0}^{M-1} \ket{n_{s,\wavepackt^e}}\\
    &= \prod_{s=0}^{M-1} \frac{1}{\sqrt{n_s!}}(\hat{a}^\dagger_{s,\wavepackt^e})^{n_s}\ket{0}\\
    &= \prod_{s=0}^{M-1}\frac{1}{\sqrt{n_s!}}\left(\frac{1}{\sqrt{N_c}}\sum_{k=0}^{N_c-1} \hat{a}_{s,\chippackt_k^{e}}^\dagger\right)^{n_s}\ket{0}\\
    &= \prod_{s=0}^{M-1}\frac{1}{\sqrt{n_s!}}\left(\frac{1}{\sqrt{N_c}}\sum_{k=0}^{N_c-1} \lambda_{s,k}\hat{a}_{s,\chippackt_k}^\dagger\right)^{n_s}\ket{0}.
\end{align}

\begin{proposition}
	\label{prop:FockDecode}
    The output of a quantum spread spectrum CDMA communication system with Fock state transmitted quantum signals after the broadcasting channel is 
    \begin{align}
        \ket{\Phi^e} &=  \hat{\mathrm{B}}\ket{\Psi^e}\\
        &= \prod_{s=0}^{M-1}\frac{1}{\sqrt{n_s!}}\left(\frac{1}{\sqrt{N_c}}\sum_{k=0}^{N_c-1}  \sum_{r=0}^{M-1}B_{rs} \lambda_{s,k}\hat{a}_{r,\chippackt_k}^\dagger\right)^{n_s}\ket{0}.\nonumber
    \end{align}
After applying the decoding operators by each user, the overall received quantum signal is 
\begin{align}
    \ket{\Phi^d} &=  \hat{\mathrm{U}}^\dagger\ket{\Phi^e}\nonumber\\
	&=\prod_{s=0}^{M-1}\frac{1}{\sqrt{n_s!}}\Bigg(\frac{1}{\sqrt{N_c}}\sum_{k=0}^{N_c-1}\Big(B_{ss}\hat{a}_{r=s,\chippackt_k}^\dagger+ \\
	&\qquad\qquad\qquad+\sum_{r\ne s} B_{rs}\lambda_{s,k}\lambda_{r,k}\hat{a}_{r,\chippackt_k}^\dagger\Big)\Bigg)^{n_s}\ket{0}\nonumber\\
	&=\prod_{s=0}^{M-1}\frac{1}{\sqrt{n_s!}}\Bigg( B_{ss}\hat{a}_{r=s,\wavepackt}^\dagger+ \sum_{r\ne s}B_{rs}\hat{a}^\dagger_{r,\wavepackt^{e_sd_r}}\Bigg)^{n_s}\ket{0}.\nonumber
\end{align}
\end{proposition}
\begin{proof}
    See Appendix \ref{appendix:coherentCoupler}.
\end{proof}
To consider the effect of the filter, we note that 
\begin{align}
    \hat{\mathrm{H}}\hat{a}^\dagger_{\wavepackt^{e_sd_r}}\hat{\mathrm{H}}^\dagger&= \sum_{\FDcoef=0}^{\infty}(D_{\FDcoef}^{s,r}\hat{a}^{'\dagger}_{r,\chippackt_\FDcoef} + {F}_{\FDcoef}^{s,r}\hat{b}^{'\dagger}_{r,\chippackt_\FDcoef}).
\end{align}

Thus the quantum signal has the following form as the result of the filter
\begin{align}
    \ket{\Phi^F}&=\hat{\mathrm{H}}\ket{\Phi^d} \\
	&=\prod_{s=0}^{M-1}\frac{1}{\sqrt{n_s!}}(\sum_{\FDcoef=0}^{\infty}\sum_{r=0}^{M-1} B_{rs}(D_{\FDcoef}^{s,r}\hat{a}^{'\dagger}_{r,\chippackt_\FDcoef}+ {F}_{\FDcoef}^{s,r}\hat{b}^{'\dagger}_{r,\chippackt_\FDcoef}))^{n_s}\ket{0}\nonumber\\
	&=\prod_{s=0}^{M-1}\frac{1}{\sqrt{n_s!}}\Bigg(\sum_{\FDcoef=0}^{\infty}\Big(B_{ss}(D_{\FDcoef}\hat{a}^{'\dagger}_{r=s,\chippackt_\FDcoef}+ {F}_{\FDcoef}\hat{b}^{'\dagger}_{r=s,\chippackt_\FDcoef})\nonumber\\
	&\qquad+\sum_{r\ne s}^{M-1} B_{rs}(D_{\FDcoef}^{s,r}\hat{a}^{'\dagger}_{r,\chippackt_\FDcoef}+ {F}_{\FDcoef}^{s,r}\hat{b}^{'\dagger}_{r,\chippackt_\FDcoef})\Big)\Bigg)^{n_s}\ket{0}.\nonumber
\end{align}
The transmitted part of the above quantum signal can be obtained using a partial trace with respect to the reflected modes.
\begin{align}
    \rho^{\text{T}} &= \Tr_R\left(\ketbra{\Phi^F}\right).
\end{align}

\subsection{Chip-time Decomposition Methodology in a Nutshell}
\label{sec::timedecompositionpicture}
As a corollary of the propositions regarding the use of the proposed chip-time interval decomposition method, we can give a complete picture of the quantum spread spectrum CDMA communication systems according to the evolution of the creation operators. The following corollary summarizes the proposed chip-time decomposition methodology for description of the quantum communication system. 

\begin{corollary}[Chip-time decomposition methodology]
    \label{col:timePicture}
    The chip-time interval decomposition results in the following evolution for creation operators
    \begin{align}
        \hat{a}^{\dagger}_{s,\wavepackt} &= \frac{1}{\sqrt{N_c}}\sum_{k=0}^{N_c-1} \hat{a}_{s,\chippackt_k}^\dagger,\\
        \hat{\mathrm{U}}\hat{a}^{\dagger}_{s,\wavepackt}\hat{\mathrm{U}}^\dagger &= \frac{1}{\sqrt{N_c}}\sum_{k=0}^{N_c-1} \lambda_{s,k}\hat{a}_{s,\chippackt_k}^\dagger = \hat{a}^{\dagger}_{s,\wavepackt^{e_s}},\\
       \hat{\mathrm{B}}\hat{a}^{\dagger}_{s,\wavepackt^{e_s}}\hat{\mathrm{B}}^\dagger  &= \frac{1}{\sqrt{N_c}} \sum_{r=0}^{M-1}\sum_{k=0}^{N_c-1}B_{rs}\lambda_{s,k}\hat{a}_{r,\chippackt_k}^\dagger = \sum_{r=0}^{M-1}B_{rs}\hat{a}^\dagger_{r,\wavepackt^{e_s}},\\
        \hat{\mathrm{U}}^\dagger\hat{a}^\dagger_{r,\wavepackt^{e_s}}\hat{\mathrm{U}}  &= \frac{1}{\sqrt{N_c}}\sum_{k=0}^{N_c-1} \lambda_{s,k}\lambda_{r,k}\hat{a}_{r,\chippackt_k}^\dagger = \hat{a}^\dagger_{r,\wavepackt^{e_s, d_r}}.
    \end{align}    
    For rectangular $\wavepackt(t)$, we further have
    \begin{align}
    	 \hat{\mathrm{H}}\hat{a}^\dagger_{r,\wavepackt^{e_s, d_r}} \hat{\mathrm{H}}^\dagger &= \sum_{\FDcoef=0}^{\infty}\left( D^{s,r}_{\FDcoef}\hat{a}_{r,\chippackt_\FDcoef}'^\dagger+F^{s,r}_{\FDcoef}\hat{b}_{r,\chippackt_\FDcoef}'^\dagger\right).
    \end{align}
\end{corollary}

According to Corollary \ref{col:timePicture}, the chip-time interval creation operators, $\hat{a}_{\chippackt_k}^\dagger$ are invariant building blocks of the quantum spread spectrum CDMA communication systems. The evolution of the field creation operators mixes these basic elements.

According to the proposed chip-time interval decomposition methodology, the transmitted signal described as a mixture of chip-time interval creation operators is modulated using the spread spectrum encoder. The coefficients of the spreading sequence act as a multiplier for the corresponding chip-time interval creation operators. The signals then pass through a quantum broadcasting channel, which transforms each individual chip-time interval creation operator. The star-coupler mixes the chip-time interval creation operators from different users resulting in inter-user interference. 

At the receiver, the creation operators admit another multiplicative factor due to the decoder. Then the optical narrow-band filter breaks this decoded signal into two parts. The reflected part is due to the rejected inter-user interference. In contrast, the transmitted part is due to the intended signal that is correctly decoded along with inevitable inter-user interference that passes through the narrow-band filter. The duration of the output quantum signal at the receiver depends on the impulse response of the optical filter. 

Corollary \ref{col:timePicture2} summarizes the overall behavior of the considered quantum communication system.
\begin{corollary}[Overall Evolution]
	\label{col:timePicture2}
	The overall evolution of the creation operators in a quantum spread spectrum CDMA communication system, where receiver $r$ is decoding the signal from transmitter $s=r$ is given by
	\begin{align}
		\hat{\mathrm{H}}\hat{\mathrm{U}}^\dagger \hat{\mathrm{B}}\hat{\mathrm{U}}&\hat{a}^{\dagger}_{s,\wavepackt}\hat{\mathrm{U}}^\dagger \hat{\mathrm{B}}^\dagger \hat{\mathrm{U}} \hat{\mathrm{H}}^\dagger =\\
		&\, B_{ss}\sum_{\FDcoef=0}^{\infty} {D}_{\FDcoef}\hat{a}_{r=s,\chippackt_\FDcoef}'^\dagger +B_{ss}\sum_{\FDcoef=0}^{\infty}{F}_{\FDcoef}\hat{b}_{r=s,\chippackt_\FDcoef}'^\dagger\nonumber\\
		& +\sum_{r\ne s}B_{rs}\sum_{\FDcoef=0}^{\infty}\left( D^{s,r}_{\FDcoef}\hat{a}_{r,\chippackt_\FDcoef}'^\dagger+F^{s,r}_{\FDcoef}\hat{b}_{r,\chippackt_\FDcoef}'^\dagger\right).\nonumber
	\end{align}
\end{corollary}
From Corollary \ref{col:timePicture2}, we observe that the term $\sum_{\FDcoef=0}^{\infty} {D}_{\FDcoef}\hat{a}_{r=s,\chippackt_\FDcoef}'^\dagger$ corresponds to the recovered quantum signal that is transmitted through the filter, $\sum_{\FDcoef=0}^{\infty}{F}_{\FDcoef}\hat{b}_{r=s,\chippackt_\FDcoef}'^\dagger$ denotes the reflected portion of the desired quantum signal due to loss induced by the frequency mismatch between the bandwidth of filter and wavepacket of the desired quantum signal, $\sum_{r\ne s}B_{rs}\sum_{\FDcoef=0}^{\infty}\left( D^{s,r}_{\FDcoef}\hat{a}_{r,\chippackt_\FDcoef}'^\dagger\right)$ denotes the interference caused by transmitter $s$ at other receivers $r\ne s$, and $\sum_{r\ne s}B_{rs}\sum_{\FDcoef=0}^{\infty}\left( F^{s,r}_{\FDcoef}\hat{b}_{r,\chippackt_\FDcoef}'^\dagger\right)$  corresponds to the part of the signal of transmitter $s$,
reached receiver $r\neq s$ and filtered out. 

From the discussion of Appendix \ref{appendix:asymptotic} on the asymptotic behavior of the quantum spread spectrum communication systems, the values of the output transmission coefficients of the filter $D^{s,r}_\FDcoef$ are negligible for interfering quantum signals. On the other hand for the correctly decoded signal of the transmitter $s$ at receiver $r=s$, the values of $F^{s,r=s}_\FDcoef={F}_{\FDcoef}$ are negligible (see Fig. \ref{fig:asymptotic}). Thus, we may argue that the results of Corollary \ref{col:timePicture2} can be further simplified asymptotically as
\begin{align}
	\hat{\mathrm{H}}\hat{\mathrm{U}}^\dagger \hat{\mathrm{B}}&\hat{\mathrm{U}}\hat{a}^{\dagger}_{s,\wavepackt}\hat{\mathrm{U}}^\dagger \hat{\mathrm{B}}^\dagger \hat{\mathrm{U}} \hat{\mathrm{H}}^\dagger  \\
	&\approx B_{ss}\sum_{\FDcoef=\FDcoef_{\text{delay}}}^{\FDcoef_{\text{delay}}+N_c} {D}_{\FDcoef}\hat{a}_{r=s,\chippackt_\FDcoef}'^\dagger +\sum_{r\ne s}B_{rs}\sum_{\FDcoef=\FDcoef_{\text{delay}}}^{\FDcoef_{\text{delay}}+N_c}F^{s,r}_{\FDcoef}\hat{b}_{r,\chippackt_\FDcoef}'^\dagger\nonumber\\
	&\approx B_{ss}\hat{a}_{r=s,\wavepackt(t-T_{\text{delay}})}'^\dagger + \sum_{r\ne s}B_{rs}\hat{b}_{r,\wavepackt^{e_sd_r}}'^\dagger,
\end{align}
which shows that the creation operator of the transmitted signal from user $s$ is approximately completely recovered at receiver $r=s$ after the filter and rejected (reflected toward the other port of the filter) at receivers $r \neq s$. .
\subsection{Signal Intensity}
\label{sec::intensity}
In this subsection we calculate the intensity of the received quantum signals. The instantaneous intensity of the quantum signal at receiver $r_0$ at time $t$ is obtained as
\begin{align}
    I_{r_0}(t)&=\ev{\hat{a}^{'\dagger}_{r_0}(t)\hat{a}'_{r_0}(t)}{\Phi^d}.
\end{align}
Also the total intensity of the chip-time $k$ can be defined as follows
\begin{align}
	I_{r_0,k}&=\int_{t_k}^{t_{k+1}}I_{r_0}(t) dt.
\end{align}

The following lemma would be usefull in our calculations.
\begin{lemma}
	\label{lem:coefficients}
	The filter coefficients have the following property
	\begin{align}
		\sum_{\FDcoef=0}^{\infty}(D_{\FDcoef}^{s,r*}{D}_{\FDcoef}^{s',r}+{F}_{\FDcoef}^{s,r*}{F}_{\FDcoef}^{s',r}) &= \frac{1}{N_c}\sum_{k=0}^{N_c-1}\lambda_{s,k}\lambda_{s',k}\\
		&=\braket{\Lambda_s}{\Lambda_{s'}},\nonumber
	\end{align}
where $\braket{\Lambda_s}{\Lambda_{s'}} $ corresponds to the normalized correlation between code sequences $\Lambda_s$ and $\Lambda_{s'}$.
\end{lemma}
\begin{proof}
    See Appendix \ref{appendix:intensity}.
\end{proof}

\begin{proposition}
	\label{prop:intensity}
    The intensity of the received quantum signal of a spread spectrum CDMA communication system incorporating either number states or coherent states at the transmitters at receiver $r_0$ is given by
    \begin{enumerate}
        \item For number state transmitted quantum signals 
        \begin{align}
            I_{r_0}(t)&=\frac{N_c}{\pulseduration} \sum_{s=0}^{M-1} n_s | B_{r_0s}D_{\FDcoef}^{s,r_0}|^2, \quad t \in [ t_\FDcoef, t_{\FDcoef+1} ),
        \end{align}
    	and the total intensity of the time interval $\FDcoef$ is 
    \begin{align}
    	I_{r_0,\FDcoef}& =  \sum_{s=0}^{M-1} n_s | B_{r_0s}D_{\FDcoef}^{s,r_0}|^2.
    \end{align}
        \item For coherent state transmitted quantum signals
        \begin{align}
            I_{r_0}(t)&=\frac{N_c}{\pulseduration} \left|\sum_{s=0}^{M-1}B_{r_0s}D_{\FDcoef}^{s, r_0}\alpha_{s}\right|^2, \quad t \in [ t_\FDcoef, t_{\FDcoef+1} ), \label{eq:coherentIntensity}
        \end{align}
    and the total intensity of the chip-time $\FDcoef$ is 
    \begin{align}
    	I_{r_0,\FDcoef}& =  \left|\sum_{s=0}^{M-1}B_{r_0s}D_{\FDcoef}^{s, r_0}\alpha_{s}\right|^2 .
    \end{align}
    \end{enumerate}
    
\end{proposition}
\begin{proof}
    See Appendix \ref{appendix:intensity}.
\end{proof}

For a balanced star coupler $|B_{rs}|^2=\frac{1}{M}$ and $r_0=0$, for number state transmission
\begin{align}
    I_{0}(t)=\frac{1}{M}\frac{N_c}{\pulseduration} \left(n_{0}|D_{\FDcoef}|^2 + \sum_{s\ne 0} n_s | D_{\FDcoef}^{s,0}|^2\right), \quad t\in [t_\FDcoef, t_{\FDcoef+1}).
\end{align}
As can be seen, the intensity is the sum of the weakened decoded signal $\frac{1}{M}\frac{N_c}{\pulseduration} n_{0}| D_{\FDcoef}|^2 $ and the multiaccess interfering signal, $\frac{1}{M}\frac{N_c}{\pulseduration}\sum_{s\ne 0} n_s | D_{\FDcoef}^{s,0}|^2$ .

For coherent states the corresponding intensity is  
\begin{align}
    I_{0}(t)=\frac{1}{M}\frac{N_c}{\pulseduration} \left|D_{\FDcoef}\alpha_{0} + \sum_{s\ne 0}D_{\FDcoef}^{s, 0}\alpha_{s}\right|^2, \quad t\in [t_\FDcoef, t_{\FDcoef+1}),
\end{align}
where the signal, $D_{\FDcoef}\alpha_{0}$, and interference, $\sum_{s\ne 0}D_{\FDcoef}^{s, 0}\alpha_{s}$, are being added coherently.

The difference between these two detection schemes, namely the coherent detection scheme for Glauber states and the incoherent detection scheme for number states, arises from the Heisenberg uncertainty principle.  Similar to the spectral QCDMA, in the context of quantum spread spectrum CDMA communication systems, the uncertainty principle leads to the unpredictability of the quantum phase angles of number states at the time of measurement, which fundamentally changes the expression of the received intensity  as discussed in \cite{rezai2021quantum}. 

\section{Two User Quantum Spread Spectrum Interference}
\label{sec:two}
As an example, we study a two-user multiple access communication system utilizing on-off keying (OOK) modulation for the transmission of one bit. Assume that the receiver with index $0$ is decoding the signal from transmitter $0$. We consider a balanced star coupler, that is
\begin{align}
    \begin{pmatrix}
        B_{00} & B_{01}\\
        B_{10} & B_{11}
    \end{pmatrix} = \frac{1}{\sqrt{2}}
    \begin{pmatrix}
        1 & 1\\
        -1 & 1
    \end{pmatrix}.
\end{align}

\subsection{Coherent Input Quantum Signals}
The output of a two-user coherent quantum spread spectrum CDMA communication system at receiver $r=0$ is a pure state given by Proposition \ref{prop:coherentFiltered} as
\begin{align}
	\rho_0^{\text{T}} &= \ketbra{\alpha_{0,\wavepackt_{\text{T}}}},
\end{align}
where
\begin{align}
	\ket{\alpha_{0,\wavepackt_{\text{T}}}} & = \prod_{\FDcoef=0}^{\infty}\Bigg|\frac{1}{\sqrt{2}}D_{\FDcoef} \alpha_{0,\chippackt_\FDcoef}+ \frac{1}{\sqrt{2}}D_{\FDcoef}^{1, 0}\alpha_{1,\chippackt_\FDcoef}\Bigg\rangle .
\end{align}
The average intensity of the received signal is given by Proposition \ref{prop:intensity} as
\begin{align}
	I_{0}(t)=\frac{1}{2}\frac{N_c}{\pulseduration} \left|D_{\FDcoef}\alpha_{0} + D_{\FDcoef}^{1, 0}\alpha_{1}\right|^2, \quad t \in [ t_\FDcoef, t_{\FDcoef+1} ).
\end{align}

Let $n_{0,\FDcoef}$ be the number of photons in time interval $\FDcoef$. Since the quantum signal is a coherent state at each chip-time, the chip-time photon statistics, $\mathbb{P}(n_{0,\FDcoef})$, has a Poisson distribution
\begin{align}
	n_{0,\FDcoef}\sim \mathrm{Poisson}\left(\frac{1}{2}\left|D_{\FDcoef} \alpha_{0}+ D_{\FDcoef}^{1, 0}\alpha_{1}\right|^2\right),
\end{align}
which gives the overall photon statistics as follows
\begin{align}
	{n}_0\sim \mathrm{Poisson}\left(\frac{1}{2}\left|\sum_\FDcoef (D_{\FDcoef} \alpha_{0}+ D_{\FDcoef}^{1, 0}\alpha_{1})\right|^2\right),
\end{align}
where ${n}_0$ is the number of photons at the receiver after the filter.

\subsection{Single Photon Input Quantum Signals}
In this subsection we consider the output of a two-user single-photon quantum spread spectrum CDMA communication system at receiver $r=0$. The details are given in Appendix \ref{appendix:2singlephoton}. In this case, due to superposition of single photons, the received quantum signal at each receiver appears as a mixed states. 

\subsubsection{One User is Transmitting}
If transmitter $s=1$ is turned off, i.e., it is in vacuum state $\ket{0}$, the received signal of receiver  $r=0$ after decoder is
\begin{align}
	\rho_0^{d} &= \frac{1}{2}\ketbra{0_0} +  \frac{1}{2} \ketbra{1_{0,\wavepackt}}.
\end{align}

Applying the filter gives
\begin{align}
	\rho_0^{\text{T}} &= \frac{1}{2}\left( 1  +   \sum_{\FDcoef=0}^{\infty}|{F}_{\FDcoef}|^2 \right)\ketbra{0_{0,\text{T}}}\nonumber\\
	&+ \frac{1}{2}\left(\sum_{\FDcoef=0}^{\infty}|D_{\FDcoef}|^2\right)\ketbra{1_{0,\wavepackt_\text{T}}},
\end{align}
where $\ket{1_{0,\wavepackt_\text{T}}}$ is the desired signal after the filter defined as
\begin{align}
	\ket{1_{0,\wavepackt_\text{T}}}\coloneqq \frac{1}{\sqrt{\sum_{\FDcoef=0}^{\infty}|D_{\FDcoef}|^2}}\sum_{\FDcoef=0}^{\infty} D_{\FDcoef} \ket{1_{0,\text{T},\chippackt_\FDcoef}}.
\end{align}
The signal intensity is obtained using Proposition \ref{prop:intensity}
\begin{align}
	I_{0}(t)=\frac{1}{2}\frac{N_c}{\pulseduration} |D_{\FDcoef}|^2, \quad t \in [ t_\FDcoef, t_{\FDcoef+1} ).
\end{align}

\subsubsection{Both Users are Transmitting} If the second transmitter is also transmitting single photons, we obtain
\begin{align}
	&\ket{\Phi^d}\Big|_{M=2, \text{balanced}}\label{eq:2userd}\\
	&= \frac{1}{2N_c}\Bigg(\sum_{k_0=0}^{N_c-1}\ket{1_{0,\chippackt_{k_0}}}\Bigg)\Bigg(\sum_{k_1=0}^{N_c-1}\lambda_{0,k_1}\lambda_{1,k_1}\ket{1_{0,\chippackt_{k_1}}}\Bigg)\nonumber\\
	&- \frac{1}{2N_c}\Bigg(\sum_{k_1=0}^{N_c-1}\ket{1_{1,\chippackt_{k_1}}}\Bigg)\Bigg(\sum_{k_0=0}^{N_c-1}\lambda_{1,k_0}\lambda_{1,k_0}\ket{1_{1,\chippackt_{k_0}}}\Bigg)\nonumber\\
	&+\frac{1}{2N_c}\Bigg(\sum_{k_0=0}^{N_c-1}\sum_{k_1=0}^{N_c-1}(1-\lambda_{0,k_0}\lambda_{1,k_0}\lambda_{0,k_1}\lambda_{1,k_1})\nonumber\\
	&\qquad\qquad\qquad\qquad\qquad\times\ket{1_{0,\chippackt_{k_0}}}\ket{1_{1,\chippackt_{k_1}}}\Bigg).\nonumber
\end{align}
The last term in summation is zero for $k_0=k_1$ regardless of the encoding sequences, 
which gives the spread spectrum version of the Hong-Ou-Mandel effect \cite{hong1987measurement} as follows. 

\begin{corollary}[Spread spectrum Hong-Ou-Mandel effect]
	The state $\ket{1_{0,\chippackt_k}}\ket{1_{1,\chippackt_k}}$ never appears in the output of the balanced two user quantum broadcasting channel for spread spectrum single photon inputs when two users are simultaneously transmitting information.
\end{corollary}
For identical codes $\Lambda _1=\Lambda_0$, (\ref{eq:2userd}) reduces to the following NOON state\footnote{NOON states belong to the class of Schrodinger cat states and have emerging applications in  quantum communications, protocol design \cite{grun2022protocol} and quantum error correction \cite{bergmann2016quantum}.}, which corresponds to the normal Hong-Ou-Mandel effect
\begin{align}
	\ket{\Phi^d}\Big|_{M=2, \text{balanced}}&=\frac{1}{\sqrt{2}}\left(\ket{2_{0,\wavepackt}}\ket{0_1}-\ket{0_0}\ket{2_{1,\wavepackt}}\right).
\end{align}

The overall received quantum signal after decoder of  receiver $r=0$ is obtained, in terms of the normalized correlation ${\braket{\Lambda_0}{\Lambda_1}}$, by taking partial trace of $\ketbra{\Phi^d}$ with respect to $r=1$, which gives
\begin{align}
	\rho_0^d & =\frac{1}{4}\left(1+\left|\braket{\Lambda_0}{\Lambda_1}\right|^2\right)\ketbra{(1+1)_{0,e_1d_0}}\label{eq:2userdensityd}\\
	&+\frac{1}{4}\left(1+\left|\braket{\Lambda_0}{\Lambda_1}\right|^2\right)\ketbra{0_0}\nonumber\\
	&+\frac{1}{2N_c}\sum_{k=0}^{N_c-1}\left(1-\lambda_{0,k}\lambda_{1,k}\braket{\Lambda_0}{\Lambda_1}\right)\ketbra{1_{0,e_1d_0,k}}.\nonumber
\end{align} 
For random (approximately orthogonal) codes with ${\braket{\Lambda_0}{\Lambda_1}\approx 0}$, (\ref{eq:2userdensityd}) reduces to
\begin{align}
	\rho_0^d & \approx \frac{1}{4}\ketbra{(1+1)_{0,e_1d_0}} + \frac{1}{4}\ketbra{0_0}\\
	&+ \frac{1}{2N_c}\sum_{k=0}^{N_c-1}\ketbra{1_{0,e_1d_0,k}},\nonumber
\end{align}
where state $\ket{(1+1)_{0,e_1d_0}}\propto  \ket{1_{0,\wavepackt}}\ket{1_{0,\wavepackt^{e_1d_0}}}$ is defined in (\ref{eq:app:K:1p1}) and is associated with the product of the desired quantum signal with interference from transmitter $s=1$. On the other hand,
$\ket{1_{0,e_1d_0,k}}$ given by
\begin{align}
	\ket{1_{0,e_1d_0,k}} = \frac{\ket{1_{0,\wavepackt}}-\lambda_{0,k}\lambda_{1,k}\ket{1_{0,\wavepackt^{e_1d_0}}}}{\sqrt{2(1-\lambda_{0,k}\lambda_{1,k}\braket{\Lambda_0}{\Lambda_1})}}
\end{align}
 contains the desired signal in superposition with the multiple access interference signal.
 
After applying the filter and taking the partial trace, we obtain the mixed received quantum signal at the output of receiver $r=0$:
\begin{align}
	\rho_0^{\text{T}} =&\constdd\ketbra{(1+1)_{0,\text{T}, e_1d_0}}\\
	&+\sum_{\FDcoef=0}^{\infty}\constfkd  \ketbra{1_{0,\text{T},e_1d_0,F_\FDcoef } }\nonumber\\
	&+ \sum_{k=0}^{N_c-1}\constdLk\ketbra{1_{0,\text{T},e_1d_0,k}}\nonumber\\
	&+\constzeroT\ketbra{0_{0,\text{T}}}.\nonumber
\end{align} 
In the above expression, $\ket{(1+1)_{0,\text{T}, e_1d_0}}$ corresponds to the case where two photons from the two transmitters have arrived at receiver $r=0$ and both have passed through the filter. Quantum signals of the form $\ket{1_{0,\text{T},e_1d_0,F_\FDcoef } }$ correspond to the cases where  two photons from the two transmitters have arrived at receiver $r=0$, one of them have passed through the filter and the reflected one corresponds to time interval~$\FDcoef$. State $\ket{1_{0,\text{T},e_1d_0,k}}$ corresponds to the case where one photon has arrived at receiver $r=0$ and it is also transmitted through the filter. The expressions for these states along with the corresponding multiplying factors are defined and calculated in Appendix \ref{appendix:2singlephoton}.   

According to Proposition \ref{prop:intensity}, the signal intensity is 
\begin{align}
	I_{0}(t)=\frac{1}{2}\frac{N_c}{\pulseduration} \left(|D_{\FDcoef}|^2 + | D_{\FDcoef}^{1,0}|^2\right), \quad t \in [tq, t{q+1}) .
\end{align}

One can also approximate the photon statistics of the received quantum signal independent from the reflection coefficients as follows.
\begin{proposition}
	\label{prop:probability}
	Consider  a two user spread spectrum QCDMA communication systems utilizing single photons with random spreading sequences and balanced broadcasting channel. Denote the set of active transmitters by $\mathcal{S}$.
	The photon statistics at receiver $r=0$ that utilizes the code $\Lambda_0$ to decode the quantum signal can be approximated as
	\begin{enumerate}
		\item $\mathcal{S}=\lbrace 0 \rbrace$
		\begin{align}
			\mathbb{P}({n}_0) \approx\left(1-\frac{1}{2}\mathfrak{D}\right)&\delta_{{n}_0,0} \\
			+ \frac{1}{2}\mathfrak{D}\times&\delta_{{n}_0,1} .\nonumber
		\end{align}
		\item $\mathcal{S}=\lbrace 1 \rbrace$
		\begin{align}
			\mathbb{P}({n}_0) \approx\left(1-\frac{1}{2}\mathfrak{D}^{0,1}\right)&\delta_{{n}_0,0} \\
			+ \frac{1}{2}\mathfrak{D}^{0,1}\times&\delta_{{n}_0,1}.\nonumber
		\end{align}
		\item $\mathcal{S}=\lbrace 0, 1 \rbrace$
		\begin{align}
			\mathbb{P}({n}_0) \approx\left(1-\frac{1}{2}(\mathfrak{D}+\mathfrak{D}^{0,1})+\frac{1}{4}\mathfrak{D}\mathfrak{D}^{0,1}\right)&\delta_{{n}_0,0} \\
			+\left(\frac{1}{2}(\mathfrak{D}+\mathfrak{D}^{0,1})-\frac{1}{2}\mathfrak{D}\mathfrak{D}^{0,1}\right)&\delta_{{n}_0,1}\nonumber\\
			+\frac{1}{4}\mathfrak{D}\mathfrak{D}^{0,1}\times&\delta_{{n}_0,2}.\nonumber
		\end{align}
	\end{enumerate}
	where the signal and interference overall transmission coefficients are respectively denoted by
	\begin{align}
		\mathfrak{D} &\coloneqq  \sum_{\FDcoef=0}^{\infty}|{D}_{\FDcoef}|^2,\\
		\mathfrak{D}^{0,1} &\coloneqq  \sum_{\FDcoef=0}^{\infty}|D^{0,1}_{\FDcoef}|^2.
	\end{align}
\end{proposition}

Proposition \ref{prop:probability} shows that the output photon statistics is approximately dominated by the transmission coefficients of the filter. As shown in Fig. \ref{fig:asymptotic} and discussed in Appendix \ref{appendix:asymptotic}, we asymptotically have $\mathfrak{D}\approx 1$ and $\mathfrak{D}^{0,1}\approx 0$.

\begin{figure}[!tb]
    \centering
    \includegraphics[width=0.9\linewidth]{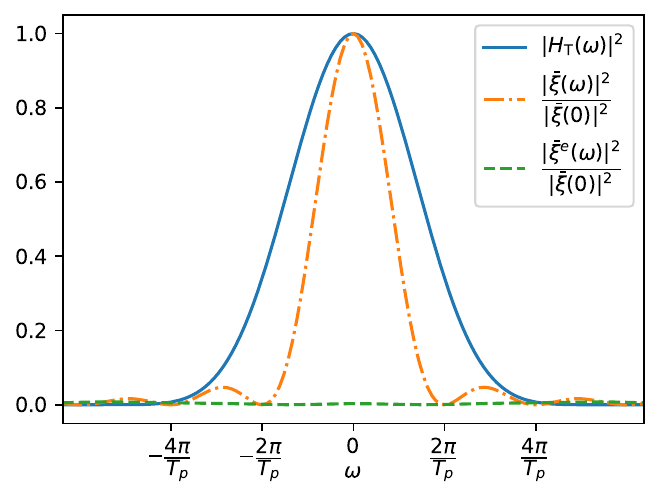}\caption{Frequency response of the filter compared to $|\bar{\wavepackt}(\omega)|^2$ and $|\bar{\wavepackt}^e(\omega)|^2$. The frequency domain wavepackets are divided by $|\bar{\wavepackt}(0)|^2$ for better illustration. Note that the original signal's temporal wavepacket is $\wavepackt(t)$ and spread spectrum signal has the temporal wavepacket of $\wavepackt^e(t)$. }
    \label{fig:frequencyResponse}
\end{figure}
\begin{figure*}[!tb]
    \centering
    \includegraphics[width=0.8\linewidth]{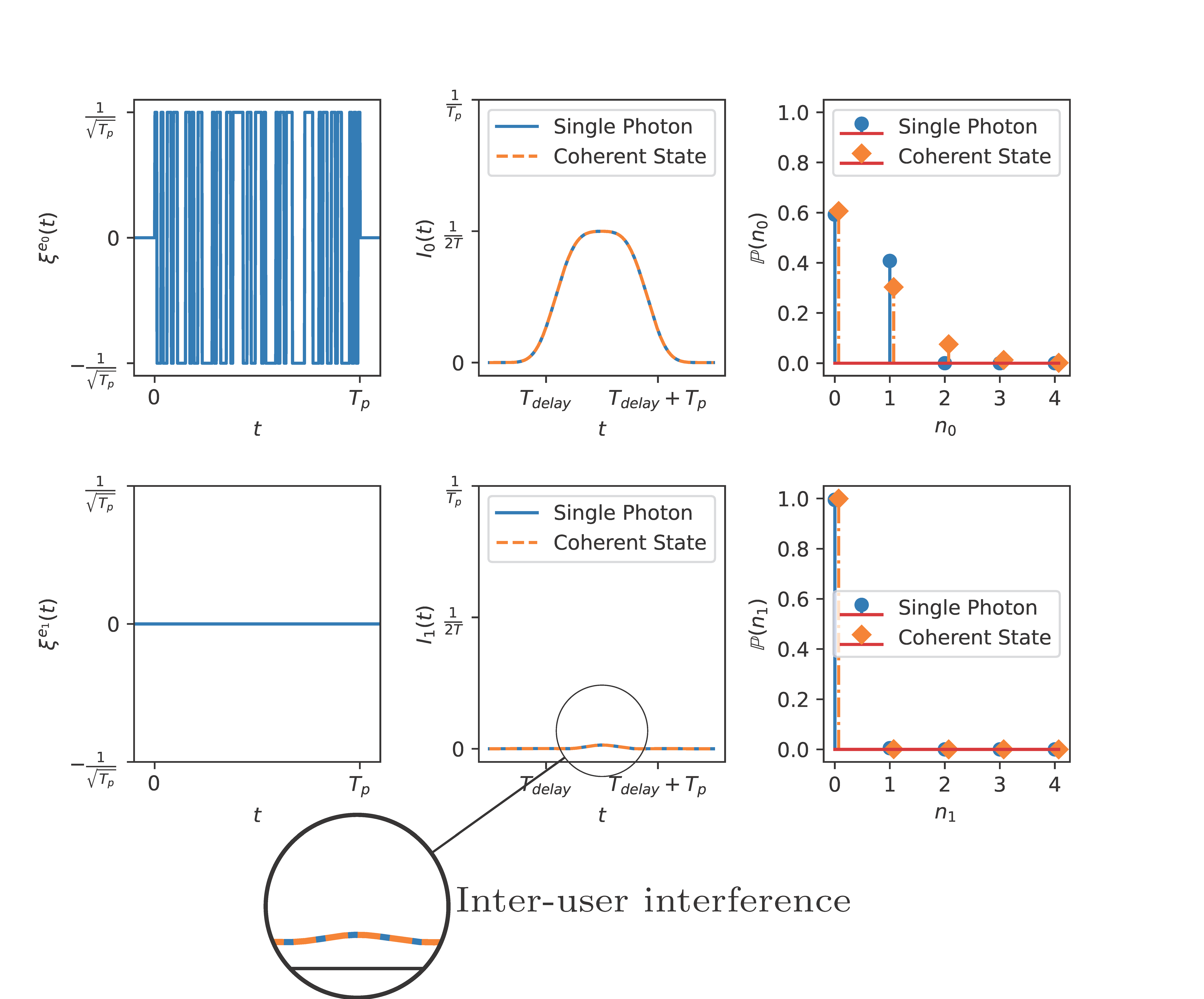}\caption{Signals for a two user system when the only first transmitter is on. The signal is reconstructed when the correct code is used at receiver $r=0$, while a small amount of inter-user interference from the first user is observed at $r=1$.}
    \label{fig:sim1}
\end{figure*}

\begin{figure*}[!tb]
    \centering
    \includegraphics[width=0.8\linewidth]{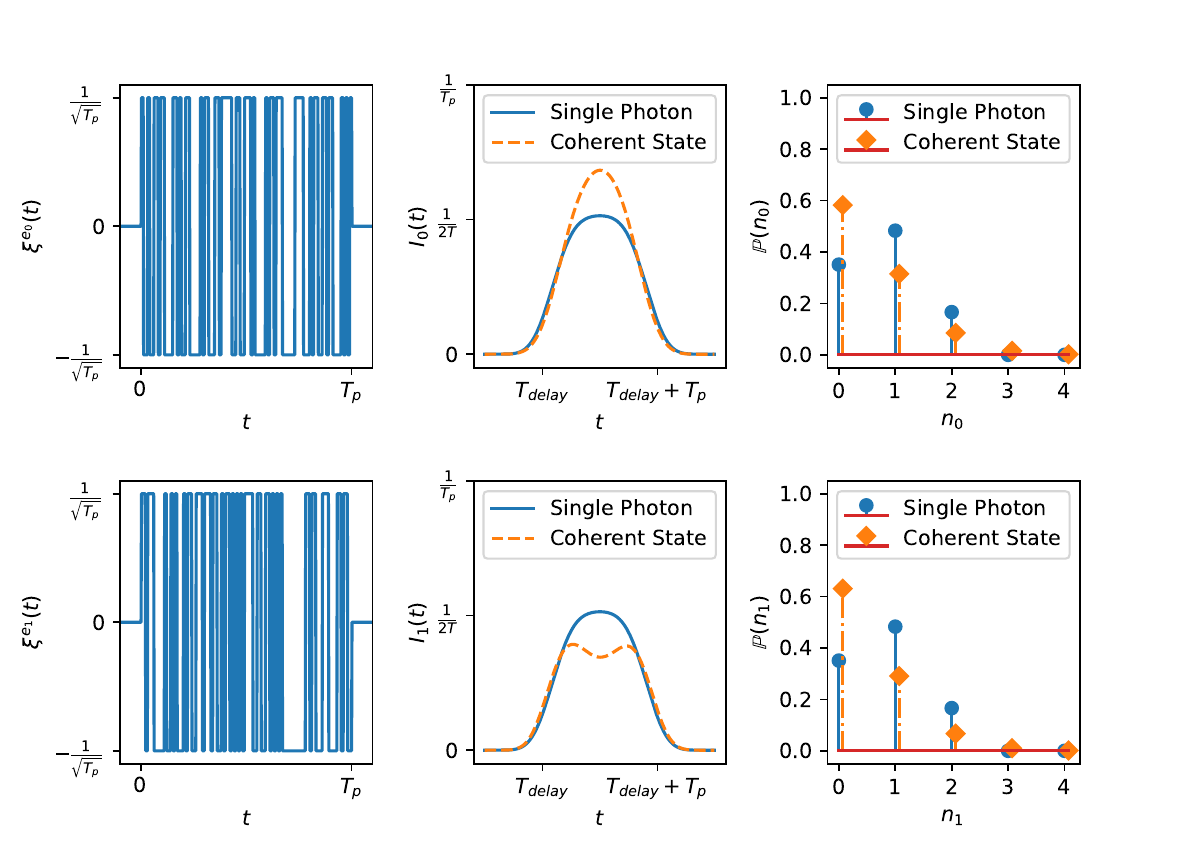}\caption{Signals for a two user system when both transmitters are on. The signals are reconstructed at the intended users.}
    \label{fig:sim2}
\end{figure*}

\subsection{Simulation Results}
In this subsection, a two-user spread spectrum QCDMA system with $N_c=100$ is simulated. The simulations are done using Python 3 and the filters are designed using the windowing method \cite{oppenheim2009discrete} available in scipy's signal processing package\footnote{Available at: https://docs.scipy.org/doc/scipy/reference/signal.html, \\accessed: 2023-01-20}. The transmitters are either transmitting single photons, $\ket{1}$, or coherent states $\ket{\alpha}$ with $\alpha=1$ and the signal duration is $\pulseduration$.  As can be seen in Fig. \ref{fig:frequencyResponse}, the encoding operator spreads the spectrum of the original signal into $\bar{\wavepackt}^e(\omega)$. The transmission response of the narrow-band filter is shown in Fig. \ref{fig:frequencyResponse}. The considered casual narrow-band filter is a FIR filter with $L=101$.

Fig. \ref{fig:sim1} shows the signals when only the first user, $s=0$, is transmitting and the second user, $s=1$, is in vacuum state. According to the received signal intensity, the signal is reconstructed when the correct code is used at receiver $r=0$. On the other hand, at receiver $r=1$, the inter-user interference from the first user is observed, which is negligible due to the use of the spread spectrum technique. This negligible interference can also be observed from the photon statistics, where almost all of the probability is concentrated around ${n}_1=0$. 
We also observe that while the average intensity of the signals for both coherent state and single photon sources are the same, their photon statistics are different.

Fig. \ref{fig:sim2} depicts the system's behavior when both users are transmitting simultaneously. In this scenario, when single photon sources are used, the average intensity of the received signal is similar for both users. In contrast, when coherent quantum signals are used, the average intensity of user $r=0$ is higher compared to receiver $r=1$ due to the positive sign of the interference term in (\ref{eq:coherentIntensity}). The photon statistics are also shifted to higher photon counts compared to when only one user was transmitting data.

\section{Conclusion}
\label{sec:conclusion}
This paper presents a comprehensive and unified approach for investigating quantum code-division multiple-access (CDMA) communication systems based on the spread spectrum technology. 

To study the system, a unique method termed chip-time interval decomposition of creation operators is developed. It is demonstrated that, similar to the concept of chips in classical CDMA communication systems, chip-time interval creation operators are the building blocks of spread spectrum quantum CDMA communications. The proposed chip-time interval decomposition approach is used to describe the evolution of quantum signals in this communication system.

 The evolution of quantum signals through the quantum communication system is studied in the Schrodinger's picture. We address two limiting scenarios for quantum transmitter signals: number (Fock) states and coherent (Glauber) states. We demonstrated that the QCDMA system's output at the receiver can be characterized using chip-time interval creation operators. Because of entanglement, particle-like single photon inputs appear as mixed outputs at each receiver, whereas wave-like coherent states result in pure quantum signals at the receivers.

A novel perspective on the quantum spread spectrum technology system is also given by applying the chip-time interval decomposition approach to describe the evolution of creation operators, resulting in the chip-time decomposition methodology for description of the QCDMA system. In this methodology, the quantum chip-time field creation operators are invariant building blocks of the communication system. The overall evolution of creation operators can be described as the weighted sum of various chip-time interval creation operators, as we have shown.

Furthermore, this study delves into the essential concepts of narrow-band filtering in quantum communication systems through the application of Fourier series and the proposed chip-time interval creation operators. The examination of the developed techniques provides a better understanding of how quantum signals can be  filtered and processed in a quantum communication system.

The mathematical models and techniques presented in this paper, related to filtering quantum signals and the application of quantum spread spectrum technology, provides a solid foundation for future research and development in the fields of quantum communications and signal processing. This study provides a strong basis for exploring new applications in areas such as quantum engineering, quantum communications, quantum signal processing, and related fields such as multiple access quantum communication systems, quantum receiver design, and quantum pulse shaping. It is expected that the novel methods developed in this study will make a significant contribution to the advancement of these fields.

\appendices
\renewcommand{\theequation}{\thesection.\arabic{equation}}
\numberwithin{equation}{section}
\section{Commutation of Chip-time interval Creation Operators}
\label{appendix:commutation}
It is evident that the function $\chippackt_k(t)$ has the required normalization property of a photon-wavepacket:
\begin{align}
    \int_0^\pulseduration |\chippackt_k(t)|^2 dt &= \int_{t_k}^{t_{k+1}}|\sqrt{N_c}\wavepackt(t)|^2 dt \\
	&= \int_{t_k}^{t_{k+1}}N_c|\wavepackt(t)|^2 dt = 1 \nonumber
\end{align}
Hence we may define the creation operator of each chip-time by letting $\hat{a}_{\chippackt_k}^\dagger$ be the creation operator for chip-time $[t_k, t_{k+1})$. Based on the wavepacket of (\ref{eq:wavepacketTimeIntervalgeneral}), we may write:
\begin{align}
    \hat{a}_{\chippackt_k}^\dagger & \coloneqq  \int_{t_k}^{t_{k+1}} dt \chippackt_k(t) \hat{a}^\dagger(t)=   \int_{t_k}^{t_{k+1}} dt \sqrt{N_c}\wavepackt(t) \hat{a}^\dagger(t).
\end{align}

We first show that according to the definition of the chip-time interval creation operators, $\comm{\hat{a}_{\chippackt_l}}{\hat{a}_{\chippackt_k}^\dagger}=\delta_{lk}$. For $l=k$, the relation $\comm{\hat{a}_{\chippackt_k}}{\hat{a}_{\chippackt_k}^\dagger}=1$ is trivial. For $l\ne k$ we may write
\begin{align}
	\hat{a}_{\chippackt_l}&\hat{a}_{\chippackt_k}^\dagger = \int_{t_l}^{t_{l+1}} dt \chippackt_{l}^*(t) \hat{a}(t) \int_{t_k}^{t_{k+1}} dt' \chippackt_k(t') \hat{a}^\dagger(t')\\
	&=\int_{t_l}^{t_{l+1}} \int_{t_k}^{t_{k+1}} dt dt' \chippackt_{l}^*(t) \chippackt_k(t') \hat{a}(t)\hat{a}^\dagger(t')\\
	&=\int_{t_l}^{t_{l+1}} \int_{t_k}^{t_{k+1}} dt dt' \chippackt_{l}^*(t) \chippackt_k(t') \left(\delta(t-t') + \hat{a}^\dagger(t')\hat{a}(t)\right) \\
	&= \int_{t_l}^{t_{l+1}} \int_{t_k}^{t_{k+1}} dt dt' \chippackt_{l}^*(t) \chippackt_k(t') \hat{a}^\dagger(t')\hat{a}(t), \ \text{since } t\neq t' \\
	&=\int_{t_k}^{t_{k+1}} dt' \chippackt_k(t') \hat{a}^\dagger(t')\int_{t_l}^{t_{l+1}} dt \chippackt_{l}^*(t) \hat{a}(t) =\hat{a}_{\chippackt_k}^\dagger \hat{a}_{\chippackt_l}\\
	\Rightarrow&\comm{\hat{a}_{\chippackt_l}}{\hat{a}_{\chippackt_k}^\dagger}=0 \label{eq:app:commute}
\end{align}
Note that since $\chippackt_l(t)$ and $\chippackt_k(t)$ are disjoint we may write:
\begin{align}
	&\int_{t_l}^{t_{l+1}} \int_{t_k}^{t_{k+1}} dt dt' \chippackt_{l}^*(t) \chippackt_k(t') \delta(t-t')\\
	&=\int_{t_l}^{t_{l+1}} dt \chippackt_{l}^*(t) \int_{t_k}^{t_{k+1}} dt' \chippackt_k(t') \delta(t-t')\\
	&= \int_{t_l}^{t_{l+1}} dt \chippackt_{l}^*(t) \chippackt_k(t) = 0 
\end{align}

Next we prove the chip-time interval decomposition theorem:
\begin{align}
    \hat{a}_{\wavepackt}^\dagger &= \int_{0}^{\pulseduration} dt \wavepackt(t) \hat{a}^\dagger(t) = \sum_{k=0}^{N_c-1} \int_{t_k}^{t_{k+1}} dt \wavepackt(t) \hat{a}^\dagger(t) \\
    &= \frac{1}{\sqrt{N_c}}\sum_{k=0}^{N_c-1} \int_{t_k}^{t_{k+1}} dt \sqrt{N_c}\wavepackt(t) \hat{a}^\dagger(t) \nonumber\\
    &= \frac{1}{\sqrt{N_c}}\sum_{k=0}^{N_c-1} \int_{t_k}^{t_{k+1}} dt \chippackt_k(t) \hat{a}^\dagger(t)\nonumber\\
    & = \frac{1}{\sqrt{N_c}}\sum_{k=0}^{N_c-1} \hat{a}_{\chippackt_k}^\dagger\nonumber
\end{align}

\section{Coherent State Decomposition}
\label{appendix:coherent}
We start by the displacement operator
\begin{align}
	D(\alpha_\wavepackt) &= \exp(\alpha \hat{a}_{\wavepackt}^\dagger - \alpha^*\hat{a}_{\wavepackt})\\
	&= \exp( \frac{\alpha}{\sqrt{N_c}}\sum_{k=0}^{N_c-1} \hat{a}_{\chippackt_k}^\dagger - \frac{\alpha^*}{\sqrt{N_c}}\sum_{k=0}^{N_c-1} \hat{a}_{\chippackt_k}) \\
	&= \exp(\sum_{k=0}^{N_c-1}  \left[\frac{\alpha}{\sqrt{N_c}}\hat{a}_{\chippackt_k}^\dagger - \frac{\alpha^*}{\sqrt{N_c}}\hat{a}_{\chippackt_k}\right]) \\
	&= \prod_{k=0}^{N_c-1}\exp(\frac{\alpha}{\sqrt{N_c}}\hat{a}_{\chippackt_k}^\dagger - \frac{\alpha^*}{\sqrt{N_c}}\hat{a}_{\chippackt_k})  \\
	&= \prod_{k=0}^{N_c-1} D\left(\frac{\alpha_{\chippackt_k}}{\sqrt{N_c}}\right)
\end{align}
where we have used the fact that for $ l\neq k$ (see (\ref{eq:app:commute}))
 \begin{align}
	\comm{\frac{\alpha}{\sqrt{N_c}}\hat{a}_{\chippackt_l}^\dagger - \frac{\alpha^*}{\sqrt{N_c}}\hat{a}_{\chippackt_l}}{\frac{\alpha}{\sqrt{N_c}}\hat{a}_{\chippackt_k}^\dagger - \frac{\alpha^*}{\sqrt{N_c}}\hat{a}_{\chippackt_k}}=0
\end{align} and 
 \begin{align}
	\text{exp}(\hat{A}+\hat{B}) = \text{exp} (\hat{A})\text{exp} (\hat{B})\text{exp} (-\frac{1}{2}\comm{\hat{A}}{\hat{B}}).
\end{align}
The decomposition of coherent state can be obtained utilizing the displacement operator
\begin{align}
	\ket{\alpha_\wavepackt} &= 	D(\alpha_\wavepackt)\ket{0} = \left(\prod_{k=0}^{N_c-1} D\left(\frac{\alpha_{\chippackt_k}}{\sqrt{N_c}}\right)\right)\ket{0} \nonumber\\
	&=  \prod_{k=0}^{N_c-1} D\left(\frac{\alpha_{\chippackt_k}}{\sqrt{N_c}}\right)\ket{0} = \prod_{k=0}^{N_c-1} \ket{\frac{\alpha_{\chippackt_k}}{\sqrt{N_c}}}
\end{align}

\section{Number State Decomposition}
\label{appendix:number}
For the number states, we may write
\begin{align}
	\ket{n_\wavepackt} &=
	\frac{ (\hat{a}^\dagger_\wavepackt)^n}{\sqrt{n!}}\ket{0} \\
	&=\frac{1}{\sqrt{n!}} \left(\frac{1}{\sqrt{N_c}}\sum_{k=0}^{N_c-1} \hat{a}_{\chippackt_k}^\dagger\right)^n\ket{0}\\
	&= \frac{1}{\sqrt{n!}\sqrt{N_c^n}}\left(\sum_{k=0}^{N_c-1} \hat{a}_{\chippackt_k}^\dagger\right)^n\ket{0}\\
	&=\frac{1}{\sqrt{n!N_c^n}}\sum_{n_0+n_1+\cdots+n_{N_c-1}=n}{n\choose n_0,n_1,\cdots,n_{N_c-1}}\nonumber\\
	&\qquad\qquad\qquad\qquad\times\prod_{k=0}^{N_c-1}( \hat{a}_{\chippackt_k}^\dagger)^{n_k}\ket{0}\\
	&=\frac{1}{\sqrt{n!N_c^n}}\sum_{n_0+n_1+\cdots+n_{N_c-1}=n}{n\choose n_0,n_1,\cdots,n_{N_c-1}}\nonumber\\
	&\qquad\qquad\qquad\qquad\times\prod_{k=0}^{N_c-1}\sqrt{n_k!}\frac{( \hat{a}_{\chippackt_k}^\dagger)^{n_k}}{\sqrt{n_k!}}\ket{0}\\
	&= \frac{1}{\sqrt{n!N_c^n}} \sum_{n_0+n_1+\cdots+n_{N_c-1}=n} {n\choose n_0,n_1,\cdots,n_{N_c-1}}\nonumber\\
	&\qquad\qquad\qquad\qquad\times\prod_{k=0}^{N_c-1} \sqrt{n_k!}\ket{n_{k,\chippackt_k}}\\
	&= \sum_{n_0+n_1+\cdots+n_{N_c-1}=n} \mathcal{C}_n(n_0,n_1,\ldots,n_{N_c-1})\nonumber\\
	&\qquad\qquad\qquad\qquad\times\prod_{k=0}^{N_c-1} \ket{n_{k,\chippackt_k}}\\
	&= \sum_{n_0+n_1+\cdots+n_{N_c-1}=n} \mathcal{C}_n(n_0,n_1,\ldots,n_{N_c-1})\nonumber\\
	&\qquad\qquad\qquad\qquad\times\ket{n_{0,\chippackt_0}, n_{1,\chippackt_1}, \cdots, n_{N_c-1,\chippackt_{N_c-1}}}
\end{align}
where $\ket{n_{k,\chippackt_k}}$ denotes that the state has $n_k$ photons with wavepacket $\chippackt_k(t)$ in interval $[t_k, t_{k+1})$. And $\prod_{k=0}^{N_c-1} \ket{n_{k,\chippackt_k}}=\ket{n_{0,\chippackt_0}, n_{1,\chippackt_1}, \cdots, n_{N_c-1,\chippackt_{N_c-1}}}$ denotes a quantum state with $n_0$ photons in $[t_0, t_1)$, $n_{N_c-1}$ photons in $[t_{N_c-1}, t_{N_c})$. This state occurs with probability $|\mathcal{C}_n(n_0,n_1,\ldots,n_{N_c-1})|^2$, where
\begin{align}
	\mathcal{C}_n(n_0,n_1,\ldots,n_{N_c-1}) & \coloneqq  \frac{1}{\sqrt{n!N_c^n}} {n\choose n_0,n_1,\cdots,n_{N_c-1}}\nonumber\\
	&\qquad\qquad\qquad\times\prod_{k=0}^{N_c-1} \sqrt{n_k!}\nonumber\\
	&=\frac{1}{\sqrt{N_c^n}} \sqrt{{n\choose n_0,n_1,\cdots,n_{N_c-1}}}\nonumber
\end{align}
and
\begin{align}
	{n\choose n_0,n_1,\cdots,n_{N_c-1}} = \frac{n!}{n_0!n_1!\cdots n_{N_c-1}!}
\end{align}
In the above derivation, we have used the multinomial theorem (for commuting operators):
\begin{align}
	\left(\sum_{k=0}^{N_c-1} \hat{x}_k\right)^n 
	&= \sum_{n_0+n_1+\cdots+n_{N_c-1}=n} {n\choose n_0,n_1,\cdots,n_{N_c-1}} \nonumber\\
	&\qquad\qquad\qquad\qquad\times\prod_{k=0}^{N_c-1}(\hat{x}_k)^{n_k}
\end{align}

\section{Decomposition of the Unitary Spreading Operator}
\label{appendix:unitary}
\subsection{Commutation Relation}
\label{appendix:unitary:commute}
The decomposition is obtained from the definition
\begin{align}
    \hat{\mathrm{U}} &= \exp({j}\sum_{k=0}^{N_c-1} \int_{t_k}^{t_{k+1}} dt \theta(t)\hat{a}^\dagger(t)\hat{a}(t))\\
    &=\prod_{k=0}^{N_c-1}\exp({j}\int_{t_k}^{t_{k+1}} dt \theta(t)\hat{a}^\dagger(t)\hat{a}(t)) \\
    &= \prod_{k=0}^{N_c-1}\exp({j}\int_{t_k}^{t_{k+1}} dt \theta_k\hat{a}^\dagger(t)\hat{a}(t))\\
    &= \prod_{k=0}^{N_c-1}\hat{\mathrm{U}}_k
\end{align}
where
\begin{equation}
    \hat{\mathrm{U}}_k \coloneqq  \exp({j}\int_{t_k}^{t_{k+1}} dt \theta_k\hat{a}^\dagger(t)\hat{a}(t))
\end{equation}
and we have used the fact that for $ l\neq k$:
\begin{equation}
    \comm{\int_{t_k}^{t_{k+1}} dt \theta(t)\hat{a}^\dagger(t)\hat{a}(t)}{\int_{t_l}^{t_{l+1}} dt \theta(t')\hat{a}^\dagger(t')\hat{a}(t')} = 0 
\end{equation}

Continuing our discussion regarding the spreading operator, we show that $\comm{\hat{\mathrm{U}}_k}{\hat{a}_{\chippackt_l}^\dagger} = 0, l\neq k$. It is sufficient to prove
\begin{align}
	\comm{\hat{\mathrm{U}}_k}{\hat{a}^\dagger(t)} &= 0,\qquad  t\notin[t_k, t_{k+1}).
\end{align}
Again we begin by the definition
\begin{align}
	\hat{\mathrm{U}}_k\hat{a}^\dagger(t)&= \exp({j}\int_{t_k}^{t_{k+1}} dt' \theta_k\hat{a}^\dagger(t')\hat{a}(t'))\hat{a}^\dagger(t)\\
	&=\sum_{l=0}^{\infty} \frac{({j})^l}{l!}\left(\int_{t_k}^{t_{k+1}} dt' \theta_k\hat{a}^\dagger(t')\hat{a}(t')\right)^l\hat{a}^\dagger(t)\\
	&=\bigg[I+{j}\int_{t_k}^{t_{k+1}} dt' \theta_k\hat{a}^\dagger(t')\hat{a}(t')\\
	&\qquad+\frac{({j})^2}{2!}\int_{t_k}^{t_{k+1}} dt' \theta_k\hat{a}^\dagger(t')\hat{a}(t')\nonumber\\
	&\qquad\qquad\times\int_{t_k}^{t_{k+1}} dt'' \theta_k\hat{a}^\dagger(t'')\hat{a}(t'')+\cdots\bigg] \hat{a}^\dagger(t)\nonumber\\
	&=\hat{a}^\dagger(t)\sum_{l=0}^{\infty} \frac{({j})^l}{l!}\left({j}\int_{t_k}^{t_{k+1}} dt' \theta_k\hat{a}^\dagger(t')\hat{a}(t')\right)^l \\
	&= \hat{a}^\dagger(t)\hat{\mathrm{U}}_k
\end{align}
The first term is trivial:
\begin{align}
	I\hat{a}^\dagger(t) &= \hat{a}^\dagger(t)I
\end{align}
Note that we always have $\comm{\hat{a}^\dagger(t)}{\hat{a}^\dagger(t')}=0$, therefore, $\hat{a}^\dagger(t)\hat{a}^\dagger(t')=\hat{a}^\dagger(t')\hat{a}^\dagger(t)$. Also since $t'\in[t_k,t_{k+1})$ and $t\notin [t_k,t_{k+1})$, we have $\hat{a}(t')\hat{a}^\dagger(t)=\delta(t'-t)+\hat{a}^\dagger(t)\hat{a}(t')=\hat{a}^\dagger(t)\hat{a}(t')$  . Hence, the second term will be:
\begin{align}
	&\int_{t_k}^{t_{k+1}} dt' \theta_k\hat{a}^\dagger(t')\hat{a}(t')\hat{a}^\dagger(t) =\nonumber\\ &\int_{t_k}^{t_{k+1}} dt' \theta_k\hat{a}^\dagger(t')\hat{a}^\dagger(t)\hat{a}(t') =\nonumber\\ &\int_{t_k}^{t_{k+1}} dt' \theta_k\hat{a}^\dagger(t)\hat{a}^\dagger(t')\hat{a}(t') =\nonumber\\
	& 
	\hat{a}^\dagger(t)\int_{t_k}^{t_{k+1}} dt' \theta_k\hat{a}^\dagger(t')\hat{a}(t')
\end{align}
With the same procedure we can prove that 
\begin{align}
	&\left(\int_{t_k}^{t_{k+1}} dt' \theta_k\hat{a}^\dagger(t')\hat{a}(t')\right)^l\hat{a}^\dagger(t) =\nonumber\\ &\qquad\hat{a}^\dagger(t)\left(\int_{t_k}^{t_{k+1}} dt' \theta_k\hat{a}^\dagger(t')\hat{a}(t')\right)^l
\end{align}
Thus we have shown that
\begin{align}
    \comm{\hat{\mathrm{U}}_k}{\hat{a}_{\chippackt_l}^\dagger} &= 0,\qquad l\neq k.
\end{align}
Hence utilizing Taylor expansion of the function $f(\cdot)$ we obtain, 
\begin{align}
    \comm{\hat{\mathrm{U}}_k}{f(\hat{a}_{\chippackt_l}^\dagger)} &= 0,\qquad l\neq k.
\end{align}
\subsection{Effect of Spreading Unitary Operator on the Creation Operator of Quantum Signals}
\label{appendix:unitary:creation}
Applying $\hat{\mathrm{U}}_k$ changes the photon wavepacket $\chippackt_k(t)$ to $\chippackt_k^{e}(t)=e^{{j}\theta_k}\chippackt_k(t)=\lambda_k\chippackt_k(t)$. Hence,
\begin{align}
	\hat{a}_{\chippackt_k^{e}}^\dagger &= \int_{t_k}^{t_{k+1}} dt \chippackt_k^{e}(t) \hat{a}^\dagger(t) = \int_{t_k}^{t_{k+1}} dt \lambda_k\chippackt_k(t) \hat{a}^\dagger(t)\nonumber\\
	&=\lambda_k \int_{t_k}^{t_{k+1}} dt \chippackt_k(t) \hat{a}^\dagger(t) = \lambda_k\hat{a}_{\chippackt_k}^\dagger
\end{align}
Thus we conclude that
 \begin{align}
 	\hat{\mathrm{U}} \hat{a}_{\chippackt_k}^\dagger \hat{\mathrm{U}}^\dagger = \hat{\mathrm{U}}_k \hat{a}_{\chippackt_k}^\dagger \hat{\mathrm{U}}_k^\dagger = \lambda_k\hat{a}_{\chippackt_k}^\dagger
 \end{align}

Therefore, we may write
\begin{align}
	\hat{\mathrm{U}} \hat{a}_{\wavepackt}^\dagger \hat{\mathrm{U}}^\dagger  = \frac{1}{\sqrt{N_c}}\sum_{k=0}^{N_c-1} \hat{\mathrm{U}}\hat{a}_{\chippackt_k}^\dagger\hat{\mathrm{U}}^\dagger = \frac{1}{\sqrt{N_c}}\sum_{k=0}^{N_c-1} \lambda_k\hat{a}_{\chippackt_k}^\dagger
\end{align}
\section{Spread Spectrum Coherent State}
\label{appendix:sscoherent}
\subsection{Encoding}
The effect of encoding operator on the coherent state:
\label{appendix:sscoherentEn}
\begin{align}
	&\ket{\alpha_{\wavepackt^e}} = \hat{\mathrm{U}}\ket{\alpha_\wavepackt} = \left(\prod_{k=0}^{N_c-1}\hat{\mathrm{U}}_k\right)\left(\prod_{k=0}^{N_c-1} \ket{\frac{\alpha_{\chippackt_k}}{\sqrt{N_c}}}\right)\\ 
	&= \left(\hat{\mathrm{U}}_0 \hat{\mathrm{U}}_1 \cdots \hat{\mathrm{U}}_{N_c-1}\right)\left( \ket{\frac{\alpha_{\chippackt_1}}{\sqrt{N_c}}} \cdots \ket{\frac{\alpha_{\chippackt_{N_c-1}}}{\sqrt{N_c}}}\right)\\
	&=\left(\hat{\mathrm{U}}_0 \hat{\mathrm{U}}_1 \cdots \hat{\mathrm{U}}_{N_c-1}\right) \left(f(\hat{a}_{\chippackt_{0}}^\dagger) f(\hat{a}_{\chippackt_{1}}^\dagger) \cdots f(\hat{a}_{\chippackt_{N_c-1}}^\dagger) \right)\ket{0}\\
	&= \hat{\mathrm{U}}_0 f(\hat{a}_{\chippackt_{0}}^\dagger)\hat{\mathrm{U}}_1 f(\hat{a}_{\chippackt_{1}}^\dagger) \cdots \hat{\mathrm{U}}_{N_c-1}f(\hat{a}_{\chippackt_{N_c-1}}^\dagger)  \ket{0}\\
	&= \prod_{k=0}^{N_c-1}\hat{\mathrm{U}}_kf(\hat{a}_{\chippackt_k}^\dagger)\ket{0}= \prod_{k=0}^{N_c-1} \hat{\mathrm{U}}_k\ket{\frac{\alpha_{\chippackt_k}}{\sqrt{N_c}}} 
\end{align}
where we have used $\comm{\hat{\mathrm{U}}_k}{f(\hat{a}_{\chippackt_l}^\dagger)} = 0, l\neq k$. 

Example for $N_c=2$:
\begin{align}
	\hat{\mathrm{U}}_0\hat{\mathrm{U}}_1 \ket{\alpha_{\wavepackt}}&= \hat{\mathrm{U}}_0\hat{\mathrm{U}}_1 f(\hat{a}_{\chippackt_{0}}^\dagger)f(\hat{a}_{\chippackt_{1}}^\dagger)\ket{0} \nonumber\\
	&= \hat{\mathrm{U}}_0 f(\hat{a}_{\chippackt_{0}}^\dagger)\hat{\mathrm{U}}_1f(\hat{a}_{\chippackt_{1}}^\dagger)\ket{0} 
\end{align}

\begin{align}
	\hat{\mathrm{U}}_k\ket{\frac{\alpha_{\chippackt_k}}{\sqrt{N_c}}} &=\ket{\frac{\alpha_{\chippackt_k^{e}}}{\sqrt{N_c}}} \\
	&=  e^{-\frac{1}{2}\left|\frac{\alpha}{\sqrt{N_c}}\right|^2}e^{\frac{\alpha}{\sqrt{N_c}}\hat{a}_{\chippackt_k^{e}}^\dagger}\ket{0}\\
	&=
	e^{-\frac{1}{2}\left|\frac{\alpha}{\sqrt{N_c}}\right|^2}e^{\frac{\alpha}{\sqrt{N_c}}\lambda_k\hat{a}_{\chippackt_k}^\dagger}\ket{0}\\
	&=e^{-\frac{1}{2}\left|\frac{\lambda_k\alpha}{\sqrt{N_c}}\right|^2}e^{\frac{\alpha}{\sqrt{N_c}}\lambda_k\hat{a}_{\chippackt_k}^\dagger}\ket{0} = \ket{\frac{\lambda_k\alpha_{\chippackt_k}}{\sqrt{N_c}}}
\end{align}
\subsection{Decoding}
The effect of decoding operator with decoding sequence $\tilde{\Lambda}$ on the coherent states encoded with sequence $\Lambda$ is derived as follows
\label{appendix:sscoherentDe}
\begin{align}
	\ket{\alpha_{\wavepackt^d}} &= 	\hat{\mathrm{U}}^\dagger\ket{\alpha_{\wavepackt^e}}= \left(\prod_{k=0}^{N_c-1}\hat{\mathrm{U}}^{\dagger}_k\right)\left(\prod_{k=0}^{N_c-1} \ket{\frac{\lambda_k\alpha_{\chippackt_k}}{\sqrt{N_c}}}\right)\\ 
	&=\left(\prod_{k=0}^{N_c-1}{\hat{\mathrm{U}}}^{\dagger}_k\right)\left(\prod_{k=0}^{N_c-1} f(\lambda_k\hat{a}_{\chippackt_k}^\dagger)\right)\ket{0}\\
	&=\prod_{k=0}^{N_c-1}  {\hat{\mathrm{U}}}^{\dagger}_k f(\lambda_k\hat{a}_{\chippackt_k}^\dagger)\ket{0}= \prod_{k=0}^{N_c-1}  {\hat{\mathrm{U}}}^{\dagger}_k\ket{\frac{\lambda_k\alpha_{\chippackt_k}}{\sqrt{N_c}}} 
\end{align}
Applying ${\hat{\mathrm{U}}}^{\dagger}_k$ changes the photon wavepacket $\chippackt_k^{e}(t)$ to $\chippackt_k^{d}(t)=e^{-j\tilde{\theta}_k}\chippackt_k^{e}(t)=\tilde{\lambda}_k\chippackt_k^{e}(t)=\tilde{\lambda}_k\lambda_k\chippackt_k(t)$, where $\tilde{\lambda}_k$ is the value of despreading code at chip-time $k$. Hence,
\begin{align}
	\hat{a}_{\chippackt_k^{d}}^\dagger &= \int_{t_k}^{t_{k+1}} dt \chippackt_k^{d}(t) \hat{a}^\dagger(t) = \int_{t_k}^{t_{k+1}} dt \tilde{\lambda}_k\chippackt_k^{e}(t) \hat{a}^\dagger(t)\nonumber\\
	&=\tilde{\lambda}_k\lambda_k \int_{t_k}^{t_{k+1}} dt \chippackt_k(t) \hat{a}^\dagger(t) = \tilde{\lambda}_k\lambda_k\hat{a}_{\chippackt_k}^\dagger
\end{align}
The rest of the proof is similar to the previous part.

\section{Spread Spectrum Number State}
\label{appendix:ssnumber}
\subsection{Encoding}
\label{appendix:ssnumber:encoding}
The effect of encoding (spreading) operator on the number states is derived as follows
\begin{align}
	\hat{\mathrm{U}}\ket{n_\wavepackt}	&=\hat{\mathrm{U}}\frac{1}{\sqrt{n!N_c^n}}\sum_{n_0+\cdots+n_{N_c-1}=n}{n\choose n_0,n_1,\cdots,n_{N_c-1}}\nonumber\\
	&\qquad\qquad\qquad\qquad\times\prod_{k=0}^{N_c-1}( \hat{a}_{\chippackt_k}^\dagger)^{n_k}\ket{0}\\
	&= \frac{1}{\sqrt{n!N_c^n}} \sum_{n_0+\cdots+n_{N_c-1}=n} {n\choose n_0,n_1,\cdots,n_{N_c-1}}\hat{\mathrm{U}}\nonumber\\
	&\qquad\qquad\qquad\qquad\times\prod_{k=0}^{N_c-1}( \hat{a}_{\chippackt_k}^\dagger)^{n_k}\ket{0}\\
	&= \frac{1}{\sqrt{n!N_c^n}} \sum_{n_0+\cdots+n_{N_c-1}=n} {n\choose n_0,n_1,\cdots,n_{N_c-1}}\nonumber\\
	&\qquad\qquad\qquad\qquad\times\left(\prod_{k=0}^{N_c-1}\hat{\mathrm{U}}_k\right)\left(\prod_{k=0}^{N_c-1}( \hat{a}_{\chippackt_k}^\dagger)^{n_k}\right)\ket{0}\\
	&= \frac{1}{\sqrt{n!N_c^n}} \sum_{n_0+\cdots+n_{N_c-1}=n} {n\choose n_0,n_1,\cdots,n_{N_c-1}}\nonumber\\
	&\qquad\qquad\qquad\qquad\times\prod_{k=0}^{N_c-1}\hat{\mathrm{U}}_k( \hat{a}_{\chippackt_k}^\dagger)^{n_k}\ket{0}\\
	&=  \frac{1}{\sqrt{n!N_c^n}} \sum_{n_0+\cdots+n_{N_c-1}=n} {n\choose n_0,n_1,\cdots,n_{N_c-1}}\nonumber\\
	&\qquad\qquad\qquad\qquad\times\prod_{k=0}^{N_c-1}\sqrt{n_k!}\hat{\mathrm{U}}_k\ket{n_{\chippackt_k}}\\
	&=  \frac{1}{\sqrt{n!N_c^n}} \sum_{n_0+\cdots+n_{N_c-1}=n} {n\choose n_0,n_1,\cdots,n_{N_c-1}}\nonumber\\
	&\qquad\qquad\qquad\qquad\times\prod_{k=0}^{N_c-1}\sqrt{n_k!}\ket{n_{\chippackt_k^{e}}}\\
	& =  \sum_{n_0+\cdots+n_{N_c-1}=n} \mathcal{C}_n( n_0,n_1,\ldots,n_{N_c-1})\prod_{k=0}^{N_c-1}\ket{n_{\chippackt_k^{e}}},
\end{align}
where,
\begin{align}
	\ket{n_{\chippackt_k^{e}}} &= \hat{\mathrm{U}}_k\ket{n_{\chippackt_k}}= \frac{( \hat{a}_{\chippackt_k^{e}}^\dagger)^{n_k}}{\sqrt{n_k!}}\ket{0}\\
	& =\frac{1}{\sqrt{n_k!}} \left(\int_{t_k}^{t_{k+1}} dt \lambda_k\chippackt_k(t) \hat{a}^\dagger(t)\right)^{n_k}\ket{0} \\
	&=\frac{1}{\sqrt{n_k!}} (\lambda_k)^{n_k}\left(\int_{t_k}^{t_{k+1}} dt \chippackt_k(t) \hat{a}^\dagger(t)\right)^{n_k}\ket{0} \\
	&=(\lambda_k)^{n_k}\frac{( \hat{a}_{\chippackt_k}^\dagger)^{n_k}}{\sqrt{n_k!}} \ket{0} = (\lambda_k)^{n_k}\ket{n_{\chippackt_k}} \label{eq:numberEncodingChip}
\end{align}
This means that the chip-time number states are eigenstates of the operator $\hat{\mathrm{U}}_k$ with eigenvalue $(\lambda_k)^{n_k}$.

Hence,
\begin{align}
	\hat{\mathrm{U}}\ket{n_\wavepackt}	&= \ket{n_{\wavepackt^e}}= \sum_{n_0+\cdots+n_{N_c-1}=n} \mathcal{C}_n( n_0,n_1,\ldots,n_{N_c-1})\nonumber\\
	&\qquad\qquad\qquad\qquad\times\prod_{k=0}^{N_c-1}(\lambda_k)^{n_k}\ket{n_{\chippackt_k}}
\end{align}

\subsubsection{Example, $n=1$}
For this case, the constraint ${n_0+\cdots+n_{N_c-1}=n=1}$ implies that only one chip-time should contain the single photon, while other chip-times are in vacuum state. Therefore, for some $k_0$ we have $n_{k_0}=1$ and $n_l=0, l\ne k_0$. The overall encoded state is thus the superposition that can be expressed as a summation over $k_0=0, 1, \cdots, N_c-1$.
\begin{align}
	\ket{1_{\wavepackt^e}}&= \sum_{n_0+\cdots+n_{N_c-1}=1} \mathcal{C}_n( n_0,n_1,\ldots,n_{N_c-1})\nonumber\\
	&\qquad\qquad\qquad\qquad\times\prod_{k=0}^{N_c-1}(\lambda_k)^{n_k}\ket{n_{\chippackt_k}}
	\\
	&=\sum_{k_0=0}^{N_c-1} \frac{1}{\sqrt{N_c}}\sqrt{\frac{1!}{\underbrace{0!\cdots 0!}_{k_0}1!\underbrace{0!\cdots 0!}_{N_c-k_0-1}}}\nonumber\\
	&\qquad\qquad\qquad\times (\lambda_{k_0})^{1}\ket{1_{\chippackt_{k_0}}} \prod_{k\ne k_0}^{N_c-1}(\lambda_k)^{0}\ket{0_{\chippackt_k}}\\
	&= \frac{1}{\sqrt{N_c}}\sum_{k_0=0}^{N_c-1}\lambda_{k_0}\ket{1_{\chippackt_{k_0}}} 
\end{align}

\subsubsection{Example, $n=2$}For this case, the constraint ${n_0+\cdots+n_{N_c-1}=n=2}$ means that either for some $k_0$ the corresponding chip-time contains two photons, or for some $k_0$ and $k_1>k_0$, the two chip-times $k_0$ and $k_1$ contain a single photon simultaneously. Therefore
\begin{align}
	\ket{2_{\wavepackt^e}}&= \sum_{n_0+\cdots+n_{N_c-1}=2} \mathcal{C}_n( n_0,n_1,\ldots,n_{N_c-1})\nonumber\\
	&\qquad\qquad\qquad\qquad\times\prod_{k=0}^{N_c-1}(\lambda_k)^{n_k}\ket{n_{\chippackt_k}}
	\\
	&=\sum_{k_0=0}^{N_c-1} \frac{1}{\sqrt{N_c^2}}\sqrt{\frac{2!}{\underbrace{0!\cdots 0!}_{k_0}2!\underbrace{0!\cdots 0!}_{N_c-k_0-1}}}\nonumber\\
	&\qquad\qquad\qquad\times (\lambda_{k_0})^{2}\ket{2_{\chippackt_{k_0}}} \prod_{k\ne k_0}^{N_c-1}(\lambda_k)^{0}\ket{0_{\chippackt_k}}\nonumber\\
	&+\underset{\text{s.t. }k_1>k_0}{\sum_{k_1=0}^{N_c-1}\sum_{k_0=0}^{N_c-1}}\frac{1}{\sqrt{N_c^2}}\sqrt{\frac{2!}{1!1!}}(\lambda_{k_0})^{1}(\lambda_{k_1})^{1} \ket{1_{\chippackt_{k_0}}}\ket{1_{\chippackt_{k_1}}} \nonumber\\
	&\qquad\qquad\qquad\times\prod_{k\ne k_0, k_1}(\lambda_k)^{0}\ket{0_{\chippackt_k}}\\
	&=  \frac{1}{N_c}\sum_{k_0=0}^{N_c-1}\ket{2_{\chippackt_{k_0}}}\\
	&\quad +\frac{1}{N_c} \underset{\text{s.t. }k_1>k_0}{\sum_{k_1=0}^{N_c-1}\sum_{k_0=0}^{N_c-1}} \sqrt{2}\lambda_{k_0}\lambda_{k_1}\ket{1_{\chippackt_{k_0}}}\ket{1_{\chippackt_{k_1}}}\nonumber
\end{align}

\subsubsection{Example, $n=3$}For this case, from constraint ${n_0+\cdots+n_{N_c-1}=n=3}$, we have three possible cases. 
\begin{enumerate}
	\item For some $k_0$ the corresponding chip-time contains three photons $\ket{3_{\chippackt_{k_0}}}$.
	\item For some $k_0$ and $k_1$, there is a single photon at chip-time $k_0$ and two photons at chip-time $k_1$, $\ket{1_{\chippackt_{k_0}}}\ket{2_{\chippackt_{k_1}}}$.
	\item For some $k_2>k_1>k_0$, there is a single photon at chip-times $k_0$, $k_1$ and $k_2$, $\ket{1_{\chippackt_{k_0}}}\ket{1_{\chippackt_{k_1}}}\ket{1_{\chippackt_{k_2}}}$.
\end{enumerate} 
We may obtain the corresponding coefficient of each case using Proposition \ref{prop:spreadnumber} as follows
\begin{align}
	&\ket{3_{\wavepackt^e}}= \sum_{n_0+\cdots+n_{N_c-1}=3} \mathcal{C}_n( n_0,n_1,\ldots,n_{N_c-1})\nonumber\\
	&\qquad\qquad\qquad\qquad\times\prod_{k=0}^{N_c-1}(\lambda_k)^{n_k}\ket{n_{\chippackt_k}}
	\\
	&=\sum_{k_0=0}^{N_c-1} \frac{1}{\sqrt{N_c^3}}\sqrt{\frac{3!}{\underbrace{0!\cdots 0!}_{k_0}3!\underbrace{0!\cdots 0!}_{N_c-k_0-1}}}\nonumber\\
	&\qquad\qquad\qquad\times (\lambda_{k_0})^{3}\ket{3_{\chippackt_{k_0}}} \prod_{k\ne k_0}^{N_c-1}(\lambda_k)^{0}\ket{0_{\chippackt_k}}\nonumber\\
	&+{\sum_{k_1=0}^{N_c-1}\sum_{k_0=0}^{N_c-1}}\frac{1}{\sqrt{N_c^3}}\sqrt{\frac{3!}{1!2!}}(\lambda_{k_0})^{1}(\lambda_{k_1})^{2} \ket{1_{\chippackt_{k_0}}}\ket{2_{\chippackt_{k_1}}} \nonumber\\
	&\qquad\qquad\qquad\times\prod_{k\ne k_0, k_1}(\lambda_k)^{0}\ket{0_{\chippackt_k}}\\
	&+\underset{\text{s.t. }k_2>k_1>k_0}{\sum_{k_2=0}^{N_c-1}\sum_{k_1=0}^{N_c-1}\sum_{k_0=0}^{N_c-1}} \frac{1}{\sqrt{N_c^3}}\sqrt{\frac{3!}{1!1!1!}}(\lambda_{k_0})^{1}(\lambda_{k_1})^{1}(\lambda_{k_2})^{1}\nonumber\\
	&\qquad\times \ket{1_{\chippackt_{k_0}}}\ket{1_{\chippackt_{k_1}}}\ket{1_{\chippackt_{k_2}}}\prod_{k\ne k_0, k_1,k_2}(\lambda_k)^{0}\ket{0_{\chippackt_k}}
	\\
	&= \frac{1}{\sqrt{N_c^3}}\sum_{k_0=0}^{N_c-1}\lambda_{k_0} \ket{3_{\chippackt_{k_0}}}\\
	&+\frac{1}{\sqrt{N_c^3}} \sum_{k_1=0}^{N_c-1}\sum_{k_0=0}^{N_c-1} \sqrt{3} \lambda_{k_0}\ket{1_{\chippackt_{k_0}}}\ket{2_{\chippackt_{k_1}}}\nonumber\\
	&+\frac{1}{\sqrt{N_c^3}} \underset{\text{s.t. }k_2>k_1>k_0}{\sum_{k_2=0}^{N_c-1}\sum_{k_1=0}^{N_c-1}\sum_{k_0=0}^{N_c-1}} \sqrt{3!} \lambda_{k_0}\lambda_{k_1}\lambda_{k_2}\ket{1_{\chippackt_{k_0}}}\ket{1_{\chippackt_{k_1}}}\ket{1_{\chippackt_{k_2}}}\nonumber
\end{align}
\subsection{Decoding}
\label{appendix:ssnumber:decoding}
Let $\hat{\mathrm{U}}^\dagger_{\tilde{\Lambda}}$ be the despreading operator associated with  decoding sequence $\tilde{\Lambda}$. Denote the chip-time decomposition of  $\hat{\mathrm{U}}^\dagger_{\tilde{\Lambda}}$ by $\prod_{k=0}^{N_c-1}\hat{\mathrm{U}}^{\dagger}_{k,\tilde{\Lambda}}$. The effect of decoding operator with decoding sequence $\tilde{\Lambda}$ on the number states encoded with sequence $\Lambda$ is derived as follows
\begin{align}
	\hat{\mathrm{U}}^\dagger_{\tilde{\Lambda}}\ket{n_{\wavepackt^e}}	&=	\hat{\mathrm{U}}^\dagger_{\tilde{\Lambda}}\sum_{n_0+\cdots+n_{N_c-1}=n} \mathcal{C}_n( n_0,n_1,\ldots,n_{N_c-1})\nonumber\\
	&\qquad\qquad\times\prod_{k=0}^{N_c-1}(\lambda_k)^{n_k}\ket{n_{\chippackt_k}}\\
	&=	\sum_{n_0+\cdots+n_{N_c-1}=n} \mathcal{C}_n( n_0,n_1,\ldots,n_{N_c-1})\hat{\mathrm{U}}^\dagger_{\tilde{\Lambda}} \nonumber\\
	&\qquad\qquad\times\prod_{k=0}^{N_c-1}(\lambda_k)^{n_k}\ket{n_{\chippackt_k}}\\
	&=	\sum_{n_0+\cdots+n_{N_c-1}=n} \mathcal{C}_n( n_0,n_1,\ldots,n_{N_c-1})\nonumber\\
	&\qquad\qquad\times\left(\prod_{k=0}^{N_c-1}\hat{\mathrm{U}}^{\dagger}_{k,\tilde{\Lambda}} \right)\left(\prod_{k=0}^{N_c-1}(\lambda_k)^{n_k}\ket{n_{\chippackt_k}}\right)\\
	&=	\sum_{n_0+\cdots+n_{N_c-1}=n} \mathcal{C}_n( n_0,n_1,\ldots,n_{N_c-1}) \nonumber\\
	&\qquad\qquad\times\prod_{k=0}^{N_c-1}(\lambda_k)^{n_k}\hat{\mathrm{U}}^{\dagger}_{k,\tilde{\Lambda}}\ket{n_{\chippackt_k}}\\
	&= \sum_{n_0+\cdots+n_{N_c-1}=n} \mathcal{C}_n( n_0,n_1,\ldots,n_{N_c-1}) \nonumber\\
	&\qquad\qquad\times\prod_{k=0}^{N_c-1}(\lambda_k)^{n_k}(\tilde{\lambda}_k)^{n_k}\ket{n_{\chippackt_k}}\\
	&= \sum_{n_0+\cdots+n_{N_c-1}=n} \mathcal{C}_n( n_0,n_1,\ldots,n_{N_c-1}) \nonumber\\
	&\qquad\qquad\times\prod_{k=0}^{N_c-1}(\tilde{\lambda}_k\lambda_k)^{n_k}\ket{n_{\chippackt_k}}
\end{align}
where the last two steps are obtained similar to (\ref{eq:numberEncodingChip}).

\section{General Filter}
\label{appendix:filter}
Now assume a general filter \cite{madsen1999optical} with the Fourier series of their transfer function given by:
\begin{align}
	H_{\text{T}}(\omega) = \sum_{\fdcoef=0}^{\infty} d_{\fdcoef} e^{-j\fdcoef\omega \tau}, \ 
	H_{\text{R}}(\omega) = \sum_{\fdcoef=0}^{\infty} f_{\fdcoef} e^{-j\fdcoef\omega \tau},
\end{align}
where $H_{\text{T}}(\omega)$ and $H_{\text{R}}(\omega)$ are the transmission response and reflection response of the filter, respectively.

Applying the filter to an input with frequency domain wavepacket $\bar{\wavepackt}(\omega)$ results in the following transmitted wavepacket
\begin{align}
	H_{\text{T}}(\omega)\bar{\wavepackt}(\omega) = \sum_{\fdcoef=0}^{\infty} d_{\fdcoef} e^{-j\fdcoef\omega \tau}\bar{\wavepackt}(\omega)
\end{align} 
Therefore
\begin{align}
	h_{\text{T}}(t)*\wavepackt(t) = \sum_{\fdcoef=0}^{\infty} d_{\fdcoef} \wavepackt(t-\fdcoef\tau)
\end{align} 
From (\ref{eq:filter:omega}) we have
\begin{align}
	\hat{\mathrm{H}}\hat{a}^\dagger(\omega)\hat{\mathrm{H}}^\dagger = H_{\text{T}}(\omega) \hat{a}'^\dagger(\omega) + H_{\text{R}}(\omega) \hat{b}'^\dagger(\omega)
\end{align} 
Therefore
\begin{align}
	\hat{\mathrm{H}}&\hat{a}^\dagger_\wavepackt\hat{\mathrm{H}}^\dagger=  \int d\omega \bar{\wavepackt}(\omega) \hat{\mathrm{H}}\hat{a}^\dagger(\omega)\hat{\mathrm{H}}^\dagger \\
	&=  \int d\omega \bar{\wavepackt}(\omega) \left[H_{\text{T}}(\omega) \hat{a}'^\dagger(\omega) + H_{\text{R}}(\omega) \hat{b}'^\dagger(\omega)\right]\\
	&= \int d\omega H_{\text{T}}(\omega)\bar{\wavepackt}(\omega)  \hat{a}'^\dagger(\omega) +  \int d\omega H_{\text{R}}(\omega)\bar{\wavepackt}(\omega) \hat{b}'^\dagger(\omega)\\
	&= \int dt (h_{\text{T}}(t)*\wavepackt(t))  \hat{a}'^\dagger(t) +  \int dt (h_{\text{R}}(t)*\wavepackt(t)) b'^\dagger(t)\\
	&= \int dt \sum_{\fdcoef=0}^{\infty} d_{\fdcoef} \wavepackt(t-\fdcoef\tau) \hat{a}'^\dagger(t) +  \int dt \sum_{\fdcoef=0}^{\infty} f_{\fdcoef} \wavepackt(t-\fdcoef\tau) \hat{b}'^\dagger(t)\\
	&=\sum_{\fdcoef=0}^{\infty} \left(d_{\fdcoef} \int dt  \wavepackt(t-\fdcoef\tau) \hat{a}'^\dagger(t) +  f_{\fdcoef} \int dt  \wavepackt(t-\fdcoef\tau) \hat{b}'^\dagger(t)\right)
\end{align} 
The case of  $\tau=T_c$ with rectangular wave-packet is of particular interest. First, we derive the evolution of chip-time creation operators $\hat{a}^\dagger_{\chippackt_k}$ through filter:
\begin{align}
	\hat{\mathrm{H}}&\hat{a}^\dagger_{\chippackt_k}\hat{\mathrm{H}}^\dagger=
	\sum_{\fdcoef=0}^{\infty} \bigg(d_{\fdcoef}\int dt  \chippackt_k(t-\fdcoef T_c) \hat{a}'^\dagger(t) \nonumber\\
	&\qquad+  f_{\fdcoef} \int  dt  \chippackt_k(t-\fdcoef T_c) \hat{b}'^\dagger(t)\bigg) \\
	&=
	\sum_{\fdcoef=0}^{\infty} \bigg( d_{\fdcoef}\int_{(\fdcoef +k)T_c}^{(\fdcoef +k+1)T_c} dt  \chippackt_k(t-\fdcoef T_c) \hat{a}'^\dagger(t) \nonumber\\
	&\qquad+  f_{\fdcoef} \int_{(\fdcoef +k)T_c}^{(\fdcoef +k+1)T_c}  dt  \chippackt_k(t-\fdcoef T_c) \hat{b}'^\dagger(t)\bigg)\\
	&=
	\sum_{\fdcoef=0}^{\infty} \bigg(d_{\fdcoef}\int_{(\fdcoef +k)T_c}^{(\fdcoef +k+1)T_c} dt  \chippackt_0(t-(\fdcoef +k)T_c) \hat{a}'^\dagger(t) \nonumber\\
	&\qquad+  f_{\fdcoef} \int_{(\fdcoef +k)T_c}^{(\fdcoef +k+1)T_c}  dt  \chippackt_0(t-(\fdcoef +k)T_c) \hat{b}'^\dagger(t)\bigg)\\
	&=
	\sum_{\fdcoef=0}^{\infty} \left(d_{\fdcoef} \hat{a}'^\dagger_{\chippackt_{k+\fdcoef }}+  f_{\fdcoef}  \hat{b}'^\dagger_{\chippackt_{k+\fdcoef }}\right)
\end{align} 

At the next step the effect of the filter on the creation operator of the quantum signal with wavepacket $\wavepackt(t)$ is 
\begin{align}
	\hat{\mathrm{H}}&\hat{a}^\dagger_\wavepackt \hat{\mathrm{H}}^\dagger=
	\sum_{\fdcoef=0}^{\infty} \bigg(d_{\fdcoef}\int_{\fdcoef T_c}^{\pulseduration+\fdcoef T_c} dt  \wavepackt(t-\fdcoef T_c) \hat{a}'^\dagger(t) \nonumber\\
	&\qquad+  f_{\fdcoef} \int_{\fdcoef T_c}^{\pulseduration+\fdcoef T_c}  dt  \wavepackt(t-\fdcoef T_c) \hat{b}'^\dagger(t)\bigg)\\
	&= \sum_{\fdcoef=0}^{\infty}\bigg( d_{\fdcoef}\sum_{k=0}^{N_c-1}\int_{t_{\fdcoef +k}}^{t_{\fdcoef +k+1}} dt  \wavepackt(t-\fdcoef T_c) \hat{a}'^\dagger(t) \nonumber\\
	&\qquad+  f_{\fdcoef} \sum_{k=0}^{N_c-1}\int_{t_{\fdcoef +k}}^{t_{\fdcoef +k+1}} dt  \wavepackt(t-\fdcoef T_c) \hat{b}'^\dagger(t)\bigg)\\
	&=  \sum_{\fdcoef=0}^{\infty} \bigg(d_{\fdcoef}\frac{1}{\sqrt{N_c}}\sum_{k=0}^{N_c-1}\int_{t_{\fdcoef +k}}^{t_{\fdcoef +k+1}} dt  \sqrt{N_c}\wavepackt(t-\fdcoef T_c) \hat{a}'^\dagger(t) \nonumber\\
	&\qquad+  f_{\fdcoef} \frac{1}{\sqrt{N_c}}\sum_{k=0}^{N_c-1}\int_{t_{\fdcoef +k}}^{t_{\fdcoef +k+1}} dt  \sqrt{N_c}\wavepackt(t-\fdcoef T_c) \hat{b}'^\dagger(t)\bigg)
\end{align} 
Assuming that $\wavepackt(t)$ has a rectangular wavepacket, we can further simplify the results as follows
\begin{align}
	 \hat{\mathrm{H}}&\hat{a}^\dagger_\wavepackt \hat{\mathrm{H}}^\dagger= \sum_{\fdcoef=0}^{\infty} \bigg(d_{\fdcoef}\frac{1}{\sqrt{N_c}}\sum_{k=0}^{N_c-1}\int_{t_{\fdcoef +k}}^{t_{\fdcoef +k+1}} dt  \sqrt{\frac{N_c}{\pulseduration}} \hat{a}'^\dagger(t) \nonumber\\
	&\qquad+  f_{\fdcoef} \frac{1}{\sqrt{N_c}}\sum_{k=0}^{N_c-1}\int_{t_{\fdcoef +k}}^{t_{\fdcoef +k+1}} dt  \sqrt{\frac{N_c}{\pulseduration}} \hat{b}'^\dagger(t)\bigg)\\
	&=  \sum_{\fdcoef=0}^{\infty}\left(\frac{1}{\sqrt{N_c}} d_{\fdcoef}\sum_{k=0}^{N_c-1}\hat{a}^{'\dagger}_{\chippackt_{k+\fdcoef }} + \frac{1}{\sqrt{N_c}} f_{\fdcoef} \sum_{k=0}^{N_c-1}\hat{b}^{'\dagger}_{\chippackt_{k+\fdcoef }} \right) \label{eq:app:filter:j}
\end{align} 
To further simplify the above relation, let $\FDcoef=k+\fdcoef$ (see Table \ref{tab:coefficients}). Note that the coefficient of $\hat{a}^{'\dagger}_{\chippackt_{\FDcoef}}$ is 
\begin{align}
	\underset{\text{s.t. }\fdcoef\leq \FDcoef\leq \fdcoef+N_c-1}{\sum_{\fdcoef=0}^\infty}\frac{1}{\sqrt{N_c}}d_{\fdcoef} &= \sum_{\fdcoef=\max(0,\FDcoef-N_c+1)}^{\FDcoef} \frac{1}{\sqrt{N_c}}d_{\fdcoef}\\
	&=\sum_{\fdcoef=0}^{\min(\FDcoef, N_c-1)}  \frac{1}{\sqrt{N_c}}d_{\FDcoef-\fdcoef}
\end{align} 
\begin{table}[!tb]
	\caption{Filter coefficients with $\FDcoef=k+\fdcoef $}
	\centering
	\begin{tabular}{c|lllllll|l}
		$\fdcoef $ & $0$ & $1$ & $2$ & $\cdots$ & $N_c-1$ & $N_c$ & $N_c+1$ & \\
		$\FDcoef$ 
		&   $\hat{a}^{'\dagger}_{\chippackt_{0}}$  
		&   $\hat{a}^{'\dagger}_{\chippackt_{1}}$ 
		&   $\hat{a}^{'\dagger}_{\chippackt_{2}}$ 
		&   $\cdots $ 
		&   $\hat{a}^{'\dagger}_{\chippackt_{N_c-1}}$ 
		&   $\hat{a}^{'\dagger}_{\chippackt_{N_c}}$ 
		&   $\hat{a}^{'\dagger}_{\chippackt_{N_c+1}}$ 
		& \\
		\hline
		& & & & & & & & \parbox[t]{3mm}{\multirow{6}{*}{\rotatebox[origin=c]{90}{ \scriptsize $\sum_{\fdcoef=0}^{\infty}\sum_{k=0}^{N_c-1}d_{\fdcoef}\hat{a}^{'\dagger}_{\chippackt_{k+\fdcoef}}$
		}}}\\
		$0$ &$d_0$ &$d_0$ &$d_0$ &$\cdots$ &$d_0$ & $0$  & $0$ & \\ 
		& & & & & & & & \\
		$1$ &  $0$ &$d_1$ &$d_1$ &$\cdots$ &$d_1$ &$d_1$ & $0$ & \\ 
		& & & & & & & & \\
		$2$ &  $0$ & $0$  &$d_2$ &$\cdots$ &$d_2$ &$d_2$ &$d_2$& \\ 
		& & &\multicolumn{5}{l|}{$\underbrace{\qquad\qquad\qquad\qquad\qquad\qquad\qquad\qquad}_{k:\,0\to N_c-1}$} & \\
		$\vdots$& $0$ &$0$ & $0$ & $\ddots$ &         &       & $\ddots$  & \\  
		\hline
		& \multicolumn{7}{c|}{$\sum_{\FDcoef=0}^{\infty}\sum_{\fdcoef=\max(0,\FDcoef-N_c+1)}^{\FDcoef}d_{\fdcoef}\hat{a}^{'\dagger}_{\chippackt_{\FDcoef}}$} &
	\end{tabular}
	\label{tab:coefficients}
\end{table}
Hence,
\begin{align}
	\hat{\mathrm{H}}&\hat{a}^\dagger_\wavepackt \hat{\mathrm{H}}^\dagger=\sum_{\FDcoef=0}^{\infty}\Bigg[\bigg(\underbrace{\sum_{\fdcoef=\max(0,\FDcoef-N_c+1)}^{\FDcoef} \frac{1}{\sqrt{N_c}}d_{\fdcoef}}_{D_{\FDcoef}}\bigg)\hat{a}^{'\dagger}_{\chippackt_{\FDcoef}} \label{eq:app:filter:k}\\
	&\qquad\qquad+ \bigg(\underbrace{\sum_{\fdcoef=\max(0,\FDcoef-N_c+1)}^{\FDcoef} \frac{1}{\sqrt{N_c}}f_{\fdcoef}}_{{F}_{\FDcoef}}\bigg)\hat{b}^{'\dagger}_{\chippackt_{\FDcoef}}\Bigg]\nonumber\\
	&\qquad\,= \sum_{\FDcoef=0}^{\infty} (D_{\FDcoef} \hat{a}^{'\dagger}_{\chippackt_{\FDcoef}}+{F}_{\FDcoef} \hat{b}^{'\dagger}_{\chippackt_{\FDcoef}})
\end{align}
Since the temporal encoding leaves the temporal width of the wavepacket intact, with a similar approach, we have
\begin{align}
	&\hat{\mathrm{H}}\hat{a}^\dagger_{\wavepackt^d}\hat{\mathrm{H}}^\dagger=
	\sum_{\fdcoef=0}^{\infty} \bigg( d_{\fdcoef}\int_{\fdcoef T_c}^{\pulseduration+\fdcoef T_c} dt  \wavepackt^d(t-\fdcoef T_c) \hat{a}'^\dagger(t) \nonumber\\
	&\qquad+  f_{\fdcoef} \int_{\fdcoef T_c}^{\pulseduration+\fdcoef T_c}  dt  \wavepackt^d(t-\fdcoef T_c) \hat{b}'^\dagger(t)\bigg)\\
	&= \sum_{\fdcoef=0}^{\infty} \bigg(d_{\fdcoef}\sum_{k=0}^{N_c-1}\int_{t_{\fdcoef +k}}^{t_{\fdcoef +k+1}} dt  \wavepackt^d(t-\fdcoef T_c) \hat{a}'^\dagger(t) \nonumber\\
	&\qquad+  f_{\fdcoef} \sum_{k=0}^{N_c-1}\int_{t_{\fdcoef +k}}^{t_{\fdcoef +k+1}} dt  \wavepackt^d(t-\fdcoef T_c) \hat{b}'^\dagger(t)\bigg)\\
	&= \sum_{\fdcoef=0}^{\infty} \bigg(d_{\fdcoef}\sum_{k=0}^{N_c-1}\tilde{\lambda}_k\lambda_k\int_{t_{\fdcoef +k}}^{t_{\fdcoef +k+1}} dt  \wavepackt(t-\fdcoef T_c) \hat{a}'^\dagger(t) \nonumber\\
	&\qquad+  f_{\fdcoef} \sum_{k=0}^{N_c-1}\tilde{\lambda}_k\lambda_k\int_{t_{\fdcoef +k}}^{t_{\fdcoef +k+1}} dt  \wavepackt(t-\fdcoef T_c) \hat{b}'^\dagger(t)\bigg)\\
	&=  \sum_{\fdcoef=0}^{\infty} \bigg(d_{\fdcoef}\frac{1}{\sqrt{N_c}}\sum_{k=0}^{N_c-1}\tilde{\lambda}_k\lambda_k\int_{t_{\fdcoef +k}}^{t_{\fdcoef +k+1}} dt  \sqrt{N_c}\wavepackt(t-\fdcoef T_c) \hat{a}'^\dagger(t) \nonumber\\
	&\qquad+  f_{\fdcoef} \frac{1}{\sqrt{N_c}}\sum_{k=0}^{N_c-1}\tilde{\lambda}_k\lambda_k\int_{t_{\fdcoef +k}}^{t_{\fdcoef +k+1}} dt  \sqrt{N_c}\wavepackt(t-\fdcoef T_c) \hat{b}'^\dagger(t)\bigg)\\
	&=  \sum_{\fdcoef=0}^{\infty}\left(\frac{1}{\sqrt{N_c}} d_{\fdcoef}\sum_{k=0}^{N_c-1}\tilde{\lambda}_k\lambda_k\hat{a}^{'\dagger}_{\chippackt_{k+\fdcoef }} + \frac{1}{\sqrt{N_c}} f_{\fdcoef} \sum_{k=0}^{N_c-1}\tilde{\lambda}_k\lambda_k\hat{b}^{'\dagger}_{\chippackt_{k+\fdcoef }} \right)
\end{align} 
Letting $\FDcoef=k+\fdcoef $ we obtain 
\begin{align}
	\hat{\mathrm{H}}&\hat{a}^\dagger_{\wavepackt^d}\hat{\mathrm{H}}^\dagger=\sum_{\FDcoef=0}^{\infty}\Bigg[\bigg(\underbrace{\sum_{\fdcoef=\max(0,\FDcoef-N_c+1)}^{\FDcoef} \frac{1}{\sqrt{N_c}}d_{\fdcoef} \tilde{\lambda}_{\FDcoef-\fdcoef}\lambda_{\FDcoef-\fdcoef}}_{\tilde{D}_{\FDcoef}}\bigg)\hat{a}^{'\dagger}_{\chippackt_{\FDcoef}} \nonumber\\
	&\quad + \bigg(\underbrace{\sum_{\fdcoef=\max(0,\FDcoef-N_c+1)}^{\FDcoef} \frac{1}{\sqrt{N_c}}f_{\fdcoef}\tilde{\lambda}_{\FDcoef-\fdcoef}\lambda_{\FDcoef-\fdcoef}}_{\tilde{F}_{\FDcoef}}\bigg)\hat{b}^{'\dagger}_{\chippackt_k}\Bigg]\label{eq:filter:p2p}\\
	&\qquad\,\,\,= \sum_{\FDcoef=0}^{\infty} (\tilde{D}_{\FDcoef} \hat{a}^{'\dagger}_{\chippackt_{\FDcoef}}+\tilde{F}_{\FDcoef} \hat{b}^{'\dagger}_{\chippackt_{\FDcoef}})
\end{align}
For Finite Impulse Response (FIR) filter:
\begin{align}
	\hat{\mathrm{H}}&\hat{a}^\dagger_{\wavepackt^d}\hat{\mathrm{H}}^\dagger=\sum_{\FDcoef=0}^{L+N_c-1}\Bigg[\bigg(\sum_{\fdcoef=\max(0,\FDcoef-N_c+1)}^{\min(\FDcoef, L)} \frac{1}{\sqrt{N_c}}d_{\fdcoef} \tilde{\lambda}_{\FDcoef-\fdcoef}\lambda_{\FDcoef-\fdcoef}\bigg)\hat{a}^{'\dagger}_{\chippackt_{\FDcoef}} \nonumber\\
	&\quad+ \bigg(\sum_{\fdcoef=\max(0,\FDcoef-N_c+1)}^{\min(\FDcoef, L)} \frac{1}{\sqrt{N_c}}f_{\fdcoef}\tilde{\lambda}_{\FDcoef-\fdcoef}\lambda_{\FDcoef-\fdcoef}\bigg)\hat{b}^{'\dagger}_{\chippackt_{\FDcoef}}\Bigg]\\
	&\qquad\,\,\,= \sum_{\FDcoef=0}^{L+N_c-1} (\tilde{D}_{\FDcoef} \hat{a}^{'\dagger}_{\chippackt_{\FDcoef}}+\tilde{F}_{\FDcoef} \hat{b}^{'\dagger}_{\chippackt_{\FDcoef}})
\end{align}
where $L$ is the number of FIR filter coefficients.

\subsection{Asymptotic Behavior}
\label{appendix:asymptotic}
It would be interesting to investigate the asymptotic behavior of the filter to give further insight into the operation of the quantum spread spectrum communication systems. We use numerical simulations to qualitatively describe the behavior of the filter. 

First, we assume that the transmission filter's bandwidth (see Fig. \ref{fig:filterlarge}) is high enough to pass the correctly decoded signal with very low loss. 
With this choice of filter, the signal wavepacket $\wavepackt(t)$ almost entirely transmits through the filter $H_{\text{T}}$. Thus in case $\tilde{\Lambda}=\Lambda$, the transmitted wavepacket is almost equal to the signal wavepacket, i.e., $\braket{\wavepackt(t)}{\wavepackt_{\text{T}}(t-T_{\text{delay}})}\approx 1$, where $T_{\text{delay}} = k_\text{delay}\times T_c$ is the delay imposed by the causal filter and $k_\text{delay}$ is the number of chip-times corresponding to $T_{\text{delay}}$. Figure \ref{fig:asymptotic} illustrates the absolute value of coefficients $D_{\FDcoef}$ and ${F}_{\FDcoef}$ corresponding to the correct decoding of the quantum spread spectrum signal.  As can be seen from this figure, with a sufficiently good filter $|D_{\FDcoef}|\approx \frac{1}{\sqrt{N_c}}, \FDcoef_{\text{delay}}\le \FDcoef < \FDcoef_{\text{delay}}+N_c$ and $|{F}_{\FDcoef}|\approx 0$.  

The other assumption required to obtain the ideal behavior corresponds to the rejection of the incorrectly decoded signals. The asymptotic behavior can be observed by assuming a high value of processing gain $N_c$, or equivalently $T_c\ll T$. This assumption implies that applying the code spreads the quantum signal in a way that a very small portion of the undesired quantum signals passes through the transmission filter, $H_{\text{T}}$. Therefore, when the two (approximately) orthogonal random codes, $\tilde{\Lambda}$ and $\Lambda$ do not match, $\braket{\wavepackt}{\wavepackt_{\text{T}}}\approx 0$. The transmission and reflection coefficients in case of incorrect decoding, $\tilde{D}_{\FDcoef}$ and $\tilde{F}_{\FDcoef}$, are shown in Fig. \ref{fig:asymptotic} for a very high processing gain. According to this figure, we asymptotically have  $|\tilde{F}_{\FDcoef}|\approx \frac{1}{\sqrt{N_c}}, \FDcoef_{\text{delay}}\le \FDcoef < \FDcoef_{\text{delay}}+N_c$ and $|\tilde{D}_{\FDcoef}|\approx 0$ which means that the incorrectly decoded signal is almost entirely reflected from the filter.

%
\section{Effect of Filter}
\label{appendix:filterEffect}
\subsection{Coherent States}
Denote the output of the filter in both modes by superscript $F$. For encoded and then decoded coherent state, we may write
\begin{align}
	\ket{\alpha_{\wavepackt^F}} &=\hat{\mathrm{H}}\ket{\alpha_{\wavepackt^d}} = \hat{\mathrm{H}}e^{-\frac{|\alpha|^2}{2}}e^{\alpha \hat{a}_{\wavepackt^d}^\dagger}\hat{\mathrm{H}}^\dagger\ket{0}\\
	&= e^{-\frac{|\alpha|^2}{2}}e^{\alpha \hat{\mathrm{H}}\hat{a}_{\wavepackt^d}^\dagger\hat{\mathrm{H}}^\dagger}\ket{0}\\ &=e^{-\frac{|\alpha|^2}{2}}\exp(\sum_{\FDcoef=0}^{\infty}\left(\tilde{D}_{\FDcoef} \alpha \hat{a}^{'\dagger}_{\chippackt_\FDcoef} + \tilde{F}_{\FDcoef} \alpha \hat{b}^{'\dagger}_{\chippackt_\FDcoef}\right))\ket{0}\\
	&=e^{-\frac{|\alpha|^2}{2}}\prod_{\FDcoef=0}^{\infty}\exp(\tilde{D}_{\FDcoef} \alpha \hat{a}^{'\dagger}_{\chippackt_\FDcoef} )\nonumber\\
	&\qquad\qquad\qquad\times\exp( \tilde{F}_{\FDcoef} \alpha \hat{b}^{'\dagger}_{\chippackt_\FDcoef})\ket{0}\\
	&=\prod_{\FDcoef=0}^{\infty}\exp(-\frac{1}{2}\left|\tilde{D}_{\FDcoef} \alpha \right|^2)\exp(\tilde{D}_{\FDcoef} \alpha \hat{a}^{'\dagger}_{\chippackt_\FDcoef} )\nonumber\\
	&\qquad\qquad\times \exp(-\frac{1}{2}\left|\tilde{F}_{\FDcoef} \alpha \right|^2)\exp( \tilde{F}_{\FDcoef} \alpha \hat{b}^{'\dagger}_{\chippackt_\FDcoef})\ket{0}\\
	&= \prod_{\FDcoef=0}^{\infty} \ket{\tilde{D}_{\FDcoef}  \alpha_{\chippackt_\FDcoef}}  \ket{\tilde{F}_{\FDcoef}  \alpha_{\chippackt_\FDcoef}}
\end{align}
Note that
\begin{align}
	\sum_{\FDcoef=0}^{\infty}\left( \left|\tilde{D}_{\FDcoef} \right|^2 + \left|\tilde{F}_{\FDcoef} \right|^2 \right)= 1
\end{align}
which is due to the unitary condition (This relation can also be derived from Lemma \ref{lem:coefficients} in subsection \ref{sec::intensity}).

Taking partial trace with respect to the reflected modes gives the following expression
\begin{align}
	\ket{\alpha_{\wavepackt_{\text{T}}}} = \prod_{\FDcoef=0}^{\infty} \ket{\tilde{D}_{\FDcoef}  \alpha_{\chippackt_\FDcoef}}  
\end{align}
%

Each term in the tensor product results in a Poisson distribution for number of photons in that specific chip-time, i.e. $n_{\FDcoef,\chippackt_\FDcoef}$. Therefore, the output statistics can be obtained by noting that sum of independent Poisson random variables is also a Poisson random variable
\begin{align}
	\mathbb{P}(n)
	&=\sum_{n_0+n_1+\cdots=n} \left|\braket{n_{0,\chippackt_0}, n_{1,\chippackt_1}, \cdots}{\alpha_{\wavepackt_{\text{T}}}}\right|^2\\
	&=\text{Poisson}\left( |\alpha|^2 \sum_{\FDcoef=0}^{\infty}\left|\tilde{D}_{\FDcoef} \right|^2 \right)
\end{align}

\subsection{Number States}
For the number state:
\begin{align}
	&\ket{n_{\wavepackt^F}} = \hat{\mathrm{H}}\ket{n_{\wavepackt^d}} =\frac{(\hat{\mathrm{H}}\hat{a}_{\wavepackt^d}^\dagger\hat{\mathrm{H}}^\dagger)^n}{\sqrt{n!}}\ket{0} = \\
	&=\frac{1}{\sqrt{n!}}\left(\sum_{\FDcoef=0}^{\infty}\left(\tilde{D}_{\FDcoef}  \hat{a}^{'\dagger}_{\chippackt_\FDcoef} + \tilde{F}_{\FDcoef}  \hat{b}^{'\dagger}_{\chippackt_\FDcoef}\right)\right)^n\ket{0} \\
	&=\sum_{
		\begin{matrix}
			n_{\text{T},0},n_{\text{T},1},\ldots, n_{\text{R},0}, n_{\text{R},1}, \ldots \ge 0  \\
			\text{s.t. }\sum_{\FDcoef=0}^{\infty} (n_{\text{T}, \FDcoef} + n_{\text{R},\FDcoef}) = n
	\end{matrix}}\\
&\qquad\quad{\sqrt{{n\choose n_{\text{T},0},n_{\text{T},1},\ldots, n_{\text{R},0}, n_{\text{R},1}, \ldots}} }\nonumber\\
	&\qquad\qquad\times\prod_{\FDcoef=0}^{\infty} \left(\left(\tilde{D}_{\FDcoef}  \right)^{n_{\text{T}, \FDcoef}} \left(\tilde{F}_{\FDcoef}  \right)^{n_{\text{R}, \FDcoef}} \ket{n_{\text{T},\FDcoef, \chippackt_\FDcoef}}\ket{n_{\text{R}, \FDcoef,\chippackt_\FDcoef}}\right),\nonumber
\end{align} 
where $n_{\text{T},\FDcoef}$ and $n_{\text{R},\FDcoef}$ denotes the number of photons at chip-time $\FDcoef$ at the transmission and reflection ports of the filter respectively. 

For single photon states the above equation reduces to
\begin{align}
	\ket{1_{\wavepackt^F}} &=	\hat{\mathrm{H}}\ket{1_{\wavepackt^d}} = \sum_{\FDcoef=0}^{\infty}\left(\tilde{D}_{\FDcoef}  \ket{1_{\text{T},\chippackt_\FDcoef}} + \tilde{F}_{\FDcoef}  \ket{1_{\text{R},\chippackt_\FDcoef}} \right)\label{eq:singlePhotonFilter}
\end{align} 
Taking the partial trace with respect to the reflected part gives the final state of the receiver
\begin{align}
	\rho^{\text{T}} &= \Tr_{R}(\rho^F) = \Tr_{R}(\ketbra{1_{\wavepackt^F}})\\
	& = {\ev{\rho^F}{0_{\text{R}}}} + \sum_{\FDcoef=0}^{\infty} {\ev{\rho^F}{1_{\text{R},\chippackt_\FDcoef}}}\\
	&=\braket{0_{\text{R}}}{1_{\wavepackt^F}}\braket{1_{\wavepackt^F}}{0_{\text{R}}} + \sum_{\FDcoef=0}^{\infty} \braket{1_{\text{R},\chippackt_\FDcoef}}{1_{\wavepackt^F}}\braket{1_{\wavepackt^F}}{1_{\text{R},\chippackt_\FDcoef}}
\end{align} 
where
\begin{align}
	\braket{0_{\text{R}}}{1_{\wavepackt^F} }&= \sum_{\FDcoef=0}^{\infty}\left(\tilde{D}_{\FDcoef}  \right)\ket{1_{\text{T},\chippackt_\FDcoef}}\\
	&=\left(\sqrt{\sum_{\FDcoef=0}^{\infty}\left|\tilde{D}_{\FDcoef}\right|^2 }\right)\ket{1_{\wavepackt_{\text{T}}}}\nonumber
	\\
	\braket{1_{\text{R},\chippackt_\FDcoef}}{1_{\wavepackt^F}} &= \left(\tilde{F}_{\FDcoef}  \right)\ket{0_\text{T}},
\end{align} 
where
\begin{align}
	\ket{1_{\wavepackt_{\text{T}}}}\coloneqq \frac{1}{\sqrt{\sum_{\FDcoef=0}^{\infty}\left|\tilde{D}_{\FDcoef}\right|^2 }}\sum_{\FDcoef=0}^{\infty}\tilde{D}_{\FDcoef}\ket{1_{\text{T},\chippackt_\FDcoef}}
\end{align}

Hence,
\begin{align}
	\rho^{\text{T}} &= \left(\sum_{\FDcoef=0}^{\infty}\tilde{D}_{\FDcoef} \ket{1_{\text{T},\chippackt_\FDcoef}}\right)
	\left(\sum_{\FDcoef=0}^{\infty}\tilde{D}_{\FDcoef}^*  \bra{1_{\text{T},\chippackt_\FDcoef}}\right)\\
	&\qquad + \sum_{\FDcoef=0}^{\infty}\left|\tilde{F}_{\FDcoef} \right|^2\ketbra{0_{\text{T}}}\nonumber\\
	&= \left(\sum_{\FDcoef=0}^{\infty}\left|\tilde{F}_{\FDcoef}\right|^2\right)\ketbra{0}+\left(\sum_{\FDcoef=0}^{\infty}\left|\tilde{D}_{\FDcoef}\right|^2\right) \ketbra{1_{\wavepackt_{\text{T}}}}
\end{align} 

\section{Quantum Broadcasting Channel}
\label{appendix:coherentCoupler}
\subsection{Coherent States}
For coherent state inputs:
\begin{align}
	\ket{\Phi}	&= \hat{\mathrm{B}}\ket{\Psi} = \hat{\mathrm{B}}\prod_{s=0}^{M-1} \ket{\alpha_{s,\wavepackt}}\\
	&= \hat{\mathrm{B}}\prod_{s=0}^{M-1} \prod_{k=0}^{N_c-1} \ket{\frac{\alpha_{s,\chippackt_k}}{\sqrt{N_c}}}\\
	&=\hat{\mathrm{B}}\prod_{s=0}^{M-1} \prod_{k=0}^{N_c-1} D\left(\frac{\alpha_{s,\chippackt_k}}{\sqrt{N_c}}\right)\ket{0}\\
	&=\hat{\mathrm{B}}  \prod_{k=0}^{N_c-1}\prod_{s=0}^{M-1} D\left(\frac{\alpha_{s,\chippackt_k}}{\sqrt{N_c}}\right)\hat{\mathrm{B}}^\dagger\ket{0} \\
	&=\hat{\mathrm{B}}  \left[\prod_{s=0}^{M-1} D\left(\frac{\alpha_{s,\chippackt_0}}{\sqrt{N_c}}\right)\right]\left[\prod_{s=0}^{M-1} D\left(\frac{\alpha_{s,\chippackt_1}}{\sqrt{N_c}}\right)\right]\cdots\nonumber\\
	&\qquad\qquad\qquad\times\left[\prod_{s=0}^{M-1} D\left(\frac{\alpha_{s,\chippackt_{N_c-1}}}{\sqrt{N_c}}\right)\right]\hat{\mathrm{B}}^\dagger\ket{0} \\
	&=\hat{\mathrm{B}}  
	\left[\prod_{s=0}^{M-1} D\left(\frac{\alpha_{s,\chippackt_0}}{\sqrt{N_c}}\right)\right] \hat{\mathrm{B}}^\dagger\hat{\mathrm{B}}  
	\left[\prod_{s=0}^{M-1} D\left(\frac{\alpha_{s,\chippackt_1}}{\sqrt{N_c}}\right)\right] \hat{\mathrm{B}}^\dagger\hat{\mathrm{B}} \cdots\nonumber\\
	&\qquad\qquad\qquad\times
	\hat{\mathrm{B}}\left[\prod_{s=0}^{M-1} D\left(\frac{\alpha_{s,\chippackt_{N_c-1}}}{\sqrt{N_c}}\right)\right]\hat{\mathrm{B}}^\dagger\ket{0} \\
	&= \prod_{k=0}^{N_c-1}\hat{\mathrm{B}} \left[\prod_{s=0}^{M-1} D\left(\frac{\alpha_{s,\chippackt_k}}{\sqrt{N_c}}\right)\right]\hat{\mathrm{B}}^\dagger\ket{0} \\
	&= \prod_{k=0}^{N_c-1}\hat{\mathrm{B}} \left[\prod_{s=0}^{M-1} D\left(\frac{\alpha_{s,\chippackt_k}}{\sqrt{N_c}}\right)\right]\ket{0} \\
	&= \prod_{k=0}^{N_c-1}\hat{\mathrm{B}} \prod_{s=0}^{M-1} \ket{\frac{\alpha_{s,\chippackt_k}}{\sqrt{N_c}}}
\end{align}
The above equation shows that the effect of star-coupler on the coherent state inputs is equivalent to the tensor product of the effect of star-coupler on each chip of the decomposed states.

According to \cite{rezai2021quantum}, 
\begin{align}
	\hat{\mathrm{B}} \prod_{s=0}^{M-1} \ket{\frac{\alpha_{s,\chippackt_k}}{\sqrt{N_c}}} = \prod_{r=0}^{M-1}\ket{\sum_{s=0}^{M-1}\frac{B_{rs}}{\sqrt{N_c}}\alpha_{s,\chippackt_k}}
\end{align}
Therefore
\begin{align}
	\ket{\Phi}	&= \hat{\mathrm{B}}\ket{\Psi} = \prod_{k=0}^{N_c-1}\prod_{r=0}^{M-1}\ket{\sum_{s=0}^{M-1}\frac{B_{rs}}{\sqrt{N_c}}\alpha_{s,\chippackt_k}} \\
	&= \prod_{r=0}^{M-1}\prod_{k=0}^{N_c-1}\ket{\sum_{s=0}^{M-1}\frac{B_{rs}}{\sqrt{N_c}}\alpha_{s,\chippackt_k}}
\end{align}
Now assume that the input of the star-coupler broadcasting channel is the encoded spread spectrum coherent states:
\begin{align}
	\ket{\Phi^e}	&= \hat{\mathrm{B}}\ket{\Psi^e} = \prod_{k=0}^{N_c-1}\prod_{r=0}^{M-1}\ket{\sum_{s=0}^{M-1}\frac{B_{rs}}{\sqrt{N_c}}\alpha_{s,\chippackt_k^e}} \\
	&= \prod_{r=0}^{M-1}\prod_{k=0}^{N_c-1}\ket{\sum_{s=0}^{M-1}\frac{B_{rs}}{\sqrt{N_c}}\lambda_{s,k}\alpha_{s,\chippackt_k}}
\end{align}
For coherent states, the received signal at receiver $r$ is a pure state that can be obtained by $\rho^e_{r}=\Tr_{r'\neq r}(\ketbra{\Phi^e}) = \ketbra{\phi^e_{r}}$.

Hence,
\begin{align}
	\ket{\phi^e_{r}}	&= \prod_{k=0}^{N_c-1}\ket{\sum_{s=0}^{M-1}\frac{B_{rs}}{\sqrt{N_c}}\lambda_{s,k}\alpha_{s,\chippackt_k}}
\end{align}

Now we apply the decoding operator on this state:
\begin{align}
	&\ket{\phi^d_{r }}	= \hat{\mathrm{U}}^\dagger \ket{\phi^e_{r }} =  \prod_{k=0}^{N_c-1}\ket{\lambda_{r,k}\sum_{s=0}^{M-1}\frac{B_{r s}}{\sqrt{N_c}}\lambda_{s,k}\alpha_{s,\chippackt_k}}\\ &=\prod_{k=0}^{N_c-1}\ket{\lambda_{r,k}\sum_{s=0}^{M-1}\frac{B_{r s}}{\sqrt{N_c}}\lambda_{s,k}\alpha_{s,\chippackt_k}} \\
	&= \prod_{k=0}^{N_c-1}\ket{(\lambda_{r,k})^2\frac{B_{r r}}{\sqrt{N_c}}\alpha_{s=r,\chippackt_k}+ \sum_{s\neq r}\lambda_{r ,k}\lambda_{s,k} \frac{B_{r s}}{\sqrt{N_c}}\alpha_{s,\chippackt_k}}\\
	&= \prod_{k=0}^{N_c-1}\ket{\frac{B_{r r}}{\sqrt{N_c}}\alpha_{s=r,\chippackt_k}+\frac{1}{\sqrt{N_c}} \sum_{s\neq r}\lambda_{r ,k}\lambda_{s,k}B_{r s}\alpha_{s,\chippackt_k}}\\
	&= \ket{\sum_{k=0}^{N_c-1}\Big(\frac{B_{r r}}{\sqrt{N_c}}\alpha_{s=r,\chippackt_k}+\frac{1}{\sqrt{N_c}} \sum_{s\neq r}\lambda_{r ,k}\lambda_{s,k}B_{r s}\alpha_{s,\chippackt_k}\Big)}\\
	&=\ket{B_{r r}\alpha_{s=r,\wavepackt}+\sum_{s\neq r}B_{r s}\alpha_{s,\wavepackt^{e_sd_r}}}
\end{align}
Note that $(\lambda_{r,k})^2=1$.
\subsection{Number (Fock) States}
Encoded signals
\begin{align}
	\ket{\Psi^e} &= \prod_{s=0}^{M-1} \ket{n_{s,\wavepackt^e}}\\
	&= \prod_{s=0}^{M-1} \frac{1}{\sqrt{n_s!}}(\hat{a}^\dagger_{s,\wavepackt^e})^{n_s}\ket{0}\\
	&= \prod_{s=0}^{M-1}\frac{1}{\sqrt{n_s!}}\left(\frac{1}{\sqrt{N_c}}\sum_{k=0}^{N_c-1} \hat{a}_{s,\chippackt_k^{e}}^\dagger\right)^{n_s}\ket{0}\\
	&= \prod_{s=0}^{M-1}\frac{1}{\sqrt{n_s!}}\left(\frac{1}{\sqrt{N_c}}\sum_{k=0}^{N_c-1} \lambda_{s,k}\hat{a}_{s,\chippackt_k}^\dagger\right)^{n_s}\ket{0}
\end{align}
After quantum broadcasting channel
\begin{align}
	\ket{\Phi^e} &=  \hat{\mathrm{B}}\ket{\Psi^e}\\
	&=\hat{\mathrm{B}}\prod_{s=0}^{M-1}\frac{1}{\sqrt{n_s!}}(\hat{a}^\dagger_{s,\wavepackt^e})^{n_s}\ket{0}\\
	&=\prod_{s=0}^{M-1}\frac{1}{\sqrt{n_s!}}(\hat{\mathrm{B}} \hat{a}^\dagger_{s,\wavepackt^e}\hat{\mathrm{B}}^\dagger)^{n_s}\ket{0}\\
	&=\prod_{s=0}^{M-1}\frac{1}{\sqrt{n_s!}}(\sum_{r=0}^{M-1} B_{rs}\hat{a}_{r,\wavepackt^e}^\dagger)^{n_s}\ket{0}\\
	&= \prod_{s=0}^{M-1}\frac{1}{\sqrt{n_s!}}\left( \frac{1}{\sqrt{N_c}}\sum_{k=0}^{N_c-1} \sum_{r=0}^{M-1} B_{rs}\hat{a}_{r,\chippackt_k^{e}}^\dagger\right)^{n_s}\ket{0}\\
	&= \prod_{s=0}^{M-1}\frac{1}{\sqrt{n_s!}}\left(\frac{1}{\sqrt{N_c}}\sum_{k=0}^{N_c-1}  \sum_{r=0}^{M-1}B_{rs} \lambda_{s,k}\hat{a}_{r,\chippackt_k}^\dagger\right)^{n_s}\ket{0}
\end{align}
Assuming that receiver $r$ intends to decode the quantum signal of transmitter $s=r$ by applying the appropriate decoding operator we obtain
\begin{align}
	\ket{\Phi^d} &=  \hat{\mathrm{U}}^\dagger\ket{\Phi^e}\\
	&=\hat{\mathrm{U}}^\dagger\prod_{s=0}^{M-1}\frac{1}{\sqrt{n_s!}}\left(\sum_{r=0}^{M-1} B_{rs}\hat{a}_{r,\wavepackt^e}^\dagger\right)^{n_s}\ket{0}\\
	&=\prod_{s=0}^{M-1}\frac{1}{\sqrt{n_s!}}\left(\sum_{r=0}^{M-1} B_{rs}\hat{\mathrm{U}}^\dagger\hat{a}_{r,\wavepackt^e}^\dagger\hat{\mathrm{U}}\right)^{n_s}\ket{0}\\
	&=\prod_{s=0}^{M-1}\frac{1}{\sqrt{n_s!}}\left(\sum_{r=0}^{M-1} B_{rs}\hat{a}_{r,\wavepackt^{e_sd_r}}^\dagger\right)^{n_s}\ket{0}
\end{align}
Form the chip-time perspective
\begin{align}
	&\ket{\Phi^d} =  \hat{\mathrm{U}}^\dagger\ket{\Phi^e}\\
	&=\hat{\mathrm{U}}^\dagger\prod_{s=0}^{M-1}\frac{1}{\sqrt{n_s!}}\left(\frac{1}{\sqrt{N_c}}\sum_{k=0}^{N_c-1}  \sum_{r=0}^{M-1}B_{rs} \lambda_{s,k}\hat{a}_{r,\chippackt_k}^\dagger\right)^{n_s}\ket{0}\\
	&=\prod_{s=0}^{M-1}\frac{1}{\sqrt{n_s!}}\left(\frac{1}{\sqrt{N_c}}\sum_{k=0}^{N_c-1}  \sum_{r=0}^{M-1}B_{rs} \lambda_{s,k}\hat{\mathrm{U}}^\dagger\hat{a}_{r,\chippackt_k}^\dagger\hat{\mathrm{U}}\right)^{n_s}\ket{0}\\
	&=\prod_{s=0}^{M-1}\frac{1}{\sqrt{n_s!}}\left(\frac{1}{\sqrt{N_c}}\sum_{k=0}^{N_c-1}  \sum_{r=0}^{M-1}B_{rs} \lambda_{s,k}\lambda_{r,k}\hat{a}_{r,\chippackt_k}^\dagger\right)^{n_s}\ket{0} \label{eq:number:decoded}\\
	&=\prod_{s=0}^{M-1}\frac{1}{\sqrt{n_s!}}\Bigg(\frac{1}{\sqrt{N_c}}\sum_{k=0}^{N_c-1}\Big(B_{ss}\hat{a}_{r=s,\chippackt_k}^\dagger+ \\
	&\qquad\qquad\qquad+\sum_{r\ne s} B_{rs}\lambda_{s,k}\lambda_{r,k}\hat{a}_{r,\chippackt_k}^\dagger\Big)\Bigg)^{n_s}\ket{0}\nonumber\\
	&=\prod_{s=0}^{M-1}\frac{1}{\sqrt{n_s!}}\Bigg( B_{ss}\hat{a}_{r=s,\wavepackt}^\dagger+ \sum_{r\ne s}B_{rs}\hat{a}^\dagger_{r,\wavepackt^{e_sd_r}}\Bigg)^{n_s}\ket{0}\nonumber
\end{align}
We apply the effect of filter, using (\ref{eq:filter:p2p}), to obtain 
\begin{align}
	\hat{\mathrm{H}}&\hat{a}^\dagger_{r,\wavepackt^{e_sd_r}}\hat{\mathrm{H}}^\dagger\\
	&=\sum_{\FDcoef=0}^{\infty}\Bigg[\Bigg(\sum_{\fdcoef=\max(0,\FDcoef-N_c+1)}^{\FDcoef} \frac{1}{\sqrt{N_c}}d_{\fdcoef} \lambda_{s,\FDcoef-\fdcoef}\lambda_{r,\FDcoef-\fdcoef}\Bigg)\hat{a}^{'\dagger}_{r,\chippackt_\FDcoef} \nonumber\\
	&+ \Bigg(\sum_{\fdcoef=\max(0,\FDcoef-N_c+1)}^{\FDcoef} \frac{1}{\sqrt{N_c}}f_{\fdcoef}\lambda_{s,\FDcoef-\fdcoef}\lambda_{r,\FDcoef-\fdcoef}\Bigg)\hat{b}^{'\dagger}_{r,\chippackt_\FDcoef}\Bigg]\nonumber\\
	&\coloneqq  \sum_{\FDcoef=0}^{\infty}\left(D_{\FDcoef}^{s,r}\hat{a}^{'\dagger}_{r,\chippackt_\FDcoef} + {F}_{\FDcoef}^{s,r}\hat{b}^{'\dagger}_{r,\chippackt_\FDcoef}\right),
\end{align}
where $D_{\FDcoef}^{s,r}$ and ${F}_{\FDcoef}^{s,r}$ are defined in (\ref{eq:Dksr}) and (\ref{eq:Fksr}).

Thus we obtain the quantum signal after the filter as
\begin{align}
	\ket{\Phi^F}&=\prod_{s=0}^{M-1}\frac{1}{\sqrt{n_s!}}\Big(\sum_{\FDcoef=0}^{\infty}\sum_{r=0}^{M-1} (B_{rs}D_{\FDcoef}^{s,r}\hat{a}^{'\dagger}_{r,\chippackt_\FDcoef} \nonumber\\
	&\qquad\qquad\qquad+ B_{rs} {F}_{\FDcoef}^{s,r}\hat{b}^{'\dagger}_{r,\chippackt_\FDcoef})\Big)^{n_s}\ket{0}\label{eq:app:numberstatePhiF}
\end{align}

\section{Intensity of the Received Quantum Signal}
\label{appendix:intensity}
First, we prove Lemma \ref{lem:coefficients} by calculating $\comm{\hat{a}_{\wavepackt^{e_sd_r}}}{\hat{a}^\dagger_{\wavepackt^{e_{s'}d_r}}}$ by two methods and comparing the results. From \cite{rezai2021quantum}, equations (A.32) and (B.55) we obtain
\begin{align}
	\comm{\hat{a}_{\wavepackt^{e_sd_r}}}{\hat{a}^\dagger_{\wavepackt^{e_{s'}d_r}}} &= \braket{\wavepackt^{e_sd_r}}{\wavepackt^{e_{s'}d_r}} = \braket{\wavepackt^{e_s}}{\wavepackt^{e_{s'}}} \\
	&= \frac{1}{N_c} \sum_{k=0}^{N_c-1}\lambda_{s,k}\lambda_{s',k}
	\label{eq:lambdasr}\\
	&=\braket{\Lambda_s}{\Lambda_{s'}}
\end{align}
Using the chip-time interval decomposition and considering the effect of filter, we may write
\begin{align}
	&\comm{\hat{a}_{\wavepackt^{e_sd_r}}}{\hat{a}^\dagger_{\wavepackt^{e_{s'}d_r}}} =\sum_{\FDcoef=0}^{\infty}\sum_{\FDcoef'=0}^{\infty} D_{\FDcoef}^{s,r*}D_{\FDcoef'}^{s',r}\comm{\hat{a}^{'}_{r,\chippackt_\FDcoef}}{\hat{a}^{'\dagger}_{r,\chippackt_{\FDcoef'}}}\nonumber\\
	&\qquad\qquad\qquad\quad+\sum_{\FDcoef=0}^{\infty}\sum_{\FDcoef'=0}^{\infty} D_{\FDcoef}^{s,r*} F_{\FDcoef'}^{s',r}\comm{\hat{a}^{'}_{r,\chippackt_\FDcoef}}{\hat{b}^{'\dagger}_{r,\chippackt_{\FDcoef'}}}\nonumber\\
	&\qquad\qquad\qquad\quad+\sum_{\FDcoef=0}^{\infty}\sum_{\FDcoef'=0}^{\infty} {F}_{\FDcoef}^{s,r*} D_{\FDcoef'}^{s',r}\comm{\hat{b}^{'}_{r,\chippackt_\FDcoef}}{\hat{a}^{'\dagger}_{r,\chippackt_{\FDcoef'}}}\nonumber\\
	&\qquad\qquad\qquad\quad+\sum_{\FDcoef=0}^{\infty}\sum_{\FDcoef'=0}^{\infty}{F}_{\FDcoef}^{s,r*} F_{\FDcoef'}^{s',r}\comm{\hat{b}^{'}_{r,\chippackt_\FDcoef}}{\hat{b}^{'\dagger}_{r,\chippackt_{\FDcoef'}}}\nonumber\\
	&\qquad= \sum_{\FDcoef=0}^{\infty}\sum_{\FDcoef'=0}^{\infty}D_{\FDcoef}^{s,r*} D_{k'}^{s',r}\delta_{\FDcoef\FDcoef'} + \sum_{\FDcoef=0}^{\infty}\sum_{\FDcoef'=0}^{\infty}{F}_{\FDcoef}^{s,r*} F_{k'}^{s',r}\delta_{\FDcoef\FDcoef'}\nonumber\\
	&\qquad= \sum_{\FDcoef=0}^{\infty}\left(D_{\FDcoef}^{s,r*} {D}_{\FDcoef}^{s',r} + {F}_{\FDcoef}^{s,r*}{F}_{\FDcoef}^{s',r}\right)\label{eq:Dsr}
\end{align}
Thus, from (\ref{eq:lambdasr}) and (\ref{eq:Dsr}) we conclude
\begin{align}
	\sum_{\FDcoef=0}^{\infty}\left(D_{\FDcoef}^{s,r*} {D}_{\FDcoef}^{s',r} + {F}_{\FDcoef}^{s,r*}{F}_{\FDcoef}^{s',r}\right) = \frac{1}{N_c} \sum_{k=0}^{N_c-1}\lambda_{s,k}\lambda_{s',k}
\end{align}
which completes the proof of Lemma \ref{lem:coefficients}.

The intensity of the decoded quantum signal after the filter can be obtained according to the following equation
\begin{align}
	I_{r_0}(t)&=\ev{\hat{a}^{'\dagger}_{r_0}(t)\hat{a}'_{r_0}(t)}{\Phi^d}
\end{align}
Since $\hat{a}'_{r}(t)\ket{0}=0$, we have $\hat{a}'_{r}(t)\ket{\psi_\wavepackt}=\comm{\hat{a}'_{r}(t)}{f(\hat{a}^\dagger_\wavepackt)}\ket{0}$. Thus we find $\hat{a}'_{r_0}(t)\ket{\Phi^F}$ for number state inputs using (\ref{eq:app:numberstatePhiF}) as follows
\begin{align}
	\hat{a}'_{r_0}(t)\ket{\Phi^F}&=	\hat{a}'_{r_0}(t)\prod_{s=0}^{M-1}\frac{1}{\sqrt{n_s!}}\Big(\sum_{\FDcoef=0}^{\infty}\sum_{r=0}^{M-1} (B_{rs}D_{\FDcoef}^{s,r}\hat{a}^{'\dagger}_{r,\chippackt_\FDcoef} \nonumber\\
	&\qquad\qquad\qquad\quad+ B_{rs} {F}_{\FDcoef}^{s,r}\hat{b}^{'\dagger}_{r,\chippackt_\FDcoef})\Big)^{n_s}\ket{0}\\
	&=\bigg[\hat{a}'_{r_0}(t),\prod_{s=0}^{M-1}\frac{1}{\sqrt{n_s!}}\Big(\sum_{\FDcoef=0}^{\infty}\sum_{r=0}^{M-1} (B_{rs}D_{\FDcoef}^{s,r}\hat{a}^{'\dagger}_{r,\chippackt_\FDcoef} \nonumber\\
	&\qquad\qquad\qquad\quad+ B_{rs} {F}_{\FDcoef}^{s,r}\hat{b}^{'\dagger}_{r,\chippackt_\FDcoef})\Big)^{n_s}\bigg]\ket{0}
\end{align}
Note that 
\begin{align}
	\comm{\hat{a}(t)}{f(\hat{a}_\wavepackt)} = \wavepackt(t)\frac{\partial}{\partial \hat{a}_\wavepackt}f(\hat{a}_\wavepackt)
\end{align}
Therefore
\begin{align}
	\comm{\hat{a}'_r(t)}{f(\hat{a}'_{r,\chippackt_\FDcoef})} &= \chippackt_\FDcoef(t)\frac{\partial}{\partial \hat{a}'_{r,\chippackt_\FDcoef}}f(\hat{a}'_{r,\chippackt_\FDcoef}) \\
	&= \left\lbrace\begin{matrix}
		\sqrt{\frac{N_c}{\pulseduration}} \frac{\partial}{\partial\hat{a}'_{r,\chippackt_\FDcoef}}f(\hat{a}'_{r,\chippackt_\FDcoef}) && t\in [t_\FDcoef, t_{\FDcoef+1})\\
		0 && t\notin [t_\FDcoef, t_{\FDcoef+1})
	\end{matrix}
	\right.
\end{align}
Thus if $t\in [t_l, t_{l+1})$, and we want to decode at receiver $r_0$:
\begin{align}
	\hat{a}'_{r_0}(t)&\ket{\Phi^d}
	=\sqrt{\frac{N_c}{\pulseduration}}\frac{\partial}{\partial\hat{a}'_{r_0,\chippackt_l}}\prod_{s=0}^{M-1}\frac{1}{\sqrt{n_s!}}\\
	&\times\Bigg(\sum_{\FDcoef=0}^{\infty}\sum_{r=0}^{M-1} (B_{rs}D_{\FDcoef}^{s,r}\hat{a}^{'\dagger}_{r,\chippackt_\FDcoef} + B_{rs} {F}_{\FDcoef}^{s,r}\hat{b}^{'\dagger}_{r,\chippackt_\FDcoef})\Bigg)^{n_s}\ket{0}\nonumber	\\
	&=\sqrt{\frac{N_c}{\pulseduration}}\sum_{s=0}^{M-1}\left( B_{r_0s}D_l^{s,r_0}\right)\frac{n_s}{\sqrt{n_s!}}\\
	&\times \Bigg(\sum_{\FDcoef=0}^{\infty}\sum_{r=0}^{M-1} (B_{rs}D_{\FDcoef}^{s,r}\hat{a}^{'\dagger}_{r,\chippackt_\FDcoef} + B_{rs} {F}_{\FDcoef}^{s,r}\hat{b}^{'\dagger}_{r,\chippackt_\FDcoef})\Bigg)^{n_s-1}\nonumber\\
	&\times \prod_{s'\neq s}\frac{1}{\sqrt{n_{s'}!}}\Bigg(\sum_{\FDcoef=0}^{\infty}\sum_{r=0}^{M-1} (B_{rs'}D_{\FDcoef}^{s',r}\hat{a}^{'\dagger}_{r,\chippackt_\FDcoef} \nonumber\\
	&\qquad\qquad\qquad+ B_{rs'} {F}_{\FDcoef}^{s',r}\hat{b}^{'\dagger}_{r,\chippackt_\FDcoef})\Bigg)^{n_{s'}}\ket{0}\nonumber
\end{align}
Define the auxiliary operator $\hat{v}^\dagger_s $ as
\begin{align}
	\hat{v}^\dagger_s = \sum_{\FDcoef=0}^{\infty}\sum_{r=0}^{M-1} B_{rs}D_{\FDcoef}^{s,r}\hat{a}^{'\dagger}_{r,\chippackt_\FDcoef} + B_{rs} {F}_{\FDcoef}^{s,r}\hat{b}^{'\dagger}_{r,\chippackt_\FDcoef}
\end{align}
We have
\begin{align}
	\hat{a}'_{r_0}(t)\ket{\Phi^d}
	&=\sqrt{\frac{N_c}{\pulseduration}}\sum_{s=0}^{M-1}\left( B_{r_0s}D_l^{s,r_0}\right)\frac{n_s}{\sqrt{n_s!}} (\hat{v}^\dagger_s)^{n_s-1}\\
	&\qquad\qquad\qquad\times\prod_{s'\neq s}\frac{1}{\sqrt{n_{s'}!}}(\hat{v}^\dagger_{s'})^{n_{s'}}\ket{0}\nonumber
\end{align}
We show that the defined auxiliary operator has the property of creation operators and thus can be considered as a virtual creation operator.
\begin{align}
	&\comm{\hat{v}_{s}}{\hat{v}^\dagger_{s'} }=\\
	&\qquad\sum_{\FDcoef=0}^{\infty}\sum_{r=0}^{M-1}\sum_{\FDcoef'=0}^{\infty}\sum_{r'=0}^{M-1} B_{rs}^*D_{\FDcoef}^{s,r*} B_{r's'}D_{\FDcoef'}^{s',r'}\comm{\hat{a}^{'}_{r,\chippackt_\FDcoef}}{\hat{a}^{'\dagger}_{r',\chippackt_{\FDcoef'}}}\nonumber\\
	&\quad+\sum_{\FDcoef=0}^{\infty}\sum_{r=0}^{M-1}\sum_{\FDcoef'=0}^{\infty}\sum_{r'=0}^{M-1} B_{rs}^*D_{\FDcoef}^{s,r*} B_{r's'}F_{\FDcoef'}^{s',r'}\comm{\hat{a}^{'}_{r,\chippackt_\FDcoef}}{\hat{b}^{'\dagger}_{r',\chippackt_{\FDcoef'}}}\nonumber\\
	&\quad+\sum_{\FDcoef=0}^{\infty}\sum_{r=0}^{M-1}\sum_{\FDcoef'=0}^{\infty}\sum_{r'=0}^{M-1} B_{rs}^*{F}_{\FDcoef}^{s,r*} B_{r's'}D_{\FDcoef'}^{s',r'}\comm{\hat{b}^{'}_{r,\chippackt_\FDcoef}}{\hat{a}^{'\dagger}_{r',\chippackt_{\FDcoef'}}}\nonumber\\
	&\quad+\sum_{\FDcoef=0}^{\infty}\sum_{r=0}^{M-1}\sum_{\FDcoef'=0}^{\infty}\sum_{r'=0}^{M-1} B_{rs}^*{F}_{\FDcoef}^{s,r*} B_{r's'}F_{\FDcoef'}^{s',r'}\comm{\hat{b}^{'}_{r,\chippackt_\FDcoef}}{\hat{b}^{'\dagger}_{r',\chippackt_{\FDcoef'}}}\nonumber\\
	&= \sum_{\FDcoef=0}^{\infty}\sum_{r=0}^{M-1}\sum_{\FDcoef'=0}^{\infty}\sum_{r'=0}^{M-1} B_{rs}^*D_{\FDcoef}^{s,r*} B_{r's'}D_{\FDcoef'}^{s',r'}\delta_{rr'}\delta_{\FDcoef\FDcoef'} \\
	&\qquad+ \sum_{\FDcoef=0}^{\infty}\sum_{r=0}^{M-1}\sum_{\FDcoef'=0}^{\infty}\sum_{r'=0}^{M-1} B_{rs}^*{F}_{\FDcoef}^{s,r*} B_{r's'}F_{\FDcoef'}^{s',r'}\delta_{rr'}\delta_{\FDcoef\FDcoef'}\nonumber\\
	&= \sum_{\FDcoef=0}^{\infty}\sum_{r=0}^{M-1} B_{rs}^*D_{\FDcoef}^{s,r*} B_{rs'}{D}_{\FDcoef}^{s',r} + \sum_{\FDcoef=0}^{\infty}\sum_{r=0}^{M-1} B_{rs}^*{F}_{\FDcoef}^{s,r*} B_{rs'}{F}_{\FDcoef}^{s',r}\\
	&=\sum_{r=0}^{M-1} B_{rs}^*B_{rs'}\left(\sum_{\FDcoef=0}^{\infty}(D_{\FDcoef}^{s,r*}{D}_{\FDcoef}^{s',r}+{F}_{\FDcoef}^{s,r*}{F}_{\FDcoef}^{s',r})\right)\\
	&=\left(\frac{1}{N_c} \sum_{k=0}^{N_c-1}\lambda_{s,k}\lambda_{s',k}\right)\sum_{r=0}^{M-1} B_{rs}^*B_{rs'}\\
	&=\delta_{ss'}
\end{align}
Since the commutation relation $\comm{\hat{v}_{s}}{\hat{v}^\dagger_{s'} }=\delta_{ss'}$ holds, the corresponding number states of these auxiliary field creation operators form orthogonal modes, hence we define 
\begin{align}
	\ket{n_{s}}_{v_s} = \frac{1}{\sqrt{n_{s'}!}}(\hat{v}^\dagger_{s'})^{n_{s'}}\ket{0}
\end{align}
which denotes a quantum state with $n_s$ photons in mode $v_s$.

Therefore
\begin{align}
	\hat{a}'_{r_0}(t)\ket{\Phi^d}
	&=\sqrt{\frac{N_c}{\pulseduration}}\sum_{s=0}^{M-1}\sqrt{n_s}\left( B_{r_0s}D_l^{s,r_0}\right)\\
	&\qquad\times\ket{n_{0}}_{v_0}\ket{n_{1}}_{v_1}\cdots\ket{n_{s}-1}_{v_s}\cdots \ket{n_{M-1}}_{v_{M-1}}\nonumber
\end{align}
Thus the intensity at time $t\in [t_\FDcoef, t_{\FDcoef+1})$ is obtained as:
\begin{align}
	I_{r_0}(t)&=\ev{\hat{a}^{'\dagger}_{r_0}(t)\hat{a}'_{r_0}(t)}{\Phi^d}\\
	&= \frac{N_c}{\pulseduration} \sum_{s=0}^{M-1} n_s | B_{r_0s}D_{\FDcoef}^{s,r_0}|^2\\
	&=\frac{N_c}{\pulseduration} n_{r_0}| B_{r_0r_0}D_{\FDcoef}|^2 + \frac{N_c}{\pulseduration} \sum_{s\ne r_0} n_s | B_{r_0s}D_{\FDcoef}^{s,r_0}|^2
\end{align}
where 
\begin{align}
	D_{\FDcoef}^{s,r_0} &= \sum_{\fdcoef=\max(0,\FDcoef-N_c+1)}^{\FDcoef} \frac{1}{\sqrt{N_c}}d_{\fdcoef} \lambda_{s,\FDcoef-\fdcoef}\lambda_{r_0,\FDcoef-\fdcoef}\\
	D_{\FDcoef}^{s=r_0,r_0} &= D_{\FDcoef} = \sum_{\fdcoef=\max(0,\FDcoef-N_c+1)}^{\FDcoef} \frac{1}{\sqrt{N_c}}d_{\fdcoef} 
\end{align}

Next we pursue the same procedure for coherent states using Proposition \ref{prop:coherentFiltered}. If $t\in [t_l, t_{l+1})$, and receiver $r_0$ decodes the data of sender $s_0=r_0$
\begin{align}
	\ket{\alpha_{\wavepackt_{\text{T}}}} & = \prod_{\FDcoef=0}^{\infty}\Bigg|B_{r_0r_0}D_{\FDcoef} \alpha_{s_0=r_0,\chippackt_k}+ \sum_{s\neq r_0}B_{r_0 s}D_{\FDcoef}^{s, r_0}\alpha_{s,\chippackt_k}\Bigg\rangle\\
	&= \prod_{\FDcoef=0}^{\infty} \exp(-\frac{1}{2}\left|\sum_{s=0}^{M-1}B_{r_0 s}D_{\FDcoef}^{s, r_0 }\alpha_{s}\right|^2)\\
	&\qquad\times\exp(\sum_{s=0}^{M-1}B_{r_0 s}D_{\FDcoef}^{s, r_0 }\alpha_{s}\hat{a}^{'\dagger}_{r,\chippackt_k})\ket{0} \nonumber
\end{align}
Based on the above equation, we may write
\begin{align}
	&\hat{a}'_{r_0 }(t)\ket{\alpha_{r_0,\wavepackt_{\text{T}}}}=\nonumber\\
	&\sqrt{\frac{N_c}{\pulseduration}}\frac{\partial}{\partial\hat{a}'_{r_0 ,\chippackt_l}}\Bigg[\prod_{\FDcoef=0}^{\infty} \exp(-\frac{1}{2}\left|\sum_{s=0}^{M-1}B_{r_0 s}D_{\FDcoef}^{s, r_0 }\alpha_{s}\right|^2)\nonumber\\
	&\qquad\qquad\times\exp(\sum_{s=0}^{M-1}B_{r_0 s}D_{\FDcoef}^{s, r_0 }\alpha_{s}\hat{a}^{'\dagger}_{r,\chippackt_\FDcoef})\Bigg]\ket{0} \nonumber	\\
	&=\sqrt{\frac{N_c}{\pulseduration}}\left(\sum_{s=0}^{M-1}B_{r_0 s}D_l^{s, r_0 }\alpha_{s}\right)\nonumber\\
	&\qquad\times\prod_{\FDcoef=0}^{\infty} \exp(-\frac{1}{2}\left|\sum_{s=0}^{M-1}B_{r_0 s}D_{\FDcoef}^{s, r_0 }\alpha_{s}\right|^2)\nonumber\\
	&\qquad\qquad\times\exp(\sum_{s=0}^{M-1}B_{r_0 s}D_{\FDcoef}^{s, r_0 }\alpha_{s}\hat{a}^{'\dagger}_{r,\chippackt_\FDcoef})\ket{0} \nonumber\\
	&=\sqrt{\frac{N_c}{\pulseduration}}\left(\sum_{s=0}^{M-1}B_{r_0 s}D_l^{s, r_0 }\alpha_{s}\right) \ket{\alpha_{r_0,\wavepackt_{\text{T}}}}
\end{align}
Then for $t\in[t_\FDcoef, t_{\FDcoef+1})$ we obtain
\begin{align}
	I_{r_0 }(t)&=\ev{\hat{a}^{'\dagger}_{r_0 }(t)\hat{a}'_{r_0 }(t)}{\Phi^d}\nonumber\\
	&= \frac{N_c}{\pulseduration} \left|\sum_{s=0}^{M-1}B_{r_0 s}D_{\FDcoef}^{s, r_0 }\alpha_{s}\right|^2
\end{align}

\section{Two-user System with Single Photons}
\label{appendix:2singlephoton}
In this section, we consider the case of $M=2$ users with single photon transmitters and on-off keying (OOK) modulation scheme over a balanced star-coupler.
\subsection{One user is transmitting}
Assume that only transmitter $s=0$ is transmitting single photon quantum signal and the other source is turned off and thus is in vacuum state.
From (\ref{eq:number:decoded}), the receiver's decoded state is
\begin{align}
	&\ket{\Phi^d}=\left(\frac{1}{\sqrt{N_c}}\sum_{k=0}^{N_c-1}  \sum_{r=0}^{1} B_{r0} \lambda_{0,k}\lambda_{r,k}\hat{a}_{r,\chippackt_k}^\dagger\right)\ket{0}\\
	&=\frac{1}{\sqrt{N_c}}\Bigg(\sum_{k=0}^{N_c-1}  B_{{0}0} \hat{a}_{{0},\chippackt_k}^\dagger \nonumber\\
	&\qquad\qquad+\sum_{k=0}^{N_c-1}B_{10} \lambda_{0,{k}}\lambda_{1,{k}}\hat{a}_{1,\chippackt_k}^\dagger\Bigg)\ket{0}
\end{align}
The first term means that the single photon has arrived at receiver 0, while in the second term stands for the case that single photon has arrived at the other receiver.

The overall density matrix is
\begin{align}
	\rho_d = \ketbra{\Phi^d}
\end{align}
In order to calculate the received state at the receiver $r=0$, we need to calculate the partial trace of $\rho_d$ with respect to $r\ne 0$.
Basis vectors of the Hilbert space of $r\ne 0$, with non-zero contribution to the partial trace, are of the form $\lbrace \ket{0_1} , \ket{1_{1,\chippackt_k}} \rbrace$.
$\ket{0_1}$ is the state where receiver $r = 1$ is in the vacuum state, and $\ket{1_{1,\chippackt_k}}$ are the cases where $1$ photon has arrived at receivers  $r=1$ at chip-time $k$. 
\begin{align}
	\rho_0^d &\coloneqq \Tr_{r\ne 0}(\rho_d) = \Tr_{r\ne 0}(\ketbra{\Phi^d}) \\
	&= \braket{0_1}{\Phi^d}\braket{\Phi^d}{0_1} + \sum_{k=0}^{N_c-1}\braket{1_{1,\chippackt_k}}{\Phi^d}\braket{\Phi^d}{1_{1,\chippackt_k}}\nonumber
\end{align}

The projection of $\ket{\Phi^d}$ with respect to $\ket{0_1}$ is 
\begin{align}
	\braket{0_1}{\Phi^d} &= \frac{1}{\sqrt{N_c}}\sum_{k=0}^{N_c-1}  B_{{0}0} \lambda_{0,{k}}\lambda_{{0},{k}}\hat{a}_{{0},\chippackt_k}^\dagger \ket{0_0}\nonumber\\
	&=  B_{{0}0} \frac{1}{\sqrt{N_c}}\sum_{k=0}^{N_c-1} \lambda_{0,{k}}\lambda_{{0},{k}}\ket{1_{{0},\chippackt_k}}\nonumber\\
	&=B_{{0}0} \ket{1_{0,\wavepackt}}
\end{align}
and the projection  with respect to $\ket{1_{1,\chippackt_k}}$ is 
\begin{align}
	\braket{1_{1,\chippackt_k}}{\Phi^d} &= \frac{1}{\sqrt{N_c}}B_{{1}0} \lambda_{0,{k}}\lambda_{{1},{k}} \ket{0_0}
\end{align}
Therefore
\begin{align}
	\rho_0^d &= |B_{{0}0}|^2 \ketbra{1_{0,\wavepackt}} + \sum_{k=0}^{N_c-1}\frac{1}{N_c}|B_{10}|^2 \ketbra{0_0} \nonumber\\
	&= |B_{10}|^2 \ketbra{0_0} +  |B_{{0}0}|^2 \ketbra{1_{0,\wavepackt}},
\end{align}
where $\ket{\psi_0}=\ket{1_{0,\wavepackt}}$ is the transmitted quantum signal by sender $s=0$. The term $\ketbra{1_{0,\wavepackt}}$ in the above statistical mixture denotes that the single photon quantum signal of transmitter $s=0$ is correctly recovered. In contrast, the term $\ketbra{0_0}$ stands for the inevitable loss induced by the broadcasting channel.

Since the evolution of the filter is represented by operator $\hat{\mathrm{H}}$, we may obtain the effect of filter using the following relation
 \begin{align}
 	\rho_0^F = \hat{\mathrm{H}}\rho_0^d \hat{\mathrm{H}}^\dagger
 \end{align}

The transmitted signal through the filter can be obtained by taking the partial trace with respect to the filter's reflection.
The basis for partial trace (with nonzero projection value) is $\lbrace \ket{0_{0,\text{R}}}, \ket{1_{0, \text{R}, \chippackt_\FDcoef}}\rbrace$. The two types of states $\ket{0_{0,\text{R}}}$ and $\ket{1_{0, \text{R}, \chippackt_\FDcoef}}$ denote vacuum state and single photon state at time interval $\FDcoef$ in the reflected output port of the filter, respectively.

In order to obtain the state after the filter we need to calculate the following partial trace.
\begin{align}
	\rho_0^{\text{T}} &= \Tr_{\text{R}}(\rho_0^F) \\
	&= \bra{0_{0,\text{R}}}\rho_0^F\ket{0_{0,\text{R}}} + \sum_{\FDcoef=0}^{\infty}\bra{1_{0, \text{R}, \chippackt_\FDcoef}}\rho_0^F\ket{1_{0, \text{R}, \chippackt_\FDcoef}}\nonumber\\
	&=\quad|B_{10}|^2 \braket{0_{0,\text{R}}}{0_0}\braket{0_0}{0_{0,\text{R}}} \\
	&\quad +  |B_{{0}0}|^2 \mel{0_{0,\text{R}}}{\hat{\mathrm{H}}}{1_{0,\wavepackt}}\mel{1_{0,\wavepackt}}{\hat{\mathrm{H}}^\dagger}{0_{0,\text{R}}} \\
	&\quad+\sum_{\FDcoef=0}^{\infty}|B_{10}|^2 \braket{1_{0, \text{R}, \chippackt_\FDcoef}}{0_0}\braket{0_0}{1_{0, \text{R}, \chippackt_\FDcoef}} \\
	&\quad +  \sum_{\FDcoef=0}^{\infty}|B_{{0}0}|^2 \mel{1_{0, \text{R}, \chippackt_\FDcoef}}{\hat{\mathrm{H}}}{1_{0,\wavepackt}}\mel{1_{0,\wavepackt}}{\hat{\mathrm{H}}^\dagger}{1_{0, \text{R}, \chippackt_\FDcoef}} 
\end{align}

The projections of $\ket{0}$ with respect to the basis functions are
\begin{align}
	\braket{1_{0, \text{R}, \chippackt_\FDcoef}}{0} &= 0\\
	\braket{0_{0,\text{R}}}{0} &= \ket{0_{0,\text{T}}}
\end{align}
From Theorem \ref{thm:filter}, the correctly decoded single photon quantum signal after filter is 
\begin{align}
	 \ket{1_{0,\wavepackt^F}}=\hat{\mathrm{H}}\ket{1_{0,\wavepackt}} = \sum_{\FDcoef=0}^{\infty}\left({D}_{\FDcoef}  \ket{1_{0,\text{T},\chippackt_\FDcoef}} + {F}_{\FDcoef}  \ket{1_{0,\text{R},\chippackt_\FDcoef}} \right)
\end{align}
The projections of this quantum signal with respect to the basis functions are obtained from 
 as follows
\begin{align}
	\mel{1_{0, \text{R}, \chippackt_k}}{\hat{\mathrm{H}}}{1_{0,\wavepackt}} &= {F}_{\FDcoef} \ket{0_{0,\text{T}}}\\
	\mel{0_{0,\text{R}}}{\hat{\mathrm{H}}}{1_{0,\wavepackt}} 
	&=\sum_{\FDcoef=0}^{\infty} D_{\FDcoef} \ket{1_{0,\text{T},\chippackt_\FDcoef}}
\end{align}
The transmitted state after the filter in case where the single photon has passed the filter is
\begin{align}
	\ket{1_{0,\wavepackt_\text{T}}}\coloneqq \frac{1}{\sqrt{\sum_{\FDcoef=0}^{\infty}|D_{\FDcoef}|^2}}\sum_{\FDcoef=0}^{\infty} D_{\FDcoef} \ket{1_{0,\text{T},\chippackt_\FDcoef}},
\end{align}
which gives the following state at the receiver end
\begin{align}
	\rho_0^{\text{T}} &= \left(|B_{10}|^2  +  |B_{{0}0}|^2 \sum_{\FDcoef=0}^{\infty}|{F}_{\FDcoef}|^2 \right)\ketbra{0_{0,\text{T}}}\label{eq:appendix:rhoT1}\\
	&\qquad+ |B_{{0}0}|^2\left(\sum_{\FDcoef=0}^{\infty}|D_{\FDcoef}|^2\right)\ketbra{1_{0,\wavepackt_\text{T}}}\nonumber
\end{align}
For a balanced star-coupler, the chip-time photon statistics can be easily calculated 
\begin{align}
	\mathbb{P}(n=0) &= \frac{1}{2}\left( 1  +   \sum_{\FDcoef=0}^{\infty}|{F}_{\FDcoef}|^2 \right)\\
	\mathbb{P}(n_k=1) &= \frac{1}{2}|D_{\FDcoef}|^2,
\end{align}
which gives the overall photon statistics
\begin{align}
	\mathbb{P}(n=0) &= \frac{1}{2}\left( 1  +   \sum_{\FDcoef=0}^{\infty}|{F}_{\FDcoef}|^2 \right)=1-\frac{1}{2}\mathfrak{D}\\
	\mathbb{P}(n=1) &= \frac{1}{2}\sum_k|D_{\FDcoef}|^2 =\frac{1}{2}\mathfrak{D}
\end{align}

Similarly when $s=1$ is transmitting single photons and $s=0$ is turned off the following mixed state is obtained at the receiver of user $s=0$:
\begin{align}
	\rho_0^{\text{T}} &= \left( |B_{10}|^2  +  |B_{{0}0}|^2 \sum_{\FDcoef=0}^{\infty}|F^{0,1}_\FDcoef|^2 \right)\ketbra{0_{0,\text{T}}}\label{eq:appendix:rhoT10}\\
	&\qquad+ |B_{{0}0}|^2\left(\sum_{\FDcoef=0}^{\infty}|D^{0,1}_\FDcoef|^2\right)\ketbra{1_{0,\wavepackt_\text{T}}},\nonumber
\end{align}
where 
\begin{align}
	\ket{1_{0,\wavepackt_\text{T}}}=\frac{1}{\sqrt{\sum_{\FDcoef=0}^{\infty}|D^{0,1}_\FDcoef|^2}}\sum_{\FDcoef=0}^{\infty} D^{0,1}_\FDcoef \ket{1_{0,\text{T},\chippackt_\FDcoef}}
\end{align}
Also
\begin{align}
	\mathbb{P}({n}_0=0) &=1-\frac{1}{2}\mathfrak{D}^{0,1}\\
	\mathbb{P}({n}_0=1) &=\frac{1}{2}\mathfrak{D}^{0,1},
\end{align}
where $\mathfrak{D}^{0,1}= \sum_{\FDcoef=0}^{\infty}|D^{0,1}_\FDcoef|^2$.

\subsection{Two Users}
Assume that transmitters $\mathcal{S}=\lbrace 0, 1\rbrace$ are transmitting single photons simultaneously. Using Proposition \ref{prop:FockDecode}, we have 
\begin{align}
	&\ket{\Phi^d}=\prod_{s=0}^1\left(\frac{1}{\sqrt{N_c}}\sum_{k=0}^{N_c-1}  \sum_{r=0}^{1}B_{rs} \lambda_{s,k}\lambda_{r,k}\hat{a}_{r,\chippackt_k}^\dagger\right)\ket{0}
\end{align}
The state $\ket{\Phi^d}$ can be expressed in terms of the chip-time number states as follows
\begin{align}
	&\ket{\Phi^d}\\
	&= \frac{1}{N_c}\Bigg(\sum_{k_0=0}^{N_c-1}B_{00}\ket{1_{0,\chippackt_{k_0}}}\Bigg)\Bigg(\sum_{k_1=0}^{N_c-1}B_{01}\lambda_{0,k_1}\lambda_{1,k_1}\ket{1_{0,\chippackt_{k_1}}}\Bigg)\nonumber\\
	&+ \frac{1}{N_c}\Bigg(\sum_{k_1=0}^{N_c-1}B_{11}\ket{1_{1,\chippackt_{k_1}}}\Bigg)\Bigg(\sum_{k_0=0}^{N_c-1}B_{10}\lambda_{1,k_0}\lambda_{0,k_0}\ket{1_{1,\chippackt_{k_0}}}\Bigg)\nonumber\\
	&+ \frac{1}{N_c}\Bigg(\sum_{k_0=0}^{N_c-1}\sum_{k_1=0}^{N_c-1}\Big(B_{00}B_{11}\ket{1_{0,\chippackt_{k_0}}}\ket{1_{1,\chippackt_{k_1}}} \nonumber\\
	&\qquad\qquad + B_{01}B_{10}\lambda_{0,k_1}\lambda_{1,k_1}\lambda_{0,k_0}\lambda_{1,k_0}\ket{1_{0,\chippackt_{k_0}}}\ket{1_{1,\chippackt_{k_1}}}\Big)\Bigg)\nonumber
\end{align}
For a balanced star-coupler
\begin{align}
	&\ket{\Phi^d}\Big|_{M=2, \text{balanced}}\\
	&= \frac{1}{2N_c}\Bigg(\sum_{k_0=0}^{N_c-1}\ket{1_{0,\chippackt_{k_0}}}\Bigg)\Bigg(\sum_{k_1=0}^{N_c-1}\lambda_{0,k_1}\lambda_{1,k_1}\ket{1_{0,\chippackt_{k_1}}}\Bigg)\nonumber\\
	&- \frac{1}{2N_c}\Bigg(\sum_{k_1=0}^{N_c-1}\ket{1_{1,\chippackt_{k_1}}}\Bigg)\Bigg(\sum_{k_0=0}^{N_c-1}\lambda_{1,k_0}\lambda_{0,k_0}\ket{1_{1,\chippackt_{k_0}}}\Bigg)\nonumber\\
	&+\frac{1}{2N_c}\Bigg(\sum_{k_0=0}^{N_c-1}\sum_{k_1=0}^{N_c-1}\Big((1-\lambda_{0,k_0}\lambda_{1,k_0}\lambda_{0,k_1}\lambda_{1,k_1})\nonumber\\
	&\qquad\qquad\qquad\qquad\qquad\times\ket{1_{0,\chippackt_{k_0}}}\ket{1_{1,\chippackt_{k_1}}}\Big)\Bigg)\nonumber
\end{align}
The last term in summation is zero for $k_0=k_1$, 
which gives the spread spectrum Hong-Ou-Mandel effect. That is the state $\ket{1_{0,\chippackt_k}}\ket{1_{1,\chippackt_k}}$ never appears in the output of the balanced star-coupler for spread spectrum single photon inputs when two users are simultaniously transmitting information.

The overall density matrix is
\begin{align}
	\rho_d = \ketbra{\Phi^d}
\end{align}
Basis vectors for the Hilbert space of $r\ne 0$ are of the form $\lbrace \ket{0_1}, \ket{1_{1,\chippackt_k}}, \ket{2_{1,\chippackt_{k}}}, \ket{1_{1,\chippackt_{k_0}}}\ket{1_{1,\chippackt_{k_1}}} (k_0< k_1)\rbrace$.
Then, we calculate the partial projections with respect to the basis vectors. The projection of $\Phi^d$ with respect to $\ket{0_1}$ is
\begin{align}
	\braket{0_1}{\Phi^d}\braket{\Phi^d}{0_1} &= \frac{1}{4}\ket{1_{0,\wavepackt}}\ket{1_{0,\wavepackt^{e_1d_0}}} \bra{1_{0,\wavepackt^{e_1d_0}}}\bra{1_{0,\wavepackt}}
\end{align} 

The product form of  $\ket{1_{0,\wavepackt}}\ket{1_{0,\wavepackt^{e_1d_0}}}$ is not normalized. Therefore, we need to obtain the normalization factor. First, we express the chip-time interval decomposition of this quantum signal
\begin{align}
	&\ket{1_{0,\wavepackt}}\ket{1_{0,\wavepackt^{e_1d_0}}}\coloneqq   \left(\frac{1}{\sqrt{N_c}}\sum_{k_0=0}^{N_c-1}\ket{1_{{0},\chippackt_{k_0}}}\right)\\
	&\qquad\qquad\times \left(\frac{1}{\sqrt{N_c}}\sum_{k_1=0}^{N_c-1}\lambda_{1,{k_1}}\lambda_{{0},{k_1}}\ket{1_{{0},\chippackt_{k_1}}}\right)\nonumber\\
	&=\frac{1}{N_c}\Bigg(\underset{\text{s.t.} k_0\ne k_1}{\sum_{k_0=0}^{N_c-1}\sum_{k_1=0}^{N_c-1}}\lambda_{1,{k_1}}\lambda_{{0},{k_1}}\ket{1_{{0},\chippackt_{k_0}}}\ket{1_{{0},\chippackt_{k_1}}}\Bigg)\nonumber\\
	&+\frac{1}{N_c}\sum_{k=0}^{N_c-1}\lambda_{1,{k}}\lambda_{{0},{k}}\sqrt{2}\ket{2_{{0},\chippackt_{k}}}\\
	&=\frac{1}{N_c}\Bigg(\underset{\text{s.t.} k_0< k_1}{\sum_{k_0=0}^{N_c-1}\sum_{k_1=0}^{N_c-1}}(\lambda_{1,{k_0}}\lambda_{{0},{k_0}}+\lambda_{1,{k_1}}\lambda_{{0},{k_1}})\ket{1_{{0},\chippackt_{k_0}}}\ket{1_{{0},\chippackt_{k_1}}}\Bigg)\nonumber\\
	&+\frac{1}{N_c}\sum_{k=0}^{N_c-1}\lambda_{1,{k}}\lambda_{{0},{k}}\sqrt{2}\ket{2_{{0},\chippackt_{k}}}
\end{align} 
Therefore
\begin{align}
	&\constplusd \coloneqq \|\frac{1}{2}\ket{1_{0,\wavepackt}}\ket{1_{0,\wavepackt^{e_1d_0}}}\|^2=\\
	&=\frac{1}{4N_c^2}\Bigg(\underset{\text{s.t.} k_0< k_1}{\sum_{k_0=0}^{N_c-1}\sum_{k_1=0}^{N_c-1}}(\lambda_{1,{k_0}}\lambda_{{0},{k_0}}+\lambda_{1,{k_1}}\lambda_{{0},{k_1}})^2\Bigg)\nonumber\\
	&\quad+\frac{1}{4N_c^2}\sum_{k=0}^{N_c-1}(\lambda_{1,{k}}\lambda_{{0},{k}}\sqrt{2})^2\nonumber\\
	&=\frac{1}{4N_c^2}\Bigg(\underset{\text{s.t.} k_0< k_1}{\sum_{k_0=0}^{N_c-1}\sum_{k_1=0}^{N_c-1}}(2+2\lambda_{1,{k_0}}\lambda_{{0},{k_0}}\lambda_{1,{k_1}}\lambda_{{0},{k_1}})\Bigg)\nonumber\\
	&\quad+\frac{1}{4N_c^2}2N_c\nonumber\\
	&=\frac{1}{4N_c^2}2\frac{N_c(N_c-1)}{2}+\frac{1}{2N_c^2}N_c\nonumber\\
	&\quad+\frac{1}{4N_c^2}\underset{\text{s.t.} k_0\ne k_1}{\sum_{k_0=0}^{N_c-1}\sum_{k_1=0}^{N_c-1}}(\lambda_{1,{k_0}}\lambda_{{0},{k_0}}\lambda_{1,{k_1}}\lambda_{{0},{k_1}})\nonumber \\
	&\quad+ \frac{1}{4N_c^2}\underbrace{\sum_{k=0}^{N_c-1}(\lambda_{1,{k}}\lambda_{{0},{k}}\lambda_{1,{k}}\lambda_{{0},{k}})}_{N_c} -\frac{1}{4N_c^2}N_c\nonumber\\
	&=\frac{1}{4N_c^2}2\frac{N_c(N_c-1)}{2}+\frac{1}{2N_c^2}N_c-\frac{1}{4N_c^2}N_c\nonumber\\
	&\quad+\frac{1}{4N_c^2}\sum_{k_0=0}^{N_c-1}\sum_{k_1=0}^{N_c-1}(\lambda_{1,{k_0}}\lambda_{{0},{k_0}}\lambda_{1,{k_1}}\lambda_{{0},{k_1}})\nonumber \\
	&= \frac{1}{4} \Bigg(1 + \Big(\frac{1}{{N_c}}\sum_{k_0=0}^{N_c-1}\lambda_{1,{k_0}}\lambda_{{0},{k_0}}\Big)\Big(\frac{1}{{N_c}}\sum_{k_1=0}^{N_c-1}\lambda_{1,{k_1}}\lambda_{{0},{k_1}}\Big)\Bigg)\nonumber\\
	&=  \frac{1}{4}\left(1+\left|\braket{\Lambda_0}{\Lambda_1}\right|^2\right)
\end{align} 
We may define the normalized state corresponding to $\frac{1}{2}\ket{1_{0,\wavepackt}}\ket{1_{0,\wavepackt^{e_1d_0}}}$ as
\begin{align}
	\ket{(1+1)_{0,e_1d_0}}&\coloneqq \frac{1}{2}(\constplusd)^{-\frac{1}{2}}\hat{a}^{\dagger}_\xi \hat{a}^{\dagger}_{\xi^{e_1d_0}}\ket{0}\label{eq:app:K:1p1}\\
	&=  \frac{1}{2}(\constplusd)^{-\frac{1}{2}}\ket{1_{0,\wavepackt}}\ket{1_{0,\wavepackt^{e_1d_0}}}\label{eq:app:oneponed}
\end{align} 
Therefore the projection of $\Phi^d$ with respect to $\ket{0_1}$ is further simplified as
\begin{align}
	\braket{0_1}{\Phi^d}\braket{\Phi^d}{0_1} &=\constplusd 	\ketbra{(1+1)_{0,e_1d_0}}
\end{align} 
This part of the received mixed quantum signal, corresponds to the case where both transmitted photons arrive at receiver 0 (with probability $|B_{00}B_{01}|^2=\frac{1}{4}$). 

The projection of $\Phi^d$ on $\ket{1_{1,\chippackt_k}}$ is calculated as follows.
\begin{align}
	&\braket{1_{1,\chippackt_k}}{\Phi^d} = \frac{1}{2N_c}\bra{1_{1,\chippackt_k}}\Bigg(\sum_{k_0=0}^{N_c-1}\sum_{k_1=0}^{N_c-1}\\
	&\qquad\qquad\Big((1-\lambda_{0,k_0}\lambda_{1,k_0}\lambda_{0,k_1}\lambda_{1,k_1})\ket{1_{0,\chippackt_{k_0}}}\ket{1_{1,\chippackt_{k_1}}}\Big)\Bigg)\nonumber\\
	&=\frac{1}{2N_c}\bra{1_{1,\chippackt_k}}\Bigg(\sum_{k_0=0}^{N_c-1}(1-\lambda_{0,k_0}\lambda_{1,k_0}\lambda_{0,k}\lambda_{1,k}) \ket{1_{0,\chippackt_{k_0}}}\ket{1_{1,\chippackt_{k}}}\\
	&+\underset{k_1\ne k}{\sum_{k_0=0}^{N_c-1}\sum_{k_1=0}^{N_c-1}}\Big((1-\lambda_{0,k_0}\lambda_{1,k_0}\lambda_{0,k_1}\lambda_{1,k_1})\ket{1_{0,\chippackt_{k_0}}}\ket{1_{1,\chippackt_{k_1}}}\Big)\Bigg)\nonumber\\
	&=\frac{1}{2N_c}\Bigg(\sum_{k_0=0}^{N_c-1}(1-\lambda_{0,k_0}\lambda_{1,k_0}\lambda_{0,k}\lambda_{1,k}) \ket{1_{0,\chippackt_{k_0}}}\underbrace{\braket{1_{1,\chippackt_k}}{1_{1,\chippackt_{k}}}}_{1}\\
	&+\underset{k_1\ne k}{\sum_{k_0=0}^{N_c-1}\sum_{k_1=0}^{N_c-1}}\Big((1-\lambda_{0,k_0}\lambda_{1,k_0}\lambda_{0,k_1}\lambda_{1,k_1})\ket{1_{0,\chippackt_{k_0}}}\underbrace{\braket{1_{1,\chippackt_k}}{1_{1,\chippackt_{k_1}}}}_{0}\Big)\Bigg)\nonumber\\
	&=\frac{1}{2N_c}\sum_{k_0=0}^{N_c-1}(1-\lambda_{0,k_0}\lambda_{1,k_0}\lambda_{0,k}\lambda_{1,k})\ket{1_{0,\chippackt_{k_0}}}
\end{align} 
Again the norm of the above equation is not one and we should find the normalization factor
\begin{align}
	\constonekd &\coloneqq \|\braket{1_{1,\chippackt_k}}{\Phi^d} \|^2 \\
	&= \frac{1}{4N_c^2}\sum_{k_0=0}^{N_c-1}|1-\lambda_{0,k_0}\lambda_{1,k_0}\lambda_{0,k}\lambda_{1,k}|^2\nonumber\\
	&=\frac{1}{4N_c^2}\sum_{k_0=0}^{N_c-1}\left(2-2\lambda_{0,k_0}\lambda_{1,k_0}\lambda_{0,k}\lambda_{1,k}\right)\nonumber\\
	&=\frac{1}{2N_c}\left(1-\frac{\lambda_{0,k}\lambda_{1,k}}{N_c}\sum_{k_0=0}^{N_c-1}\lambda_{0,k_0}\lambda_{1,k_0}\right)\nonumber\\
	&=\frac{1}{2N_c}\left(1-\lambda_{0,k}\lambda_{1,k}\braket{\Lambda_0}{\Lambda_1}\right)\nonumber
\end{align} 
Next, we define the normalized state as 
\begin{align}
	\ket{1_{0,e_1d_0,k}} &\coloneqq  \frac{(\constonekd)^{-\frac{1}{2}}}{2N_c}\sum_{k_0=0}^{N_c-1}(1-\lambda_{0,k_0}\lambda_{1,k_0}\lambda_{0,k}\lambda_{1,k})\ket{1_{0,\chippackt_{k_0}}}\nonumber\\
	&=\frac{(\constonekd)^{-\frac{1}{2}}}{2\sqrt{N_c}}\frac{1}{\sqrt{N_c}}\sum_{k_0=0}^{N_c-1}\ket{1_{0,\chippackt_{k_0}}}\\
	&-\frac{(\constonekd)^{-\frac{1}{2}}}{2\sqrt{N_c}} \lambda_{0,k}\lambda_{1,k}\frac{1}{\sqrt{N_c}}\sum_{k_0=0}^{N_c-1}\lambda_{0,k_0}\lambda_{1,k_0}\ket{1_{0,\chippackt_{k_0}}}\nonumber\\
	&=\frac{(\constonekd)^{-\frac{1}{2}}}{2\sqrt{N_c}}\left(\ket{1_{0,\wavepackt}}-\lambda_{0,k}\lambda_{1,k}\ket{1_{0,\wavepackt^{e_1d_0}}}\right)\label{eq:app:k:0e0d1k}\\
	&=\frac{\ket{1_{0,\wavepackt}}-\lambda_{0,k}\lambda_{1,k}\ket{1_{0,\wavepackt^{e_1d_0}}}}{\sqrt{2(1-\lambda_{0,k}\lambda_{1,k}\braket{\Lambda_0}{\Lambda_1})}},
\end{align} 
which is the superposition of the original signal and the 
multiple access interference signal.

The projection of $\Phi^d$ on $\ket{1_{1,\chippackt_{k_0}}}\ket{1_{1,\chippackt_{k_1}}} $ with $k_0< k_1$ is calculated as follows.
\begin{align}
	&(\bra{1_{1,\chippackt_{k_1}}}\bra{1_{1,\chippackt_{k_0}}} )\ket{\Phi^d}=(\bra{1_{1,\chippackt_{k_1}}}\bra{1_{1,\chippackt_{k_0}}} )\nonumber\\
	&\qquad\times - \frac{1}{2N_c}\Bigg(\sum_{k'_1=0}^{N_c-1}\ket{1_{1,\chippackt_{k'_1}}}\Bigg)\nonumber\\
	&\qquad\times\Bigg(\sum_{k'_0=0}^{N_c-1}\lambda_{1,k'_0}\lambda_{0,k'_0}\ket{1_{1,\chippackt_{k'_0}}}\Bigg)\nonumber\\
	&=-\frac{1}{2N_c}  (\bra{1_{1,\chippackt_{k_1}}}\bra{1_{1,\chippackt_{k_0}}} )\nonumber\\
	&\times\Bigg(\ket{1_{1,\chippackt_{k_0}}}+\ket{1_{1,\chippackt_{k_1}}}+\underset{\text{s.t.} k'_1\ne k_0, k_1}{\sum_{k'_1=0}^{N_c-1}}\ket{1_{1,\chippackt_{k'_1}}}\Bigg)\nonumber\\
	&\times\Bigg(\lambda_{1,k_0}\lambda_{0,k_0}\ket{1_{1,\chippackt_{k_0}}}+\lambda_{1,k_1}\lambda_{0,k_1}\ket{1_{1,\chippackt_{k_1}}}\nonumber\\
	&\qquad+\underset{\text{s.t.} k'_0\ne k_0, k_1}{\sum_{k'_0=0}^{N_c-1}}\lambda_{1,k'_0}\lambda_{0,k'_0}\ket{1_{1,\chippackt_{k'_0}}}\Bigg)\nonumber\\
	&= -\frac{1}{2N_c} (\lambda_{0,{k_0}}\lambda_{1,{k_0}}+ \lambda_{0,{k_1}}\lambda_{1,{k_1}})\ket{0_0}
\end{align} 
Therefore
\begin{align}
&	(\bra{1_{1,\chippackt_{k_1}}}\bra{1_{1,\chippackt_{k_0}}})\ketbra{\Phi^d}(\ket{1_{1,\chippackt_{k_0}}}\ket{1_{1,\chippackt_{k_1}}})\label{eq:app:k:0d1}\\
	&=\frac{1}{4N_c^2} \left|\lambda_{0,{k_0}}\lambda_{1,{k_0}}+ \lambda_{0,{k_1}}\lambda_{1,{k_1}}\right|^2\ketbra{0_0}\nonumber\\
	&= \frac{1}{2N_c^2} \left(1 + \lambda_{0,{k_0}}\lambda_{1,{k_0}} \lambda_{0,{k_1}}\lambda_{1,{k_1}}\right)\ketbra{0_0}\nonumber
\end{align} 
The projection of $\Phi^d$ on $\ket{2_{1,\chippackt_{k}}}$ is
\begin{align}
	&\braket{2_{1,\chippackt_{k}}}{\Phi^d} =  - \frac{1}{2N_c}\bra{2_{1,\chippackt_{k}}}\Bigg(\sum_{k'_1=0}^{N_c-1}\ket{1_{1,\chippackt_{k'_1}}}\Bigg)\\
	&\qquad\times\Bigg(\sum_{k'_0=0}^{N_c-1}\lambda_{1,k'_0}\lambda_{0,k'_0}\ket{1_{1,\chippackt_{k'_0}}}\Bigg)\nonumber\\
	&=  - \frac{1}{2N_c}\bra{2_{1,\chippackt_{k}}}\Bigg(\ket{1_{1,\chippackt_{k}}}+\underset{\text{s.t.} k'_1\ne k}{\sum_{k'_1=0}^{N_c-1}}\ket{1_{1,\chippackt_{k'_1}}}\Bigg)\nonumber\\
	&\times\Bigg(\lambda_{1,k}\lambda_{0,k}\ket{1_{1,\chippackt_{k}}}+\underset{\text{s.t.} k'_1\ne k}{\sum_{k'_1=0}^{N_c-1}}\lambda_{1,k'_1}\lambda_{0,k'_1}\ket{1_{1,\chippackt_{k'_1}}}\Bigg)\nonumber\\
	&=- \frac{1}{2N_c}\lambda_{1,k}\lambda_{0,k}\sqrt{2}\braket{2_{1,\chippackt_{k}}}\nonumber\\
	&=- \frac{1}{2N_c}\lambda_{1,k}\lambda_{0,k}\sqrt{2} \ket{0_0}\nonumber
\end{align}
Therefore
\begin{align}
	\braket{2_{1,\chippackt_{k}}}{\Phi^d}\braket{\Phi^d}{2_{1,\chippackt_{k}}} &= \frac{1}{2N_c^2}\ketbra{0_0}\label{eq:app:k:0d2}
\end{align}
From (\ref{eq:app:k:0d1}) and (\ref{eq:app:k:0d2}), the coefficient of $ \ketbra{0_0}$ is
\begin{align}
	\constzerod&\coloneqq\frac{1}{2N_c^2}\underset{\text{s.t.} k_0< k_1}{\sum_{k_0=0}^{N_c-1}\sum_{k_1=0}^{N_c-1}} (1 + \lambda_{0,{k_0}}\lambda_{1,{k_0}} \lambda_{0,{k_1}}\lambda_{1,{k_1}})\nonumber\\
	&+\sum_{k=0}^{N_c-1} \frac{1}{2N_c^2}\nonumber\\
	&=\frac{1}{4N_c^2}\underset{\text{s.t.} k_0\ne k_1}{\sum_{k_0=0}^{N_c-1}\sum_{k_1=0}^{N_c-1}} (1 + \lambda_{0,{k_0}}\lambda_{1,{k_0}} \lambda_{0,{k_1}}\lambda_{1,{k_1}})\nonumber\\
	&+ \frac{1}{4N_c^2}\underbrace{\sum_{k=0}^{N_c-1}(\lambda_{1,{k}}\lambda_{{0},{k}}\lambda_{1,{k}}\lambda_{{0},{k}})}_{N_c} -\frac{1}{4N_c} + \frac{1}{2N_c}\nonumber\\
	&=\frac{1}{4N_c^2}N_c(N_c-1) +\frac{1}{4N_c}+ \frac{1}{4}\left|\braket{\Lambda_0}{\Lambda_1}\right|^2\nonumber\\
	&=\frac{1}{4}+ \frac{1}{4}\left|\braket{\Lambda_0}{\Lambda_1}\right|^2\nonumber\\
	&=  \frac{1}{4}\left(1+\left|\braket{\Lambda_0}{\Lambda_1}\right|^2\right)
	\\
	&=\constplusd
\end{align}
The overall received quantum signal is
\begin{align}
	\rho_0^d & = \constplusd\ketbra{(1+1)_{0,e_1d_0}} \label{eq:app:k:densityd2} \\
	&+ \sum_{k=0}^{N_c-1}\constonekd\ketbra{1_{0,e_1d_0,k}}+ \constzerod\ketbra{0_0}\nonumber\\
	&=\frac{1}{4}\left(1+\left|\braket{\Lambda_0}{\Lambda_1}\right|^2\right)\ketbra{(1+1)_{0,e_1d_0}}\\
	&+\frac{1}{4}\left(1+\left|\braket{\Lambda_0}{\Lambda_1}\right|^2\right)\ketbra{0_0}\nonumber\\&+\frac{1}{2N_c}\sum_{k=0}^{N_c-1}\left(1-\lambda_{0,k}\lambda_{1,k}\braket{\Lambda_0}{\Lambda_1}\right)\ketbra{1_{0,e_1d_0,k}}\nonumber
\end{align} 
For random (approximately orthogonal) codes  $\braket{\Lambda_0}{\Lambda_1}\approx 0$ therefore $\constplusd = \constzerod \approx \frac{1}{4}$ and $\constonekd\approx \frac{1}{2N_c}$. Consequently,
\begin{align}
	\rho_0^d & \approx \frac{1}{4}\ketbra{(1+1)_{0,e_1d_0}} + \frac{1}{4}\ketbra{0_0}\\
	&+ \frac{1}{2N_c}\sum_{k=0}^{N_c-1}\ketbra{1_{0,e_1d_0,k}}\nonumber
\end{align} 
For similar codes $\Lambda_0=\Lambda_1$, $\constplusd = \constzerod = \frac{1}{2}$, $\constonekd=0$ and $\ket{(1+1)_{0,e_1d_0}}=\ket{2_0}$  . Therefore, the density matrix (\ref{eq:app:k:densityd2}), due to the Hong-Ou-Mandel interference, reduces to 
\begin{align}
	\rho_0^d & = \frac{1}{2}\ketbra{2_0} +\frac{1}{2}\ketbra{0_0}
\end{align} 

Next, we apply the filter to $\rho^d$ and then we take the trace with respect to the reflected signals. The states $\ket{1_{0,e_1d_0,k}}$ correspond to the case when only a single photon arrives at the receiver, therefore the derivation is simliar to the single photon transmitter in (\ref{eq:appendix:rhoT1}). Therefore, utilizing Theorem \ref{thm:filter} and Proposition \ref{prop:filter} for (\ref{eq:app:k:0e0d1k}) we may write
\begin{align}
	&\hat{\mathrm{H}}\ket{1_{0,e_1d_0,k}}=\frac{(\constonekd)^{-\frac{1}{2}}}{2\sqrt{N_c}}\left(\hat{\mathrm{H}}\ket{1_{0,\wavepackt}}-\lambda_{0,k}\lambda_{1,k}\hat{\mathrm{H}}\ket{1_{0,\wavepackt^{e_1d_0}}}\right)\nonumber\\
	&=\frac{(\constonekd)^{-\frac{1}{2}}}{2\sqrt{N_c}}\Bigg(\sum_{\FDcoef=0}^{\infty}\Big({D}_{\FDcoef}  \ket{1_{0,\text{T},\chippackt_\FDcoef}} + {F}_{\FDcoef}  \ket{1_{0,\text{R},\chippackt_\FDcoef}} \Big)\nonumber\\
	&-\lambda_{0,k}\lambda_{1,k}\sum_{\FDcoef=0}^{\infty}\Big({D}^{0,1}_{\FDcoef}  \ket{1_{0,\text{T},\chippackt_\FDcoef}} + {F}^{0,1}_{\FDcoef}  \ket{1_{0,\text{R},\chippackt_\FDcoef}} \Big)\Bigg)\nonumber
\end{align} 
The projection with respect to $\ket{0_{0, \text{R}}}$ is
\begin{align}
	&(\constonekd)^{\frac{1}{2}}\bra{0_{0, \text{R}}}\hat{\mathrm{H}}\ket{1_{0,e_1d_0,k}}\\
	&=\frac{1}{2\sqrt{N_c}}\Bigg(\sum_{\FDcoef=0}^{\infty}D_{\FDcoef}\ket{1_{0,\text{T},\chippackt_{\FDcoef}}}-{\lambda_{0,k}\lambda_{1,k}}\sum_{\FDcoef=0}^{\infty}D^{0,1}_{\FDcoef}\ket{1_{0,\text{T},\chippackt_{\FDcoef}}} \nonumber\Bigg)
\end{align}
We define the normalized state as
\begin{align}
	&\ket{1_{0,\text{T},e_1d_0,k}} \coloneqq  \left( \constdLk\right)^{-\frac{1}{2}} \times\Bigg(\sum_{\FDcoef=0}^{\infty}D_{\FDcoef}\ket{1_{0,\text{T},\chippackt_{\FDcoef}}}\nonumber\\
	&\qquad-{\lambda_{0,k}\lambda_{1,k}}\sum_{\FDcoef=0}^{\infty}D^{0,1}_{\FDcoef}\ket{1_{0,\text{T},\chippackt_{\FDcoef}}}\Bigg)\\
	&=\left( \constdLk\right)^{-\frac{1}{2}} \times\Bigg[\Big({\sum_{\FDcoef=0}^{\infty}|D_{\FDcoef}|^2}\Big)^{\frac{1}{2}}\ket{1_{0,\wavepackt_\text{T}}}\nonumber\\
	&\qquad-{\lambda_{0,k}\lambda_{1,k}}\Big({\sum_{\FDcoef=0}^{\infty}|D^{0,1}_{\FDcoef}|^2}\Big)^{\frac{1}{2}}\ket{1_{0,\wavepackt_\text{T}^{e_1d_0}}}\Bigg],
\end{align} 
where the corresponding normalization coefficient is 
\begin{align}
	\constdLk\coloneqq &\frac{1}{4N_c}\sum_{\FDcoef=0}^{\infty}|D_{\FDcoef}-\lambda_{0,k}\lambda_{1,k}D^{0,1}_{\FDcoef}|^2. \label{eq:app:k:constdLk}
\end{align}  
The projection with respect to $\ket{1_{0, \text{R}, \chippackt_\FDcoef}}$ is
\begin{align}
	(\constonekd)^{\frac{1}{2}}\bra{1_{0, \text{R}, \chippackt_\FDcoef}}\hat{\mathrm{H}}\ket{1_{0,e_1d_0,k}} &= \frac{1}{2\sqrt{N_c}}(F_{\FDcoef}-\lambda_{0,k}\lambda_{1,k}F^{0,1}_{\FDcoef})\ket{0_{0,\text{T}}}
	\end{align}
We define the following coefficient to simplify the terms involving  $\ketbra{0_{0,\text{T}}}$
\begin{align}
	\constfLk\coloneqq & \frac{1}{4N_c}\sum_{\FDcoef=0}^{\infty}|F_{\FDcoef}-\lambda_{0,k}\lambda_{1,k}F^{0,1}_{\FDcoef}|^2 . \label{eq:app:k:constfLk}
\end{align}   
Therefore
\begin{align}
	&\Tr_{\text{R}}(\constonekd\hat{\mathrm{H}}\ketbra{1_{0,e_1d_0,k}}\hat{\mathrm{H}}^\dagger) =  \label{eq:app:k:R11}\\
	&=\constonekd\bra{0_{0,\text{R}}}\hat{\mathrm{H}}\ket{1_{0,e_1d_0,k}}\bra{1_{0,e_1d_0,k}}\hat{\mathrm{H}}^\dagger\ket{0_{0,\text{R}}}\nonumber\\
	&\quad+\constonekd\sum_{\FDcoef=0}^{\infty}\bra{1_{0, \text{R}, \chippackt_\FDcoef}}\hat{\mathrm{H}}\ket{1_{0,e_1d_0,k}}\bra{1_{0,e_1d_0,k}}\hat{\mathrm{H}}^\dagger\ket{1_{0, \text{R}, \chippackt_\FDcoef}}\nonumber\\
	&=\constfLk\ketbra{0_{0, \text{T}}} +	\constdLk\ketbra{1_{0,\text{T},e_1d_0,k}}\nonumber
\end{align}

Using Theorem \ref{thm:filter} and Proposition \ref{prop:filter}, the effect of filter on the two photon state of (\ref{eq:app:oneponed}) is obtained as
\begin{align}
	&(\constplusd)^{\frac{1}{2}}\hat{\mathrm{H}}\ket{(1+1)_{0,e_1d_0}} \label{eq:app:K:filtertphoton} \\
	&\quad= \frac{1}{2} \left(\sum_{\FDcoef_0=0}^{\infty}(D_{\FDcoef_0}\ket{1_{0,\text{T},\chippackt_{\FDcoef_0}}}+F_{\FDcoef_0}\ket{1_{0, \text{R},\chippackt_{\FDcoef_0}}})\right)\nonumber\\
	&\qquad\times \left(\sum_{\FDcoef_1=0}^{\infty}(D^{0,1}_{\FDcoef_1}\ket{1_{0,\text{T},\chippackt_{\FDcoef_1}}}+F^{0,1}_{\FDcoef_1}\ket{1_{0, \text{R},\chippackt_{\FDcoef_1}}})\right).\nonumber
\end{align} 

When we have two photons at the receiver, the basis for partial trace (with nonzero projection value) is $\lbrace \ket{0_{0, \text{R}}},  \ket{1_{0,\text{R}, \chippackt_\FDcoef}}, \ket{2_{0,\text{R},\chippackt_\FDcoef}}, \ket{1_{0,\text{R},\chippackt_{\FDcoef_0}}}\ket{1_{0,\text{R},\chippackt_{\FDcoef_1}}} (\FDcoef_0< \FDcoef_1)\rbrace$.

In order to obtain the quantum signal corresponding to $\hat{\mathrm{H}}\ket{(1+1)_{0,e_1d_0}}$ after the filter we need to compute $\Tr_{\text{R}}(\hat{\mathrm{H}}\ketbra{(1+1)_{0,e_1d_0}}\hat{\mathrm{H}}^\dagger)$.
The projection of $\hat{\mathrm{H}}\ket{(1+1)_{0,e_1d_0}}$ with respect to $\ket{0_{0, \text{R}}}$ is
\begin{align}
	&(\constplusd)^{\frac{1}{2}}\bra{0_{0, \text{R}}}\hat{\mathrm{H}}\ket{(1+1)_{0,e_1d_0}}=\\
	&=\frac{1}{2} \left(\sum_{\FDcoef_0=0}^{\infty}D_{\FDcoef_0}\ket{1_{0,\text{T},\chippackt_{\FDcoef_0}}}\right)\left(\sum_{\FDcoef_1=0}^{\infty}D^{0,1}_{\FDcoef_1}\ket{1_{0,\text{T},\chippackt_{\FDcoef_1}}}\right)\nonumber\\
	&= \frac{1}{2}\underset{\text{s.t. }\FDcoef_1>\FDcoef_0}{\sum_{\FDcoef_1=0}^{\infty}\sum_{\FDcoef_0=0}^{\infty}}(D^{0,1}_{\FDcoef_0}D_{\FDcoef_1}+D_{\FDcoef_0}D^{0,1}_{\FDcoef_1})\ket{1_{0,\text{T},\chippackt_{\FDcoef_0}}}\ket{1_{0,\text{T},\chippackt_{\FDcoef_1}}}\nonumber\\
	&\qquad+\frac{1}{2}\sum_{\FDcoef=0}^{\infty}\sqrt{2}D_{\FDcoef}D^{0,1}_{\FDcoef} \ket{2_{0,\text{T},\chippackt_{\FDcoef}}},
\end{align}
which corresponds to the case where two photons have arrived at receiver $r=0$ and both have passed through the filter. The norm of the above expression is
\begin{align}
	&\constdd\coloneqq\frac{1}{4}\underset{\text{s.t. }\FDcoef_1>\FDcoef_0}{\sum_{\FDcoef_1=0}^{\infty}\sum_{\FDcoef_0=0}^{\infty}}|D^{0,1}_{\FDcoef_0}D_{\FDcoef_1}+D_{\FDcoef_0}D^{0,1}_{\FDcoef_1}|^2\nonumber\\
	&\qquad\qquad\qquad\qquad+\frac{1}{4}2\sum_{\FDcoef_0=0}^{\infty}|D_{\FDcoef_0}D^{0,1}_{\FDcoef_0}|^2 \label{eq:app:k:constdd}
\end{align}
We obtain the normalized transmitted two photon state as
\begin{align}
	&\ket{(1+1)_{0,\text{T}, e_1d_0}} \coloneqq  \frac{1}{2}\left(\constdd\right)^{-\frac{1}{2}}\times  \nonumber\\
	&\left(\sum_{\FDcoef_0=0}^{\infty}D_{\FDcoef_0}\ket{1_{0,\text{T},\chippackt_{\FDcoef_0}}}\right)\left(\sum_{\FDcoef_1=0}^{\infty}D^{0,1}_{\FDcoef_1}\ket{1_{0,\text{T},\chippackt_{\FDcoef_1}}}\right)\\
	&=\left(\constdd\right)^{-\frac{1}{2}}\Big({\sum_{\FDcoef_0=0}^{\infty}|D_{\FDcoef_0}|^2}\Big)^{\frac{1}{2}}\Big({\sum_{\FDcoef_0=0}^{\infty}|D^{0,1}_{\FDcoef_0}|^2}\Big)^{\frac{1}{2}}\nonumber\\
	&\qquad\qquad\qquad\times \ket{1_{0,\wavepackt_\text{T}}}\ket{1_{0,\wavepackt_\text{T}^{e_1d_0}}},
\end{align}
Hence, 
\begin{align}
	&(\constplusd)^{\frac{1}{2}}\bra{0_{0, \text{R}}}\hat{\mathrm{H}}\ket{(1+1)_{0,e_1d_0}}=(\constdd)^{\frac{1}{2}} \ket{(1+1)_{0,\text{T}, e_1d_0}}\label{eq:app:k:R222}
\end{align}
Another term in the expression of the partial trace is the projection of (\ref{eq:app:K:filtertphoton}) with respect to $\ket{1_{0,\text{R}, \chippackt_\FDcoef}}$
\begin{align}
	 &(\constplusd)^{\frac{1}{2}}\bra*{1_{0,\text{R}, \chippackt_\FDcoef}}\hat{\mathrm{H}}\ket{(1+1)_{0,e_1d_0}}=\nonumber\\
	&=\frac{1}{2} F^{0,1}_{\FDcoef}\sum_{\FDcoef_0=0}^{\infty}D_{\FDcoef_0}\ket{1_{0,\text{T},\chippackt_{\FDcoef_0}}}+\frac{1}{2} {F}_{\FDcoef}\sum_{\FDcoef_1=0}^{\infty}D^{0,1}_{\FDcoef_1}\ket{1_{0,\text{T},\chippackt_{\FDcoef_1}}}\nonumber\\
	&=\frac{1}{2} \sum_{\FDcoef_0=0}^{\infty}(F^{0,1}_{\FDcoef}D_{\FDcoef_0}+{F}_{\FDcoef}D^{0,1}_{\FDcoef_0})\ket{1_{0,\text{T},\chippackt_{\FDcoef_0}}}, \label{eq:app:k:R21}
\end{align}
which corresponds to the case where two photons have arrived at receiver $r=0$ and only one has passed through the filter. The norm of the above expression is obtained as
\begin{align}
	\constfkd &\coloneqq\frac{1}{4}\sum_{\FDcoef_0=0}^{\infty}|F^{0,1}_{\FDcoef}D_{\FDcoef_0}+{F}_{\FDcoef}D^{0,1}_{\FDcoef_0}|^2 \label{eq:app:k:constfkd}
\end{align}
Therefore, the normalized state corresponding to $\braket{1_{0,\text{R}, \chippackt_k}}{(1+1)_{0,\text{T}, e_1d_0}}$ can be defined as
\begin{align}
	\ket{1_{0,\text{T},e_1d_0,F_\FDcoef } }&\coloneqq \frac{1}{2}\left(\constfkd\right)^{-\frac{1}{2}}\sum_{\FDcoef_0=0}^{\infty}(F^{0,1}_{\FDcoef}D_{\FDcoef_0}+{F}_{\FDcoef}D^{0,1}_{\FDcoef_0})\ket{1_{0,\text{T},\chippackt_{\FDcoef_0}}}\\
	&=\frac{1}{2}\left(\constfkd\right)^{-\frac{1}{2}}\Bigg[F^{0,1}_{\FDcoef}\Big({\sum_{\FDcoef_0=0}^{\infty}|D_{\FDcoef_0}|^2}\Big)^{\frac{1}{2}}\ket{1_{0,\wavepackt_\text{T}}}\\
	&\qquad\qquad\quad+{F}_{\FDcoef}\Big({\sum_{\FDcoef_0=0}^{\infty}|D^{0,1}_{\FDcoef_0}|^2}\Big)^{\frac{1}{2}}\ket{1_{0,\wavepackt_\text{T}^{e_1d_0}}}\Bigg]\nonumber
\end{align}
Hence, we may write
\begin{align}
	&(\constplusd)^{\frac{1}{2}}\bra{1_{0,\text{R}, \chippackt_\FDcoef}}\hat{\mathrm{H}}\ket{(1+1)_{0,e_1d_0}}\nonumber\\
	&\qquad= \constfkd 	\ketbra{1_{0,\text{T},e_1d_0,F_\FDcoef } }
\end{align}

The final term in the partial trace corresponds to zero photons at the intended receiver due to the loss induced by the filter's frequency mismatch obtained from (\ref{eq:app:K:filtertphoton}) as
\begin{align}
	(\constplusd)^{\frac{1}{2}}&(\bra{1_{0,\text{R},\chippackt_{\FDcoef_1}}}\bra{1_{0,\text{R},\chippackt_{\FDcoef_0}}})\hat{\mathrm{H}}\ket{(1+1)_{0,e_1d_0}}\nonumber\\
	&\qquad=
		\frac{1}{2}(F^{0,1}_{\FDcoef_0}F_{\FDcoef_1}+F_{\FDcoef_0}F^{0,1}_{\FDcoef_1})\ket{0_{0,\text{T}}}, \label{eq:app:k:R211}
\end{align}
and
\begin{align}
	\bra{2_{0,\text{R},\chippackt_\FDcoef}}\hat{\mathrm{H}}\ket{(1+1)_{0,e_1d_0}}&=\frac{1}{2}
		\sqrt{2}F^{0,1}_{\FDcoef}F_{\FDcoef}\ket{0_{0,\text{T}}}, \label{eq:app:k:R22}
\end{align}
which corresponds to the case where two photons have arrived at receiver $r=0$ and none of them has passed through the filter.
For this case, we define the following coefficient, for summation of coefficients of  $\ketbra{0_{0,\text{T}}}$, to simplify the final result.
  \begin{align}
  	&\constff \coloneqq\frac{1}{4}\underset{\text{s.t. }\FDcoef_1>\FDcoef_0}{\sum_{\FDcoef_1=0}^{\infty}\sum_{\FDcoef_0=0}^{\infty}}|F^{0,1}_{\FDcoef_0}F_{\FDcoef_1}+F_{\FDcoef_0}F^{0,1}_{\FDcoef_1}|^2+\frac{1}{4}2\sum_{\FDcoef_0=0}^{\infty}|F_{\FDcoef_0}F^{0,1}_{\FDcoef_0}|^2 \label{eq:app:k:constff}
  \end{align}
Also based on (\ref{eq:app:k:constdLk}), (\ref{eq:app:k:constfLk}) and (\ref{eq:app:k:constfkd}), we define
  \begin{align}
	&\constfL \coloneqq\sum_{k=0}^{N_c-1}\constfLk=\frac{1}{4N_c}\sum_{k=0}^{N_c-1}\sum_{\FDcoef=0}^{\infty}|F_{\FDcoef}-\lambda_{0,k}\lambda_{1,k}F^{0,1}_{\FDcoef}|^2\label{eq:app:k:constfL}\\
	&\constdL \coloneqq\sum_{k=0}^{N_c-1}\constdLk=\frac{1}{4N_c}\sum_{k=0}^{N_c-1}\sum_{\FDcoef=0}^{\infty}|D_{\FDcoef}-\lambda_{0,k}\lambda_{1,k}D^{0,1}_{\FDcoef}|^2\label{eq:app:k:constdL}\\
	&\constfd \coloneqq\sum_{\FDcoef=0}^{\infty}\constfkd =\frac{1}{4}\sum_{\FDcoef=0}^{\infty}\sum_{\FDcoef_0=0}^{\infty}|F^{0,1}_{\FDcoef}D_{\FDcoef_0}+{F}_{\FDcoef}D^{0,1}_{\FDcoef_0}|^2\label{eq:app:k:constfd}
\end{align}

Using the density matrix in (\ref{eq:app:k:densityd2}), and the projections in (\ref{eq:app:k:R11}), (\ref{eq:app:k:R222}), (\ref{eq:app:k:R21}), (\ref{eq:app:k:R211}) and (\ref{eq:app:k:R22}), the transmitted quantum signal after the filter can be described  as
\begin{align}
	\rho_0^{\text{T}} &= \Tr_{\text{R}}(\hat{\mathrm{H}}\rho_0^d\hat{\mathrm{H}}^\dagger)\\
	&= \bra{0_{0, \text{R}}}\hat{\mathrm{H}}\rho_0^d\hat{\mathrm{H}}^\dagger\ket{0_{0, \text{R}}}\nonumber\\ 
	&\quad+ \sum_{\FDcoef=0}^{\infty}\bra{1_{0,\text{R}, \chippackt_\FDcoef}}\hat{\mathrm{H}}\rho_0^d\hat{\mathrm{H}}^\dagger\ket{1_{0,\text{R}, \chippackt_\FDcoef}}\nonumber \\
	&\quad+\sum_{\FDcoef=0}^{\infty}\bra{2_{0,\text{R}, \chippackt_\FDcoef}}\hat{\mathrm{H}}\rho_0^d\hat{\mathrm{H}}^\dagger\ket{2_{0,\text{R}, \chippackt_\FDcoef}} \nonumber\\
	&\quad+\sum_{\FDcoef_0=0}^{\infty}\sum_{\FDcoef_1=0}^{\infty}\bra{1_{0,\text{R},\chippackt_{\FDcoef_1}}}\bra{1_{0,\text{R},\chippackt_{\FDcoef_0}}}\hat{\mathrm{H}}\rho_0^d\hat{\mathrm{H}}^\dagger\ket{1_{0,\text{R},\chippackt_{\FDcoef_0}}}\ket{1_{0,\text{R},\chippackt_{\FDcoef_1}}} \nonumber\\
	&=\constdd\ketbra{(1+1)_{0,\text{T}, e_1d_0}}\\
	&\quad+\sum_{\FDcoef=0}^{\infty}\constfkd  \ketbra{1_{0,\text{T},e_1d_0,F_\FDcoef } }\nonumber\\
	&\quad+ \sum_{k=0}^{N_c-1}\constdLk\ketbra{1_{0,\text{T},e_1d_0,k}}\nonumber\\
	&\quad+\constzeroT\ketbra{0_{0,\text{T}}},\nonumber
\end{align} 
where
\begin{align}
	\constzeroT&\coloneqq \constzerod+ \constff+\constfL
\end{align}


Therefore the photon statistics of the output of receiver ${r=0}$, i.e. ${n}_0$, are calculated as follows
\begin{align}
	\mathbb{P}&({n}_0=0) =\constzerod+ \constff+\constfL\label{eq:app:k:probs}\\
	\mathbb{P}&({n}_0=1) =\constfd + \constdL\nonumber\\
	\mathbb{P}&({n}_0=2) =\constdd\nonumber
\end{align}

Let us further simplify the above expressions in terms of the signal and interference amplitudes defined by:
\begin{align}
	\mathfrak{D} &\coloneqq \sum_{\FDcoef=0}^{\infty}|{D}_{\FDcoef}|^2\\
	\mathfrak{D}^{0,1} &\coloneqq \sum_{\FDcoef=0}^{\infty}|D^{0,1}_{\FDcoef}|^2
\end{align}

We use the narrow-band filter approximation and the approximate orthogonality of the spreading sequences, i.e. $\frac{1}{N_c}\sum_{k=0}^{N_c-1} \lambda_{0,k}\lambda_{1,k}\approx 0$, $\sum_{\FDcoef=0}^{\infty}F^{0,1}_{\FDcoef}{F}_{\FDcoef}\approx 0 $ and $\sum_{\FDcoef=0}^{\infty}D^{0,1}_{\FDcoef}{D}_{\FDcoef}\approx 0$ (due to the random phase of the coefficients), to simplify the results.

Note that from (\ref{eq:app:k:constff})
\begin{align}
	\constff&=\frac{1}{4}\left(\sum_{\FDcoef_0=0}^{\infty}|F^{0,1}_{\FDcoef_0}|^2\right)\left(\sum_{\FDcoef_1=0}^{\infty}|F_{\FDcoef_1}|^2\right) +\frac{1}{4} \left|\sum_{\FDcoef=0}^{\infty}F^{0,1}_{\FDcoef}{F}_{\FDcoef}\right|^2\\
	& = \frac{1}{4}(1-\mathfrak{D})(1-\mathfrak{D}^{0,1}) + \frac{1}{4}\left|\sum_{\FDcoef=0}^{\infty}F^{0,1}_{\FDcoef}{F}_{\FDcoef}\right|^2\\
	&\approx \frac{1}{4}(1-\mathfrak{D})(1-\mathfrak{D}^{0,1})
\end{align}
Similarly from (\ref{eq:app:k:constdd}),
\begin{align}
	\constdd&=\frac{1}{4}\left(\sum_{\FDcoef_0=0}^{\infty}|D^{0,1}_{\FDcoef_0}|^2\right)\left(\sum_{\FDcoef_1=0}^{\infty}|D_{\FDcoef_1}|^2\right) + \frac{1}{4}\left|\sum_{\FDcoef=0}^{\infty}D^{0,1}_{\FDcoef}{D}_{\FDcoef}\right|^2\\
	&=\frac{1}{4}\mathfrak{D}\mathfrak{D}^{0,1} + \frac{1}{4}\left|\sum_{\FDcoef=0}^{\infty}D^{0,1}_{\FDcoef}{D}_{\FDcoef}\right|^2\\
	&\approx \frac{1}{4}\mathfrak{D}\mathfrak{D}^{0,1}
\end{align}
Furthermore from (\ref{eq:app:k:constfL}),
\begin{align}
	&\constfL=\frac{1}{4N_c}\sum_{k=0}^{N_c-1}\sum_{\FDcoef=0}^{\infty}|F_{\FDcoef}-\lambda_{0,k}\lambda_{1,k}F^{0,1}_{\FDcoef}|^2 \nonumber\\
	= &\frac{1}{4N_c}\sum_{k=0}^{N_c-1}\sum_{\FDcoef=0}^{\infty}(|F_{\FDcoef}^{0,1}|^2+|F_{\FDcoef}|^2-2\lambda_{0,k}\lambda_{1,k}\Re{F_{\FDcoef}F_{\FDcoef}^{0,1}})\nonumber\\
	=&\frac{1}{4N_c}\sum_{k=0}^{N_c-1}\left(2-\sum_{\FDcoef=0}^{\infty}(|D_{\FDcoef}^{0,1}|^2+|D_{\FDcoef}|^2)\right)\nonumber\\
	&\quad  -\frac{1}{2N_c}\left(\sum_{k=0}^{N_c-1} \lambda_{0,k}\lambda_{1,k}\right)\Re{\sum_{\FDcoef=0}^{\infty} F_{\FDcoef}F_{\FDcoef}^{0,1}}\\
	\approx& \frac{1}{2}-\frac{1}{4}\sum_{\FDcoef=0}^{\infty}(|D_{\FDcoef}^{0,1}|^2+|D_{\FDcoef}|^2)\nonumber\\
	=&\frac{1}{2}-\frac{1}{4}(\mathfrak{D}+\mathfrak{D}^{0,1})
\end{align}
Similarly from (\ref{eq:app:k:constdL}),
\begin{align}
	&\constdL=\frac{1}{4N_c}\sum_{k=0}^{N_c-1}\sum_{\FDcoef=0}^{\infty} |D_{\FDcoef}-\lambda_{0,{k}}\lambda_{1,{k}}D^{0,1}_{\FDcoef}|^2 \nonumber\\
	=&\frac{1}{4N_c}\sum_{k=0}^{N_c-1}\sum_{\FDcoef=0}^{\infty}(|D_{\FDcoef}^{0,1}|^2+|D_{\FDcoef}|^2-2\lambda_{0,k}\lambda_{1,k}\Re{D_{\FDcoef}D_{\FDcoef}^{0,1}})\nonumber\\
	\approx& \frac{1}{4}\sum_{\FDcoef=0}^{\infty}(|D_{\FDcoef}^{0,1}|^2+|D_{\FDcoef}|^2)\nonumber\\
	=&\frac{1}{4}(\mathfrak{D}^{0,1}+\mathfrak{D})
\end{align}
Also from (\ref{eq:app:k:constfd}),
\begin{align}
\constfd&=\frac{1}{4}\sum_{\FDcoef_0=0}^{\infty}\sum_{\FDcoef_1=0}^{\infty}|F^{0,1}_{\FDcoef_0}{D}_{\FDcoef_1}+F_{\FDcoef_0}D^{0,1}_{\FDcoef_1}|^2 \nonumber\\
	&=\frac{1}{4}\left(\sum_{\FDcoef_1=0}^{\infty}|{D}_{\FDcoef_1}|^2\right)\left(\sum_{\FDcoef_0=0}^{\infty}|F^{0,1}_{\FDcoef_0}|^2\right)\nonumber\\
	&\ +\frac{1}{4}\left(\sum_{\FDcoef_0=0}^{\infty}|{F}_{\FDcoef_0}|^2\right)\left(\sum_{\FDcoef_1=0}^{\infty}|D^{0,1}_{\FDcoef_1}|^2\right)\nonumber\\
	&\ +\frac{1}{2}\Re{\left(\sum_{\FDcoef_0=0}^{\infty} F_{\FDcoef_0}F^{0,1}_{\FDcoef_0}\right)\left(\sum_{\FDcoef_1=0}^{\infty}{D}_{\FDcoef_1}D^{0,1}_{\FDcoef_1}\right)}\nonumber\\
	&\approx \frac{1}{4}\left(\sum_{\FDcoef_1=0}^{\infty}|{D}_{\FDcoef_1}|^2\right)\left(1-\sum_{\FDcoef_1=0}^{\infty}|D^{0,1}_{\FDcoef_1}|^2\right)\nonumber\\
	&\ +\frac{1}{4}\left(1-\sum_{\FDcoef_1=0}^{\infty}|{D}_{\FDcoef_1}|^2\right)\left(\sum_{\FDcoef_1=0}^{\infty}|D^{0,1}_{\FDcoef_1}|^2\right)\nonumber\\
	&=\frac{1}{4}\left(\mathfrak{D}(1-\mathfrak{D}^{0,1})+\mathfrak{D}^{0,1}(1-\mathfrak{D})\right)\nonumber\\
	&=\frac{1}{4}(\mathfrak{D}^{0,1}+\mathfrak{D}-2\mathfrak{D}\mathfrak{D}^{0,1})
\end{align}
Substituting approximate values of the coefficients in (\ref{eq:app:k:probs}) gives
\begin{align}
	\mathbb{P}({n}_0=0) &\approx \frac{1}{4} + \frac{1}{2}-\frac{1}{4}(\mathfrak{D}+\mathfrak{D}^{0,1}) +  \frac{1}{4}(1-\mathfrak{D})(1-\mathfrak{D}^{0,1}) \nonumber\\
	&=1-\frac{1}{2}(\mathfrak{D}+\mathfrak{D}^{0,1})+\frac{1}{4}\mathfrak{D}^{0,1}\mathfrak{D}\\
	\mathbb{P}({n}_0=1)& \approx \frac{1}{4}(\mathfrak{D}^{0,1}+\mathfrak{D}-2\mathfrak{D}\mathfrak{D}^{0,1})+ \frac{1}{4}(\mathfrak{D}^{0,1}+\mathfrak{D})\nonumber\\
	&=\frac{1}{2}(\mathfrak{D}^{0,1}+\mathfrak{D})-\frac{1}{2}\mathfrak{D}\mathfrak{D}^{0,1}\\
	\mathbb{P}({n}_0=2) &\approx \frac{1}{4}\mathfrak{D}\mathfrak{D}^{0,1}
\end{align}

\section{Tensor Product Notation}
\label{appendix:tensor}
We use the symbol $\prod$ for both the tensor-product of quantum states and ordinary product of operators. Since some authors use the symbol $\otimes$, it is beneficial to discuss this notation. For complex numbers,  the symbol $\prod$ denotes the ordinary product. For example $$ \prod_{k=0}^{N_c-1} \lambda_k = \lambda_0 \lambda_1 \cdots \lambda_{N_c-1}.$$ For quantum states, the tensor product of vectors $\ket{\psi_{\chippackt_k}}\in\mathcal{H}_k$, i.e. $$\otimes_{k=0}^{N_c-1} \ket{\psi_{\chippackt_k}} = \prod_{k=0}^{N_c-1} \ket{\psi_{\chippackt_k}},$$ results in a vector in the Hilbert space $\mathcal{H}$ with higher dimensions.  
Also note that, as a convention, vacuum states are often not shown in the tensor product state \cite{beck2012quantum}. This convention, for example, implies that:
\begin{align*}
	\ket{1_{\xi_k}}\coloneqq&\ket{0_{\xi_0}}\ket{0_{\xi_1}}\cdots\ket{1_{\xi_k}}\cdots \ket{0_{\xi_{N_c-1}}}\\
	=&	\ket{1_{\xi_k}} \prod_{l\ne k} \ket{0_{\xi_l}}
\end{align*}

For quantum operators, the tensor product is utilized to bring an operator that acts on states inside a chip-time Hilbert space $\mathcal{H}_k$ into the larger Hilbert space $\mathcal{H}$. As an example of quantum operators, note that the chip-time interval spreading operator $\hat{\mathrm{U}}_k$ is defined in $\mathcal{H}$ but its action only affects quantum signals with components inside  $\mathcal{H}_k$ and corresponds to unity for other chip-times. 
 Then we have $$\hat{\mathrm{U}}_k = \underbrace{I\otimes I \otimes\cdots \otimes I\otimes}_{k}\hat{\mathrm{U}}'_k \underbrace{\otimes I\otimes I \otimes\cdots \otimes I }_{N_c-k-1},$$ where $\hat{\mathrm{U}}'_k$ is represented in the chip-time Hilbert space $\mathcal{H}_k$, and we have 
\begin{align*}
\prod_{k=0}^{N_c-1}\hat{\mathrm{U}}_k &= \prod_{k=0}^{N_c-1} \underbrace{I\otimes I \otimes\cdots \otimes I\otimes}_{k}\hat{\mathrm{U}}'_k \underbrace{\otimes I\otimes I \otimes\cdots \otimes I }_{N_c-k-1} \\
&= \hat{\mathrm{U}}'_0 \otimes \hat{\mathrm{U}}'_1 \otimes \cdots \hat{\mathrm{U}}'_k \otimes  \cdots \hat{\mathrm{U}}'_{N_c-1}
 \end{align*}

\bibliographystyle{IEEEtran}
\bibliography{IEEEabrv,resources}
%

\begin{IEEEbiography}[{\includegraphics[width=1in,height=1.25in,clip,keepaspectratio]{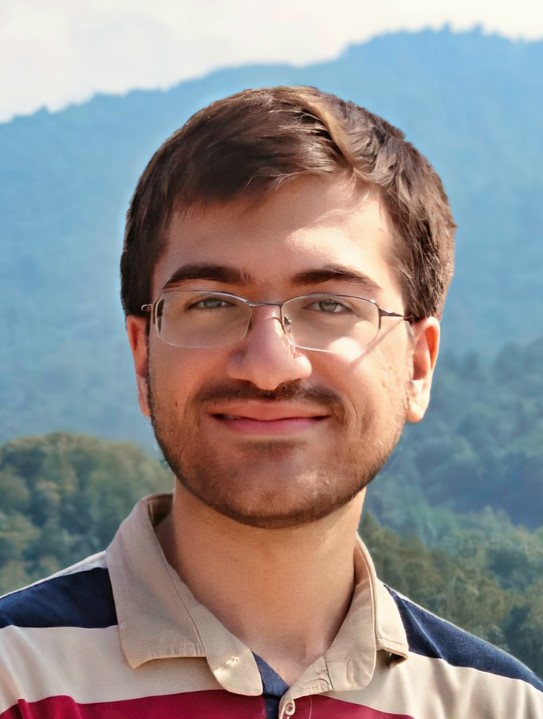}}]{Mohammad Amir Dastgheib}
	was born in Shiraz, Iran, in 1994. b received his B.Sc.
	from Shiraz University in 2016, and his M.Sc. and
	Ph.D. from Sharif University of Technology (SUT)
	in 2018 and 2023, all in Electrical Engineering
	with honors and as an outstanding student. Since 2016, he has been a member of the Optical Networks Research Lab (ONRL) at SUT. He is currently a postdoctoral researcher at the Sharif Quantum Center (SQC) and a Member of the Technical Staff of the Advanced Quantum Communications Lab since 2021. 
    
    He received National Khwarizmi Youth Award in mathematics in 2011, the Award of Research Endeavors among the electrical engineering department's M.Sc. students from the Sharif University of Technology in 2018, and the 2020 Outstanding M.Sc. Thesis Award from the IEEE Iran section. 
    
    His current research interests include quantum communications signals and systems, quantum communication networks, optical wireless communications, statistical learning, and machine learning applications in communication networks.
\end{IEEEbiography}

\begin{IEEEbiography}[{\includegraphics[width=1in,height=1.25in,clip,keepaspectratio]{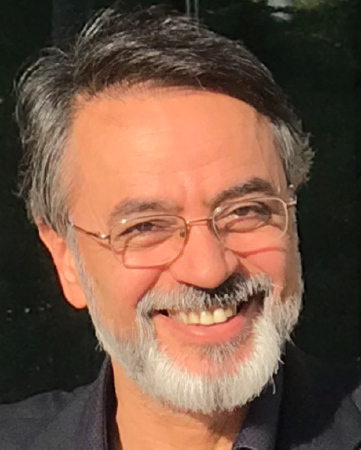}}]{Jawad A. Salehi}
	(M'84-SM'07-F'11) was born in Kazemain, Iraq, in 1956. He received the
	B.Sc.degree from the University of California at
	Irvine in 1979, and the M.Sc. and Ph.D. degrees
	in electrical engineering from the University of
	Southern California (USC), in 1980 and 1984,
	respectively. From 1984 to 1993, he was a Member
	of the Technical Staff of the Applied Research
	Area, Bell Communications Research (Bellcore),
	Morristown, New Jersey. In 1990, he was with the
	Laboratory of Information and Decision Systems, Massachusetts Institute of Technology, as a visiting research scientist conducting research on optical multiple-access networks. He was an Associate Professor from 1997 to
	2003 and currently he is a Distinguished Professor with the Department
	of Electrical Engineering (EE), Sharif University of Technology (SUT),
	Tehran, Iran.

	From 2003 to 2006, he was the Director of the National Center of
	Excellence in Communications Science at the EE department of SUT. In
	2003, he founded and directed the Optical Networks Research Laboratory
	for advanced theoretical and experimental research in futuristic all-optical networks. Currently, he is the Head of Sharif Quantum Center and the quantum group of the Institute for Convergence Science \& Technology emphasizing in advancing quantum communication systems, quantum optical
	signal processing and quantum information science.

	His current research interests include quantum optics, quantum communications signals and systems, quantum CDMA, quantum Fourier optics, and optical wireless communication (indoors and underwater). He is the holder of 12 U.S. patents on optical CDMA.

	Dr. Salehi was named as among the 250 preeminent and most influential
	researchers worldwide by the Institute for Scientific Information Highly Cited in the Computer-Science Category, 2003. He is a recipient of the
	Bellcore's Award of Excellence, the Outstanding Research Award of the EE Department of SUT in 2002 and 2003, the Nationwide Outstanding Research
	Award 2003, and the Nation's Highly Cited Researcher Award 2004.
	From 2001 to 2012, he was an Associate Editor of the Optical CDMA of
	the IEEE TRANSACTIONS ON COMMUNICATIONS. Professor Salehi is
	a member of the Iran Academy of Science and a Fellow of the Islamic World Academy of Science, Amman, Jordan.
\end{IEEEbiography}


\begin{IEEEbiography}[{\includegraphics[width=1in,height=1.25in,clip,keepaspectratio]{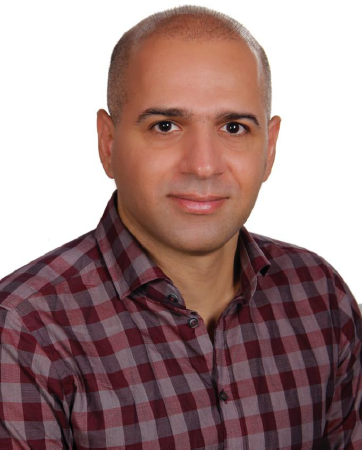}}]{Mohammad Rezai}
	was born in Firoozabad,
	Iran in 1983. He received the B.S. degree from the
	University of Sistan and Baluchestan in 2006, the
	M.S. degree in physics from the Sharif University
	of Technology in 2009, and the Ph.D. degree in
	physics from the University of Stuttgart, Germany,
	in 2018. From 2010 to 2013, he was a member
	of the International Max Planck Research School
	for Advanced Materials and a member of research
	staff in the field of condensed matter physics in
	the Institute for Theoretical Physics III, University of Stuttgart, Germany.

	In 2013 he joined the 3rd Physikalisches Institut, University of Stuttgart,	Germany, where he engaged in optical quantum information processing experiments. From 2019 to 2022, he was a postdoctoral researcher with Sharif Quantum Center and Electrical Engineering Department, Sharif University of Technology, Tehran, Iran.
	Since 2022, he has been the Head of Research Planning Department of
	Sharif Quantum Center and a faculty member of the Institute for Convergence Science \& Technology, Sharif University of Technology, Tehran, Iran.

	His current research interests include quantum holography, quantum Fourier optics, quantum multiple access communication systems and quantum coherence in photosynthetic systems.
	Dr. Rezai was elected to the Iran National Elite Foundation in 2019 and a recipient of the Max Planck scholarship in 2010.
\end{IEEEbiography}




\end{document}